\documentclass[english]{revtex4-1}
\usepackage[T1]{fontenc}
\usepackage[latin9]{inputenc}
\usepackage{color}
\usepackage{array}
\usepackage{float}
\usepackage{textcomp}
\usepackage{multirow}
\usepackage{amstext}
\usepackage{amssymb}
\usepackage{graphicx}

\makeatletter

\providecommand{\tabularnewline}{\\}

\makeatother

\usepackage{babel}
\begin{document}

\title{Pressure Induced Thermodynamically Stable and Mechanically Robust
Li-rich Unknown Li-Sn Compounds: A Step Towards Improvement of Li-Sn
Batteries}

\author{Raja Sen}

\author{Priya Johari}
\email{priya.johari@snu.edu.in, psony11@gmail.com}

\selectlanguage{english}%

\affiliation{Department of Physics, School of Natural Sciences, Shiv Nadar University,
Greater Noida, Gautam Budhha Nagar, UP 201 314, India.}
\begin{abstract}
Volume expansion and elastic softening of Sn anode on lithiation result
in mechanical degradation and pulverization of Sn, affecting the overall
performance of Li-Sn batteries. It can however be overcome by using
exotic high pressure quenched phase as prelithiated reagent. Moreover,
it is known that under pressure many unusual stoichiometric which
are basically impossible at ambient pressure, can be synthesized,
that may even survive the decompression from high to ambient pressure.
We therefore have performed a comprehensive study using evolutionary
algorithm and density functional theory based simulations to understand
the lithiation of Sn anode at pressure ranging from 1 atm to 20 GPa.
The ground state structures of all stable and metastable Li-Sn compounds
have been identified at ambient and moderate pressures and their properties
have been studied to understand the role of pressure in re-defining
the reaction mechanism during charging-discharging process in Li-ion
batteries. Besides the well-known existing Li-Sn compounds, our studies
reveal the existence of five unreported stoichiometries ($\mathrm{Li_{8}Sn_{3}}$,
$\mathrm{Li_{3}Sn_{1}}$, $\mathrm{Li_{4}Sn_{1}}$, $\mathrm{Li_{5}Sn_{1}}$,
and $\mathrm{Li_{7}Sn_{1}}$) and their associated crystal structures
at ambient and high pressure. While $\mathrm{Li_{8}Sn_{3}}$ has been
identified as one of the most stable Li-Sn compound in the entire
pressure range (1 atm\textendash 20 GPa), the pressure induced Li-rich
compounds like $\mathrm{Li_{5}Sn_{1}}$ and $\mathrm{Li_{7}Sn_{1}}$
have been classified as providing higher theoretical gravimetric capacity
of 1129 and 1580 mA h g$^{-1}$, respectively, than the capacity of
the known most lithiated phase, i.e., $\mathrm{Li_{17}Sn_{4}}$ (960
mA h g$^{-1}$). Most importantly, our calculations show reduction
in volume expansion by $\sim$50\% at 20 GPa, and reveal that the
application of pressure can reduce the chance of Li plating and improve
the mechanical properties, which are desired to make the battery safer
and its life longer.
\end{abstract}

\maketitle

\section{Introduction}

\textcolor{black}{In recent years, rechargeable lithium-ion batteries
(LIBs) have become a robust candidate by playing crucial role in the
field of energy conversion and storage systems. The superior properties
of LIBs such as high energy density, longer lifespan, lower toxicity,
better safety, }and design flexibility, have helped them to distinguish
from their competitors such as nickel-cadmium and lead acid batteries,
and establish them as the prominent choice for the portable electronics
devices.\cite{Nitta_2015,Tarascon_2001,Etacheri_2011} However, to
expand the horizon of their use in green transportations (\textit{e.g.};
electric vehicles, hybrid electric vehicles, and plug-in hybrid vehicles)
and smart grid applications, a further optimization of the energy
and power density, with an improved cycle life is demanded. This has
initiated the search for the next generation electrode materials that
can enhance the efficiency as well as durability of the LIBs.\cite{Etacheri_2011,Kang_2006,Armand_2008}
As per current scenario, the majority of commercial LIBs employ graphitic
anode for their outstanding cyclic performance, relative low expenditure,
and stable electrochemical properties.\cite{Nitta_2015,Tarascon_2001}
But, low specific capacity of graphite ($372$ mA h g$^{-1}$) greatly
restricts the potential of LIBs in the applications seeking high capacity
and power. The low capacity of graphite anode is primarily due to
the geometric restriction of graphite\textquoteright s structure in
which six carbon can accommodate only one Li-ion to form the intercalated
compound, LiC$_{6}$.\cite{Nitta_2015,Tarascon_2001,Whittingham_2014}
To improve the capacity, metals and semi-metals that can electrochemically
form alloys with lithium have therefore been extensively investigated
in the recent years, to move from the era of intercalation to integration
chemistry, where compounds are built by formation/cleavage of covalent
bonds during charge-discharge process.\cite{Park_csr_2010,Zhang_2011,Obrovac_2014,Tian_2015,Chou_2011,Priya_2011}

In search of new anode materials, Sn and Sn based composites have
received significant attention because of their higher capacity (990
mAh g$^{-1}$ for $\mathrm{\mathit{\mathrm{Li_{4.4}Sn}}}$) and abundant
availability in nature.\cite{Park_csr_2010,Chou_2011} Although the
neighbors of Sn in group IV, \textit{i.e.}, Si and Ge, are also well
known for their extraordinary theoretical capacities (4200 mAh g$^{-1}$
for $\mathrm{Li_{4.4}Si}$ and 1600 mAh g$^{-1}$ for $\mathrm{Li_{4.4}Ge}$),\cite{Park_csr_2010,Chou_2011}
but several studies reveal that Sn would be a better choice because
of following reasons: (i) Owing to the large interstitial space in
Sn, Li diffusion is more favorable in Sn as compared to Si and Ge;\cite{Chou_2011}
(ii) Li-Sn compounds are enthalpically more stable, followed by Li-Ge
and Li-Si compounds;\cite{Chou_2011} (iii) \textcolor{black}{Sn is
ductile in nature while Si and Ge are brittle, which makes the mechanical
integrity of Sn better than the preceding group IV elements};\cite{Larcher@2007,Idota@1997}
(iv) Being a metal, electrical conductivity of Sn is better than Si
and Ge,\cite{Na@2010} \textit{etc}. Besides having several advantages,
the application of Sn as an anode is still far from commercialization
because of huge irreversible capacity loss, particle fracture, and
electrochemical pulverization due to drastic variation in volume during
lithiation-delithiation process (\textasciitilde{} 257\%).\cite{Zhang_2011,Chou_2011,Tian_2015,Zeng_2015}
The microscopic mechanisms underlying these phenomena are still not
entirely understood. Thus, in order to gain a deep insight into the
\textcolor{black}{failure mechanism of a Li-Sn battery during charging
and discharging, and to find a possible way to improve it's mechanical
properties, it is required to understand the process at the atomistic
level. This can be done by studying }the variation in the atomic structure
and properties of Sn anode (``fingerprint'') during lithiation process,
in detail.

The Li-Sn binary system has a very rich phase diagram. In an increasing
order of lithium content, the following phases are experimentally
reported: $\mathrm{Li_{2}Sn_{5}}$ (\emph{$\mathrm{P4/mbm}$},\emph{$\mathrm{Z=2}$}),\cite{Robert_2007,Dunlap_1999,Li2Sn5}
$\mathrm{Li_{1}Sn_{1}}$ \{(\emph{$\mathrm{P2/m}$},\emph{$\mathrm{Z=3}$}),
($\mathrm{I4_{1}/amd}$,$\mathrm{Z=12}$)\},\cite{Robert_2007,Dunlap_1999,alpha_LiSn,beta_LiSn}
$\mathrm{Li_{7}Sn_{3}}$ ($\mathrm{P2_{1}/m}$,$\mathrm{Z=2}$),\cite{Robert_2007,Dunlap_1999,Li7Sn3}
$\mathrm{Li_{5}Sn_{2}}$ ($\mathrm{R\overline{3}m}$,$\mathrm{Z=2}$),\cite{Robert_2007,Dunlap_1999,Li5Sn2}
$\mathrm{Li_{13}Sn_{5}}$ ($\mathrm{P\overline{3}m1}$, $\mathrm{Z=}1$),\cite{Robert_2007,Dunlap_1999,Li13Sn5}
$\mathrm{Li_{7}Sn_{2}}$ ($\mathrm{Cmmm}$ ,$\mathrm{Z=}2$),\cite{Robert_2007,Dunlap_1999,Li7Sn2}
$\mathrm{Li_{17}Sn_{4}}$($\mathrm{F\overline{4}3m}$, $\mathrm{Z=20}$).\cite{Robert_2007,Dunlap_1999,Li17Sn4_Gowar,Li17Sn4_Lipu}
The crystal structure of the most lithiated phase $\mathrm{Li_{17}Sn_{4}}$,
was previously thought to be of stoichiometry $\mathrm{Li_{22}Sn_{5}}$,\cite{Li22Sn5}
but this has been corrected by Goward et al. and Lupu et al\cite{Li17Sn4_Gowar,Li17Sn4_Lipu}
in the beginning of this century. The current phase diagram of Li-Sn
systems has been made at atmospheric pressure and high temperature
and there still lies a large uncertainty in the Li-rich part of Li-Sn
phase diagram at high pressure. While, it has already been demonstrated
that the structure and properties of both end members of Li-Sn systems
can largely be modified with the use of pressure.\cite{Matsuoka_2009,Vnuk_1984}
Thus, pressure can be used as an extra variable for engineering the
structure and properties of Li-Sn compounds. Apart from this, there
remains other significant advantages to study the Li-Sn phase diagram
at moderately high pressure for high performance LIBs, such as: (i)
Pressure can significantly diminish the melting point of materials
and enrich the the structural chemistry of compounds. Under pressure
many unusual stoichiometric which are basically impossible at ambient
pressure, can be synthesized with very exotic chemical behavior,\cite{Zhang_2013,Stearns@2003,Zhang@2016}
which may even survive the decompression from high to ambient pressure;
(ii) \textcolor{black}{High pressure phases when quenched at ambient
pressure, often possess superior mechanical properties};\cite{Zeng_2015}
(iii) Quenched high pressure phases can directly be used as pre-lithiated
anode materials for void space manipulation during lithiation-delithiation
process and hence, for providing the better performance of anode in
LIBs.\cite{Zhao_2014,Cloud_2014}

In past few years, with the evolution of powerful computational algorithms
and methods such as evolutionary algorithm, random sampling, minima
hopping etc., stable structures of several unknown and novel materials
at ambient and high pressure conditions have been discovered, which
have also been later confirmed by experiments.\cite{Oganov_2006,Oganov_2011,Lyakhov_2013,Pickard_2011,Goedecker_2004}
Very recently, such studies have been performed for studying Li-Si
and Li-Ge compounds at ambient conditions.\cite{Tipton_2013,Morris_2014,Valencia-Jaime_2016}
These studies confirmed the existence of experimentally reported phases
of the system, and at the same time enriched the phase diagram of
both Li-Si and Li-Ge with the discovery of several new stable and
metastable structures. Furthermore, Zeng et al.\cite{Zeng_2015} predicted
and synthesized a new phase of $\mathrm{Li_{15}Si_{4}}$ ($\beta$-phase)
at 7 GPa, while very recently, Zhang et al.\cite{Zhang@2016} also
revealed the possibility of existence of several Li-rich Li-Si compounds
at high pressure. However, to the best of our knowledge, no such study
has been done to date on the Li-Sn system, which is one of the potential
candidate for the LIBs. Therefore, in current work, we aim to investigate
the Li-Sn binary phase diagram at ambient and moderate high pressure
(up to 20 GPa) by performing an extensive search on the Li-Sn systems
using the evolutionary algorithm code USPEX, in conjunction with first-principles
density functional theory (DFT) based calculations.\cite{Oganov_2006,Oganov_2011,Lyakhov_2013}
Besides the experimentally reported well known structures, our study
reveals several new stable and metastable compounds of Li-Sn with
quite diverse and unusual crystal structures. This includes prediction
of one of the most stable compound, i.e., $\mathrm{Li_{8}Sn_{3}}$
at ambient pressure, and several Li-rich metastable and stable phases
at ambient and high pressure, respectively. It is noteworthy to mention
that possibly not all but several low-energy metastable compound can
be synthesized and they can exhibit superior properties than their
corresponding stable phases.\cite{Badding@1995,Sun@2016} Furthermore,
in our quest for a better understanding of the Li-Sn cell mechanism,
we have also calculated electrochemical, mechanical, and electronic
properties of all stable and metastable Li-Sn compounds. We believe
that our study will provide a basis for future experimental work.

\section{METHODS}

We used evolutionary algorithm based technique as implemented in the
USPEX code,\cite{Oganov_2006,Oganov_2011,Lyakhov_2013} together with
DFT to find stable and metastable $\mathrm{Li-Sn}$ compounds as well
as their ground state structures, at ambient and high pressure. USPEX
has already been demonstrated as a powerful tool to determine the
lowest energy structure \textit{via} global minimization of t\textcolor{black}{he
surface fre}e-energy, with great success.\cite{Zhang_2013,Zeng_2015}
Our calculations to search for stable and metastable structures of
Li-Sn compounds were carried out using USPEX in two steps. Firstly,
most promising compositions of $\mathrm{Li-Sn}$ compounds were explored
through variable composition method at 1 atm, 5, 10, and 20 GPa pressure
considering up to 40 atoms per unit cell. The calculations were carried
out over the course of 50-60 generations where, the first generation
was blossomed randomly with 150-200 structures by sampling 20-25 different
compositions. For the subsequent generations, 40 child structures
were produced, through different symmetry generators, namely heredity,
transmutation, softmutation, and random, with probabilities of 40\%,
20\%, 20\%, and 20\%, respectively. In second step, a fixed composition
search of each ambiguous stoichiometry of $\mathrm{Li-Sn}$ compounds
with different number of ``cell formula'' units (depending upon
the stoichiometry, it varies from 1 to 5) was carried out for over
30-40 generations, with 30-35 different structures in each generation.
The population of first generation in fixed composition search was
also set with 80-120 structures, in order to densely map the configuration
space of random search. Negative enthalpy of formation which is the
basic criterion for finding the energetically stable compounds was
calculated \textit{via} below equation:

\begin{equation}
\mathrm{\triangle H_{_{f}}(Li_{x}Sn_{y})}=\mathrm{\frac{[H(Li_{x}Sn_{y})-xH(Li)-yH(Sn)]}{(x+y)}}\label{eq:1}
\end{equation}

where, $\mathrm{H=U+PV}$ is the enthalpy of each compositions in
their crystal form and $\mathrm{\triangle H_{_{f}}}$ is the formation
enthalpy of the compound per atom. In the expression of enthalpy,
$\textrm{U}$, $\textrm{P}$, and $\mathrm{V}$ represent the internal
energy, pressure, and volume, respectively.

It should be noted that for a given stoichiometry and pressure, the
structure having lowest negative value of formation enthalpy can be
considered as most favorable phase as compared to others. But, in
order to judge the thermodynamic stability of a particular $\mathrm{Li-Sn}$
compound with respect to two nearest neighbor compositions and/or
pure elements at ambient or high pressure, it is important to draw
the thermodynamics convex hull. The thermodynamics convex hull which,
is a representater of compound's formation enthalpy with its respective
stoichiometric ratio at given pressure, is also able to elucidate
the all possible decomposition routes. Basically, a phase can be identified
as thermodynamically stable ground state phase if it lies on the convex
hull. Hence, the stoichiometries that do not have a representative
structure on the convex hull, are considered as either metastable
or unstable. However, in order to govern the local (meta)stability,
we further investigated the phonon dispersion curves of all respective
$\mathrm{Li-Sn}$ compounds. A structure can be considered as dynamically
stable, if and only if, no imaginary phonon frequencies are detected
through out the Brillouin zone in phonon dispersion curve. In principle,
a metastable compound can also be synthesized.\cite{Badding@1995,Sun@2016}

The structure relaxation of all compositions generated by USPEX were
accomplished using first-principles density functional theory (DFT)
as implemented in the Vienna Ab-initio Simulation Package (VASP).\cite{vasp-1,vasp-2}
A plane-wave basis set was employed to expand the valence electronic
wave functions, while the projector augmented-wave (PAW) type pseudo-potentials
were considered to account the interactions with nuclei and core electrons.\cite{PAW2}
For electron-electron exchange and correlation interactions, the functional
of Perdew-Burke-Ernzerhof (PBE), a form of the generalized gradient
approximation (GGA) was used, in the current work.\cite{PBE} The
plane-wave kinetic energy cutoff was considered as $700$ eV, while
the reciprocal space resolution for \textit{k}-points generation in
final structures relaxation and enthalpy calculations was set to $0.02\text{\ensuremath{\times}}2\Pi$
$\mathrm{\mathring{A}}^{-1}$ with uniform $\Gamma-$ centered meshes.
All the above mentioned parameters ensure that the enthalpy calculations
are well converged with fluctuation in enthalpy to be less than 1
meV/atom. In order to probe the dynamical stability, phonon frequencies
throughout the whole Brillouin Zone were calculated using density
functional perturbation theory (DFPT) as implemented in the VASP code
along with the Phonopy package.\cite{Phonopy} Depending upon the
primitive cells structure, $l\times m\times n$ supercells (where,
$\mathit{\mathrm{\mathrm{2}\leq\mathit{l,m,n}\leq}\mathrm{3}}$) were
used to calculate the phonon dispersion curve.

In order to analyse electrochemical properties of Li-Sn compounds
with lithiation at different pressure value, we calculated the Li-insertion
voltage. For this, we considered only the stable compositions of $\textrm{Li-Sn}$
that lie on the convex hull tie line at a given pressure, and assumed
that the reaction proceeds from pure $\mathrm{Sn}$ to subsequent
Li-rich phases by exchange of Li atoms only. Therefore, for a general
reaction $\mathrm{Li_{x_{0}}Sn+(x_{1}-x_{0})Li\rightarrow Li_{x_{1}}Sn}$,
the cell voltage between these two compositions was computed using
a well-established formula:\cite{Ceder@1997} 

\begin{equation}
\mathrm{V=\frac{G(Li_{x_{1}}Sn)-G(Li_{x_{0}}Sn)-(x_{1}-x_{0})G(Li)}{(x_{1}-x_{0})}}\label{eq:2}
\end{equation}

where, $\mathrm{G(Li_{x}Sn)}$ and $\mathrm{G(Li})$ are the Gibbs
frees energy of $\mathrm{Li_{x}Sn}$ per $\mathrm{Sn}$ atom and the
Gibbs frees energy of $\mathrm{Li}$ per $\mathrm{Li}$ atom, respectively.
Here, we neglected the contribution of zero-point energy and lattice
vibrational energy to the Gibbs free energy as it is relatively very
small and further reduces when used in the above formula. Thus, we
approximated $\mathrm{G}$ with the enthalpy of each composition.

In addition, the theoretical gravimetric capacity, $\mathrm{GC}$,
was also determined using following equation:

\begin{equation}
\mathrm{\mathrm{GC}=\frac{xF}{M_{Sn}}}\label{eq:3}
\end{equation}

where, $\mathrm{x}$ is the number of Li atoms present in $\mathrm{Li_{x}Sn}$,
$\mathrm{F}$ is the Faraday constant, and $\mathrm{M_{Sn}}$ represents
the Molar-Mass of Sn. 

To examine mechanical properties, elastic constants were calculated
using the VASP code, while the bulk modulus (B), shear modulus (G),
Young modulus (Y), and Poisson's ratio ($\nu$) were estimated using
the Voigt-Reuss-Hill approximation (Section 12 of SI).\cite{Hill_1951}

\section{RESULTS AND DISCUSSIONS}

\subsection{Thermodynamic Stability and Phase Diagram of $\mathbf{\mathrm{\mathbf{Li-Sn}}}$
Compounds.}

\begin{figure}
\begin{centering}
\includegraphics[width=8cm]{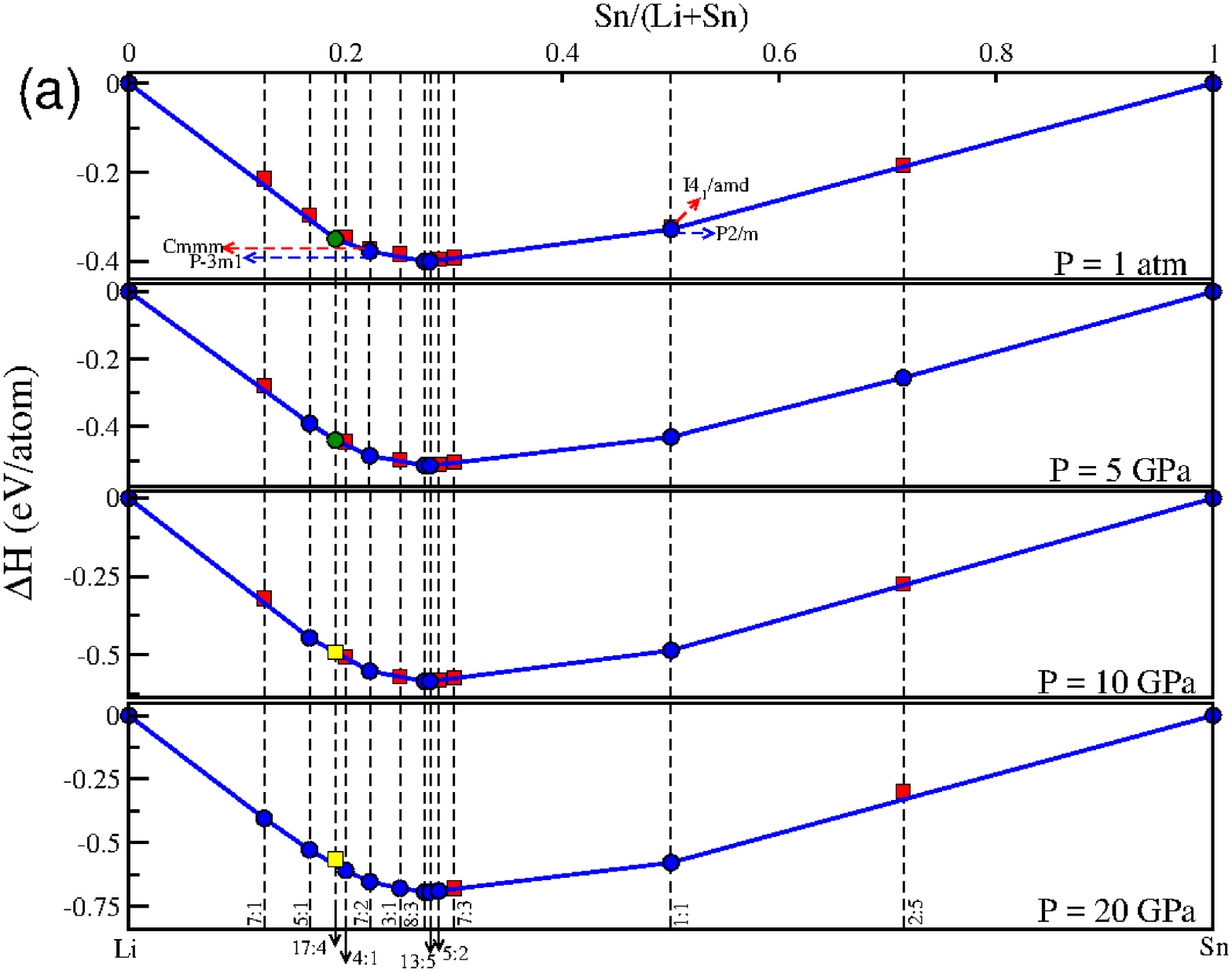}\includegraphics[width=8cm]{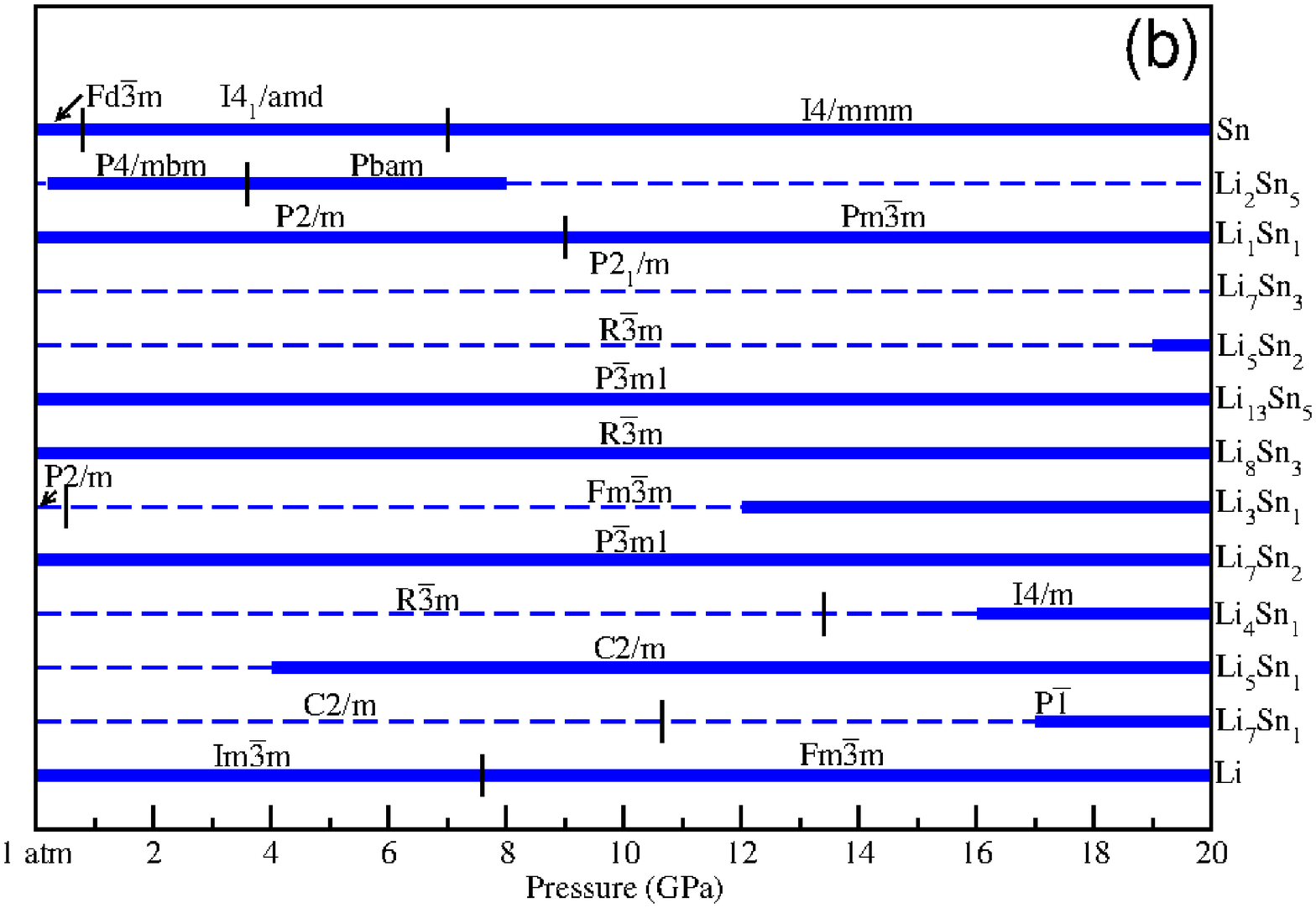}
\par\end{centering}
\caption{(a) Convex hull for the Li-Sn system at 1atm, 5, 10, and 20 GPa pressure.
Thermodynamically stable phases of Li-Sn (except $\textrm{L\ensuremath{i_{17}}S\ensuremath{n_{4}}}$)
are represented with blue circles, while red squares are used to identify
the low-energy metastable phases. Meanwhile, the same is represented
by green circle and yellow square, respectively, for the $\textrm{L\ensuremath{\textrm{i}_{17}}S\ensuremath{\textrm{n}_{4}}}$.
(b) Pressure-composition phase diagram of the Li-Sn system ranging
from 1atm to 20 GPa. The stable and metastable phases are shown by
solid bold and thin dash lines, respectively.\label{Convex-hull-1}}
\end{figure}

To explore the stable and metastable Li-Sn compounds in the pressure
ranging from $1$ atm to $20$ GPa, we computed convex hull and pressure-composition
phase diagram, depicted in \textcolor{black}{Fig.\ref{Convex-hull-1}}.
It can be noted from the phase diagram (Fig.1-(b)) that within the
given pressure range, Li exhibits two different stable phases: $\mathrm{Im\overline{3}m}$
(1 atm\textendash 7.6 GPa) and $\mathrm{Fm\overline{3}m}$ (7.6 GPa\textendash 20
GPa), while Sn possesses three stable phases viz. $\mathrm{Fd\overline{3}m}$
(1 atm\textendash 0.8 GPa), $\mathrm{I4_{1}/amd}$ (0.8 GPa\textendash 7.2
GPa), and $\mathrm{I4/mmm}$ (7.2 GPa\textendash 20 GPa). This phase
sequence for pure $\mathrm{Li}$ and $\mathrm{Sn}$ are also in accordance
with the previous experimental and theoretical data.\cite{Hanfland@1999,Neaton@1999,Yu@2006,Barnett@1963}
In order to draw the thermodynamic convex hull at different pressure,
enthalpy of appropriate phase of pure Li and Sn are considered. Here,
it should also be noted that in our calculations we considered maximum
of 40 atoms per unit cell. This restricts us to obtain the stoichiometry
$\textrm{L\ensuremath{\textrm{i}_{17}}S\ensuremath{\textrm{n}_{4}}}$
($\mathrm{F\overline{4}3m}$, Z = 20 ) in our search as $\mathrm{Li_{17}Sn_{4}}$
contains 420 atoms in the conventional cell, which are beyond our
limit to consider. However, in order to account the effect of $\mathrm{Li_{17}Sn_{4}}$
on the thermodynamical stability of other compounds, we explicitly
considered the formation enthalpy of $\textrm{L\ensuremath{\textrm{i}_{17}}S\ensuremath{\textrm{n}_{4}}}$
(calculated separately) to draw convex hull at the pressure values
of 1 atm, 5, 10, and 20 GPa. In figure showing convex hull (Fig.\ref{Convex-hull-1}(a)),
thermodynamically stable phases of Li-Sn (except $\textrm{L\ensuremath{\textrm{i}_{17}}S\ensuremath{\textrm{n}_{4}}}$)
are represented with blue circles, while red squares are used to identify
the low-lying local minima on all the convex hull. Meanwhile, the
same is represented by green circle and yellow square, respectively,
for the $\textrm{L\ensuremath{\textrm{i}_{17}}S\ensuremath{\textrm{n}_{4}}}$.

On analysing the convex hull for 1 atm pressure (Fig.\ref{Convex-hull-1}(a)\&(b)),
we found that our ab \textit{initio} evolutionary search correctly
predicts most of the experimentally known Li-Sn compositions and their
respective phases, except for $\mathrm{Li_{7}Sn_{2}}$. In case of
$\mathrm{Li_{7}Sn_{2}}$ experimentally known phase is $\mathrm{\mathrm{Cmmm}}$,
while we found $\mathrm{P\overline{3}m1}$ phase to be more stable
than $\mathrm{\mathrm{Cmmm}}$. However, the difference in the enthalpy
for both phases is just $6$ meV/atoms, which is in agreement with
results of Geneser et al.\cite{Genser_2001} From our calculations,
the experimentally known stoichiometries(phases) such as $\mathrm{Li_{1}Sn_{1}}$
($\mathrm{P2/m}$), $\mathrm{Li_{13}Sn_{5}}$ ($\mathrm{P\bar{3}m1}$),
$\mathrm{Li_{8}Sn_{3}}$ ($\mathrm{R\bar{3}m}$), $\mathrm{Li_{7}Sn_{2}}$
($\mathrm{P\bar{3}m1}$), and $\mathrm{Li_{17}Sn_{4}}$ ($\mathrm{F\bar{4}3m}$),
are found to lie on convex-hull tie-line, while the other experimentally
known stoichiometries like $\mathrm{Li_{2}Sn_{5}}$ ($\mathrm{P4/mbm}$),
$\mathrm{Li_{7}Sn_{3}}$ ($\mathrm{P2_{1}/m}$), and $\mathrm{Li_{5}Sn_{2}}$
($\mathrm{R\bar{3}m}$) are found marginally above the convex-hull
tie-line ($\sim$1 \textendash{} 3 meV/atom), representing these phases
to be metastable at 1 atm pressure, as none of the structures possess
dynamical anharmonicity in their respective phonon dispersion curves
(Fig.S1 in SI). The decomposition energy of these experimentally identified
compounds in terms of measured vertical length from the convex hull
are tabulated in \textcolor{black}{Table T1 of SI}. Finally, in order
to give a firm insight in our theoretical findings and to cross check
the thermodynamics stability of experimentally reported Li-Sn compounds,
we compared our results with the convex hull build upon the experimental
structures available in open quantum material database (OQMD).\cite{Saal@2013,Kirklin@2015}
Except for $\mathrm{Li_{2}Sn_{5}}$ (after inclusion of formation
enthalpy of $\mathrm{Li_{17}Sn_{4}}$ in the convexhull), our results
are found to be in good agreement with the OQMD based convex hull.
In case of $\mathrm{Li_{2}Sn_{5}}$, the decomposition energy is less
than 3 meV/atom. This energy difference lies close to the boundary
line of numerical accuracy in our calculations. Therefore, we believe
that such discrepancy can be ignored.

Besides the above mentioned known compositions in the Li-Sn system,
very interestingly, our calculations revealed one of the most stable
stoichiometry (next to $\mathrm{Li_{13}Sn_{5}}$) of Li-Sn, i.e.,
$\mathrm{Li_{8}Sn_{3}}$ ($\mathrm{R\overline{3}m}$) which currently
does not exist in the phase-diagram for Li-Sn, neither at ambient
conditions nor at high temperature or pressure. While, our calculations
predict this composition to be stable through out the investigated
pressure range. This composition, however, was previously predicted
by Gasior et al. in 1996 using electromotive force methods,\cite{Gasior_1996}
but it's structure and phase was not known until now and thus, this
composition is still missing in the Li-Sn phase diagram. But, our
results strongly convince us for its existence and thus, we believe
that careful measurements at ambient conditions can perhaps discover
this composition experimentally as well. Furthermore, our calculations
also acknowledge four yet unknown Li-rich compositions: $\mathrm{Li_{3}Sn_{1}}$
($\mathrm{P2/m}$), $\mathrm{Li_{4}Sn_{1}}$ ($\mathrm{R\bar{3}m}$),
$\mathrm{Li_{5}Sn_{1}}$ ($\mathrm{C2/m}$), and $\mathrm{Li_{7}Sn_{1}}$
($\mathrm{C2/m}$). In case of 1 atm pressure, all these stoichiometries
do not lie on the convex hull tie-line but the absence of imaginary
frequencies in phonon dispersion curves \textcolor{black}{(Fig. S1
in SI)} and calculation of mechanical properties (which we will discuss
in later section), disclose these compounds to be metastable at 1
atm pressure. Moreover, not to ignore the crucial role played by ion-dynamics
in deciding the ground state energy of materials, especially of light
materials, we re-calculated the formation enthalpies at 1 atm pressure
by considering the vibrational contributions (zero-point energy).
The results presented in \textcolor{black}{Table T2 of SI}, clearly
indicate the negligible effect of zero-point energy (ZPE) on the formation
enthalpies for all Li-Sn compounds ($\sim$smaller by three orders
of magnitude), which is not even changing the order of stability of
the compounds. We therefore have neglected the contribution of ZPE
while addressing the relative stability of Li-Sn compounds at higher
pressure. It should be noted that since the metastable compounds,
especially the low-energy ones, are synthesizable under certain thermodynamical
conditions,\cite{Badding@1995,Sun@2016} we believe that most of above
discussed Li-rich compositions can be synthesized. Moreover, if synthesized,
these metastable/stable compounds should be recoverable as well, when
decompressed to ambient pressure, again because of non-existence of
negative frequencies in their phonon spectra at 1 atm.\textbf{\cite{Lu@2016}
}Also, one can not exclude \textcolor{black}{the} possibility of formation
of these phases during lithiation of Sn anode since at the end of
the discharge of the Sb/Na cell Darwiche et al.\textbf{\cite{Darwiche@2012}
}observed the presence of cubic \textbf{$\mathrm{Na_{3}Sb}$}, which
is metastable at ambient pressure and can only be synthesized under
high pressure. Thus, a careful experimental study is indeed needed
to focus on the formation of metastable phases during lithiation-delithiation.

To study the effect of high pressure ($1$atm$\leq$P$\leq$20 GPa),
we computed the convex-hull at $5$, $10,$ and $20$ GPa. It can
be clearly noticed from the of Fig.\ref{Convex-hull-1}(a) that at
$5$ and $10$ GPa pressure, except $\mathrm{Li_{2}Sn_{5}}$ and $\mathrm{Li_{5}Sn_{1}}$,
all other stoichiometries show similar stability as in case of 1 atm
pressure. Our calculations reveal $\mathrm{Li_{2}Sn_{5}}$ to be thermodynamically
stable between pressure range $0.2$ GPa\textendash $8$ GPa, while
$\mathrm{Li_{5}Sn_{1}}$ is found to be stable from 4 GPa onwards.
Additionally, a pressure induced phase transformation is also observed
in the stoichiometry $\mathrm{Li_{2}Sn_{5}}$ and $\mathrm{Li_{1}Sn_{1}}$.
In case of $\mathrm{Li_{2}Sn_{5}}$, the $\mathrm{P4/mbm}$ phase
gets transformed to $\mathrm{Pbam}$ above $3.9$ GPa, while $\mathrm{Li_{1}Sn_{1}}$
transforms from $\mathrm{P2/m}$ to $\mathrm{Pm\bar{3}m}$ at around
$9$ GPa, which is also in agreement with the study of Genser et al.\cite{Genser_2001}
Also, it is found that the experimentally most lithiated phase, $\mathrm{Li_{17}Sn_{4}}$,
losses its thermodynamic stability at around 9.8 GPa. Above this pressure,
$\mathrm{Li_{17}Sn_{4}}$ shall decompose into $\mathrm{Li_{5}Sn_{1}}$
$\mathrm{(C2/m)}$ and $\mathrm{Li_{7}Sn_{2}}$ $(\mathrm{P\bar{3}m1)}$
(i.e., $\mathrm{Li_{17}Sn_{4}(F\bar{4}3m)}\mathrm{\stackrel{>9.8GPa}{\longrightarrow}}\mathrm{Li_{5}Sn_{1}(C2/m)+Li_{7}Sn_{2}(P\bar{3}m1})$),
as shown in \textcolor{black}{Fig. S3 in SI.} The most fascinating
results are observed at $20$ GPa pressure. Our calculations disclose
a diverse chemistry of Li-Sn compounds by predicting the stability
of some novel exotic Li-rich Li-Sn compounds. Except for $\mathrm{Li_{2}Sn_{5}}$
and $\mathrm{Li_{7}Sn_{3}}$, all metastable stoichiometries such
as $\mathrm{Li_{5}Sn_{2}}$, $\mathrm{Li_{3}Sn_{1}}$, $\mathrm{Li_{4}Sn_{1}}$,
and $\mathrm{Li_{7}Sn_{1}}$ become stable at this range of pressure.
Moreover, it should be noted here that the compositions: $\mathrm{Li_{5}Sn_{1}}$
and $\mathrm{Li_{7}Sn_{1}}$, can accommodate more lithium as compared
to the presently known highest lithiated compound, $\mathrm{Li_{17}Sn_{4}}$.
\textcolor{black}{Though it is understandable from the perspective
of LIBs, that the pressure range where these Li-rich compounds get
stable is quite high, but not to forget that these Li-rich compounds
should be recoverable when quenched to ambient pressure, as discussed
above.}\textbf{ }

The phase diagram of Li-Sn compounds presented in Fig.\ref{Convex-hull-1}(b)
provides a clear picture of different phases acquired by individual
composition in the pressure range of 1 atm to 20 GPa. Among several
investigated Li-Sn compositions, robust stability and no phase transition
are observed in three of the compounds, i.e., $\mathrm{Li_{13}Sn_{5}}$,
$\mathrm{Li_{8}Sn_{3}}$, and $\mathrm{Li_{7}Sn_{2}}$. Through out
the investigated pressure range (1 atm\textendash 20 GPa) they exhibit
$\mathrm{P\bar{3}m1}$, $\mathrm{R\bar{3}m}$, and $\mathrm{P\bar{3}m1}$
space group symmetry, respectively. Interestingly, $\mathrm{Li_{13}Sn_{5}}$
and $\mathrm{Li_{8}Sn_{3}}$ are also identified as the most stable
Li-Sn compounds (Fig.\ref{Convex-hull-1}(a)). Other than these, $\mathrm{Li_{7}Sn_{3}}$,
$\mathrm{Li_{5}Sn_{2}}$, and $\mathrm{Li_{5}Sn_{1}}$ are also found
to remain in a single space group symmetry viz. $\mathrm{P2_{1}/m}$,
$\mathrm{R\bar{3}m}$, and $\mathrm{C2/m}$, respectively, in the
entire pressure range. However, like former compositions ($\mathrm{Li_{13}Sn_{5}}$,
$\mathrm{Li_{8}Sn_{3}}$, and $\mathrm{Li_{7}Sn_{2}}$), these materials
($\mathrm{Li_{7}Sn_{3}}$, $\mathrm{Li_{5}Sn_{2}}$, and $\mathrm{Li_{5}Sn_{1}}$)
are not found to be stable through out the investigated pressure range.
Our calculations reveal $\mathrm{Li_{7}Sn_{3}}$ to be metastable
in the given pressure range, while $\mathrm{Li_{5}Sn_{2}}$ and $\mathrm{Li_{5}Sn_{1}}$
get stable above 18.2 and 4 GPa, respectively. Other novel stoichiometries
like $\mathrm{Li_{3}Sn_{1}}$, $\mathrm{Li_{4}Sn_{1}}$, and $\mathrm{Li_{7}Sn_{1}}$
are found to be stable at 12, 16, and 17 GPa, respectively. These
compounds are also found to change phase from $\mathrm{P2/m}$ to
$\mathrm{Fm\bar{3}m}$ at 0.8 GPa, $\mathrm{R\bar{3}m}$ to $\mathrm{I4/m}$
at 13.2 GPa, and $\mathrm{C2/m}$ to $\mathrm{P\bar{1}}$ at 10.8
GPa, respectively (Fig.\ref{Convex-hull-1}(b)).

\subsection{Crystal Structure of $\mathbf{\mathrm{\mathbf{Li-Sn}}}$ Compounds}

\hspace{-2cm}
\begin{table}[h]
\caption{The space group and calculated equilibrium lattice parameters a (Å),
b (Å), and c (Å), $\alpha$(deg), $\beta$(deg), $\gamma$(deg), and
volume (Å$^{3}$/atom) at given pressure for Li, Sn and Li-Sn compounds
are given. Gravimetric capacities of corresponding Li-Sn compounds
are written insight. \label{tab:structural-parameters-1} }

\centering{}\noindent\resizebox{\textwidth}{!}{
\begin{tabular}{|c|c|c|c|c|c|c|c|c|c|c|c|c|c|c|c|c|c|c|c|}
\hline 
\textbf{\tiny{}System (GC)} & \textbf{\tiny{}Pressure} & \textbf{\tiny{}Space } & \textbf{\tiny{}a}{\tiny{} } & \textbf{\tiny{}b } & \textbf{\tiny{}c } & \textbf{\tiny{}$\alpha$} & \textbf{\tiny{}$\beta$ } & \textbf{\tiny{}$\gamma$ } & \textbf{\tiny{}Volume} & \textbf{\tiny{}System (GC)} & \textbf{\tiny{}Pressure} & \textbf{\tiny{}Space } & \textbf{\tiny{}a} & \textbf{\tiny{}b} & \textbf{\tiny{}c} & \textbf{\tiny{}$\alpha$} & \textbf{\tiny{}$\beta$ } & \textbf{\tiny{}$\gamma$} & \textbf{\tiny{}Volume}\tabularnewline
{\tiny{}(mAh/g)} &  & \textbf{\tiny{}Group (f.u)} & \textbf{\tiny{}($\mathrm{\mathring{A}}$)} & \textbf{\tiny{}($\mathrm{\mathring{A}}$)} & \textbf{\tiny{}($\mathrm{\mathring{A}}$)} & \textbf{\tiny{}($^{\circ}$)} & \textbf{\tiny{}($^{\circ}$)} & \textbf{\tiny{}($^{\circ}$)} & \textbf{\tiny{}(}{\tiny{}Å$^{3}$}\textbf{\tiny{}/atom)} & {\tiny{}(mAh/g)} &  & \textbf{\tiny{}Group (f.u)} & \textbf{\tiny{}($\mathrm{\mathring{A}}$)} & \textbf{\tiny{}($\mathrm{\mathring{A}}$)} & \textbf{\tiny{}($\mathrm{\mathring{A}}$)} & \textbf{\tiny{}($^{\circ}$)} & \textbf{\tiny{}($^{\circ}$)} & \textbf{\tiny{}($^{\circ}$)} & \textbf{\tiny{}(}{\tiny{}Å$^{3}$}\textbf{\tiny{}/atom)}\tabularnewline
\hline 
 & {\tiny{}1 atm} & {\tiny{}$\mathrm{Fd\bar{3m}}$(Z=8)} & {\tiny{}6.651} & {\tiny{}6.651} & {\tiny{}6.651} & {\tiny{}90.00} & {\tiny{}90.00} & {\tiny{}90.00} & {\tiny{}36.78} &  & {\tiny{}1 atm} & {\tiny{}$\mathrm{Im\bar{3}m}$(Z=2)} & {\tiny{}3.439} & {\tiny{}3.439} & {\tiny{}3.439} & {\tiny{}90.00} & {\tiny{}90.00} & {\tiny{}90.00} & {\tiny{}20.33}\tabularnewline
\cline{2-10} \cline{12-20} 
\textbf{\tiny{}$\mathrm{\mathbf{Sn}}$} & {\tiny{}5 GPa} & {\tiny{}$\mathrm{I4_{1}/amd}$(Z=4)} & {\tiny{}5.948} & {\tiny{}5.948} & {\tiny{}3.201} & {\tiny{}90.00} & {\tiny{}90.00} & {\tiny{}90.00} & {\tiny{}28.31} & {\tiny{}$\mathrm{\mathbf{Li}}$} & {\tiny{}5 GPa} & {\tiny{}$\mathrm{Im\bar{3}m}$(Z=2)} & {\tiny{}3.175} & {\tiny{}3.175} & {\tiny{}3.175} & {\tiny{}90.00} & {\tiny{}90.00} & {\tiny{}90.00} & {\tiny{}16.00}\tabularnewline
\cline{2-10} \cline{12-20} 
\textbf{\tiny{}(-)} & {\tiny{}10 GPa} & {\tiny{}$I\mathrm{4/mmm}$(Z=2)} & {\tiny{}3.778} & {\tiny{}3.778} & {\tiny{}3.349} & {\tiny{}90.00} & {\tiny{}90.00} & {\tiny{}90.00} & {\tiny{}23.91} & \textbf{\tiny{}(-)} & {\tiny{}10 GPa} & {\tiny{}$\mathrm{Fm\bar{3}m}$(Z=4)} & {\tiny{}3.814} & {\tiny{}3.814} & {\tiny{}3.814} & {\tiny{}90.00} & {\tiny{}90.00} & {\tiny{}90.00} & {\tiny{}13.87}\tabularnewline
\cline{2-10} \cline{12-20} 
 & {\tiny{}20 GPa} & {\tiny{}$\mathrm{I4/mmm}$(Z=2)} & {\tiny{}3.647} & {\tiny{}3.647} & {\tiny{}3.285} & {\tiny{}90.00} & {\tiny{}90.00} & {\tiny{}90.00} & {\tiny{}21.84} &  & {\tiny{}20 GPa} & {\tiny{}$\mathrm{Fm\bar{3}m}$(Z=4)} & {\tiny{}3.591} & {\tiny{}3.591} & {\tiny{}3.591} & {\tiny{}90.00} & {\tiny{}90.00} & {\tiny{}90.00} & {\tiny{}11.57}\tabularnewline
\hline 
 & {\tiny{}1 atm} & {\tiny{}$\mathrm{P4/mbm}$(Z=2)} & {\tiny{}10.378} & {\tiny{}10.378} & {\tiny{}3.141} & {\tiny{}90.00} & {\tiny{}90.00} & {\tiny{}90.00} & {\tiny{}24.17} &  & {\tiny{}1 atm} & {\tiny{}$\mathrm{C2/m}$(Z=4)} & {\tiny{}15.678} & {\tiny{}4.826} & {\tiny{}8.058} & {\tiny{}90.00} & {\tiny{}102.67} & {\tiny{}90.00} & {\tiny{}18.59}\tabularnewline
\cline{2-10} \cline{12-20} 
{\tiny{}$\mathrm{\mathbf{Li_{2}Sn_{5}}}$} & {\tiny{}5 GPa} & {\tiny{}$\mathrm{Pbam}$(Z=2)} & {\tiny{}10.365} & {\tiny{}9.745} & {\tiny{}3.045} & {\tiny{}90.00} & {\tiny{}90.00} & {\tiny{}90.00} & {\tiny{}21.97} & {\tiny{}$\mathrm{\mathbf{Li_{7}Sn_{1}}}$} & {\tiny{}5 GPa} & {\tiny{}$\mathrm{C2/m}$(Z=4)} & {\tiny{}14.762} & {\tiny{}4.556} & {\tiny{}7.598} & {\tiny{}90.00} & {\tiny{}102.40} & {\tiny{}90.00} & {\tiny{}15.60}\tabularnewline
\cline{2-10} \cline{12-20} 
\textbf{\tiny{}(90)} & {\tiny{}10 GPa} & {\tiny{}$\mathrm{Pbam}$(Z=2)} & {\tiny{}10.247} & {\tiny{}9.407} & {\tiny{}2.981} & {\tiny{}90.00} & {\tiny{}90.00} & {\tiny{}90.00} & {\tiny{}20.52} & \textbf{\tiny{}(1580)} & {\tiny{}10 GPa} & {\tiny{}$\mathrm{C2/m}$(Z=4)} & {\tiny{}14.208} & {\tiny{}4.392} & {\tiny{}7.324} & {\tiny{}90.00} & {\tiny{}102.14} & {\tiny{}90.00} & {\tiny{}13.96}\tabularnewline
\cline{2-10} \cline{12-20} 
 & {\tiny{}20 GPa} & {\tiny{}$\mathrm{Pbam}$(Z=2)} & {\tiny{}10.090} & {\tiny{}8.955} & {\tiny{}2.885} & {\tiny{}90.00} & {\tiny{}90.00} & {\tiny{}90.00} & {\tiny{}18.62} &  & {\tiny{}20 GPa} & {\tiny{}$\mathrm{P\overline{1}}$(Z=2)} & {\tiny{}4.609} & {\tiny{}6.693} & {\tiny{}6.697} & {\tiny{}67.57} & {\tiny{}79.92} & {\tiny{}80.25} & {\tiny{}11.68}\tabularnewline
\hline 
 & \multirow{2}{*}{{\tiny{}1 atm}} & {\tiny{}$\mathrm{P2/m}$(Z=3)} & {\tiny{}5.178} & {\tiny{}3.225} & {\tiny{}7.812} & {\tiny{}90.00} & {\tiny{}105.25} & {\tiny{}90.00} & {\tiny{}20.98} &  & \multirow{2}{*}{{\tiny{}1 atm}} & \multirow{2}{*}{{\tiny{}$\mathrm{C2/m}$(Z=8)}} & \multirow{2}{*}{{\tiny{}15.918}} & \multirow{2}{*}{{\tiny{}5.738}} & \multirow{2}{*}{{\tiny{}12.043}} & \multirow{2}{*}{{\tiny{}90.00}} & \multirow{2}{*}{{\tiny{}128.86}} & \multirow{2}{*}{{\tiny{}90.00}} & \multirow{2}{*}{{\tiny{}17.84}}\tabularnewline
\cline{3-10} 
 &  & {\tiny{}$\mathrm{I4_{1}/amd}$(Z=12)} & {\tiny{}4.454} & {\tiny{}4.454} & {\tiny{}26.043} & {\tiny{}90.00} & {\tiny{}90.00} & {\tiny{}90.00} & {\tiny{}21.53} &  &  &  &  &  &  &  &  &  & \tabularnewline
\cline{2-10} \cline{12-20} 
{\tiny{}$\mathrm{\mathbf{Li_{1}Sn_{1}}}$} & {\tiny{}5 GPa} & {\tiny{}$\mathrm{P2/m}$(Z=3)} & {\tiny{}4.940} & {\tiny{}3.145} & {\tiny{}7.526} & {\tiny{}90.00} & {\tiny{}106.18} & {\tiny{}90.00} & {\tiny{}18.72} & {\tiny{}$\mathrm{\mathbf{Li_{5}Sn_{1}}}$} & {\tiny{}5 GPa} & {\tiny{}$\mathrm{C2/m}$(Z=8)} & {\tiny{}15.135} & {\tiny{}5.474} & {\tiny{}11.453} & {\tiny{}90.00} & {\tiny{}128.87} & {\tiny{}90.00} & {\tiny{}15.39}\tabularnewline
\cline{2-10} \cline{12-20} 
\textbf{\tiny{}(226)} & {\tiny{}10 GPa} & {\tiny{}$\mathrm{Pm\overline{3}m}$(Z=1)} & {\tiny{}3.248} & {\tiny{}3.248} & {\tiny{}3.248} & {\tiny{}90.00} & {\tiny{}90.00} & {\tiny{}90.00} & {\tiny{}17.14} & \textbf{\tiny{}(1129)} & {\tiny{}10 GPa} & {\tiny{}$\mathrm{C2/m}$(Z=8)} & {\tiny{}14.631} & {\tiny{}5.308} & {\tiny{}11.076} & {\tiny{}90.00} & {\tiny{}128.83} & {\tiny{}90.00} & {\tiny{}13.96}\tabularnewline
\cline{2-10} \cline{12-20} 
 & {\tiny{}20 GPa} & {\tiny{}$\mathrm{Pm\overline{3}m}$(Z=1)} & {\tiny{}3.135} & {\tiny{}3.135} & {\tiny{}3.135} & {\tiny{}90.00} & {\tiny{}90.00} & {\tiny{}90.00} & {\tiny{}15.41} &  & {\tiny{}20 GPa} & {\tiny{}$\mathrm{C2/m}$(Z=8)} & {\tiny{}13.948} & {\tiny{}5.089} & {\tiny{}10.568} & {\tiny{}90.00} & {\tiny{}128.71} & {\tiny{}90.00} & {\tiny{}12.20}\tabularnewline
\hline 
 & {\tiny{}1 atm} & {\tiny{}$P2_{1}/m$(Z=2)} & {\tiny{}8.562} & {\tiny{}4.738} & {\tiny{}9.491} & {\tiny{}90.00} & {\tiny{}106.10} & {\tiny{}90.00} & {\tiny{}18.50} &  & {\tiny{}1 atm} & {\tiny{}$\mathrm{F\overline{4}3m}$(Z=20)} & {\tiny{}19.714} & {\tiny{}19.714} & {\tiny{}19.714} & {\tiny{}90.00} & {\tiny{}90.00} & {\tiny{}90.00} & {\tiny{}18.24}\tabularnewline
\cline{2-10} \cline{12-20} 
{\tiny{}$\mathrm{\mathbf{Li_{7}Sn_{3}}}$} & {\tiny{}5 GPa} & {\tiny{}$\mathrm{P2_{1}/m}$(Z=2)} & {\tiny{}8.225} & {\tiny{}4.528} & {\tiny{}9.092} & {\tiny{}90.00} & {\tiny{}106.07} & {\tiny{}90.00} & {\tiny{}16.27} & {\tiny{}$\mathrm{\mathbf{Li_{17}Sn_{4}}}$} & {\tiny{}5 GPa} & {\tiny{}$\mathrm{F\overline{4}3m}$(Z=20)} & {\tiny{}18.789} & {\tiny{}18.789} & {\tiny{}18.789} & {\tiny{}90.00} & {\tiny{}90.00} & {\tiny{}90.00} & {\tiny{}15.79}\tabularnewline
\cline{2-10} \cline{12-20} 
\textbf{\tiny{}(527)} & {\tiny{}10 GPa} & {\tiny{}$\mathrm{P2_{1}/m}$(Z=2)} & {\tiny{}8.009} & {\tiny{}4.387} & {\tiny{}8.341} & {\tiny{}90.00} & {\tiny{}106.05} & {\tiny{}90.00} & {\tiny{}14.91} & \textbf{\tiny{}(960)} & {\tiny{}10 GPa} & {\tiny{}$\mathrm{F\overline{4}3m}$(Z=20)} & {\tiny{}18.198} & {\tiny{}18.198} & {\tiny{}18.198} & {\tiny{}90.00} & {\tiny{}90.00} & {\tiny{}90.00} & {\tiny{}14.35}\tabularnewline
\cline{2-10} \cline{12-20} 
 & {\tiny{}20 GPa} & {\tiny{}$\mathrm{P2_{1}/m}$(z=2)} & {\tiny{}7.710} & {\tiny{}4.196} & {\tiny{}8.489} & {\tiny{}90.00} & {\tiny{}106.03} & {\tiny{}90.00} & {\tiny{}13.20} &  & {\tiny{}20 GPa} & {\tiny{}$\mathrm{F\overline{4}3m}$(Z=20)} & {\tiny{}17.398} & {\tiny{}17.398} & {\tiny{}17.398} & {\tiny{}90.00} & {\tiny{}90.00} & {\tiny{}90.00} & {\tiny{}12.54}\tabularnewline
\hline 
 & {\tiny{}1 atm} & {\tiny{}$\mathrm{R\overline{3}m}$(Z=3)} & {\tiny{}4.732} & {\tiny{}4.732} & {\tiny{}19.810} & {\tiny{}90.00} & {\tiny{}90.00} & {\tiny{}120.00} & {\tiny{}18.29} &  & {\tiny{}1 atm} & {\tiny{}$\mathrm{R\bar{3}m}$(Z=3)} & {\tiny{}4.742} & {\tiny{}4.742} & {\tiny{}13.932} & {\tiny{}90.00} & {\tiny{}90.00} & {\tiny{}120.00} & {\tiny{}18.08}\tabularnewline
\cline{2-10} \cline{12-20} 
{\tiny{}$\mathrm{\mathbf{Li_{5}Sn_{2}}}$} & {\tiny{}5 GPa} & {\tiny{}$\mathrm{R\overline{3}m}$(Z=3)} & {\tiny{}4.520} & {\tiny{}4.520} & {\tiny{}19.079} & {\tiny{}90.00} & {\tiny{}90.00} & {\tiny{}120.00} & {\tiny{}16.07} & {\tiny{}$\mathrm{\mathbf{Li_{4}Sn_{1}}}$} & {\tiny{}5 GPa} & {\tiny{}$\mathrm{R\bar{3}m}$(Z=3)} & {\tiny{}4.524} & {\tiny{}4.524} & {\tiny{}13.173} & {\tiny{}90.00} & {\tiny{}90.00} & {\tiny{}120.00} & {\tiny{}15.57}\tabularnewline
\cline{2-10} \cline{12-20} 
\textbf{\tiny{}(564)} & {\tiny{}10 GPa} & {\tiny{}$\mathrm{R\overline{3}m}$(Z=3)} & {\tiny{}4.376} & {\tiny{}4.376} & {\tiny{}18.629} & {\tiny{}90.00} & {\tiny{}90.00} & {\tiny{}120.00} & {\tiny{}14.72} & \textbf{\tiny{}(903)} & {\tiny{}10 GPa} & {\tiny{}$\mathrm{R\bar{3}m}$(Z=3)} & {\tiny{}4.386} & {\tiny{}4.386} & {\tiny{}12.701} & {\tiny{}90.00} & {\tiny{}90.00} & {\tiny{}120.00} & {\tiny{}14.11}\tabularnewline
\cline{2-10} \cline{12-20} 
 & {\tiny{}20 GPa} & {\tiny{}$\mathrm{R\overline{3}m}$(Z=3)} & {\tiny{}4.187} & {\tiny{}4.187} & {\tiny{}17.994} & {\tiny{}90.00} & {\tiny{}90.00} & {\tiny{}120.00} & {\tiny{}13.01} &  & {\tiny{}20 GPa} & {\tiny{}$\mathrm{I4/m}$(Z=10)} & {\tiny{}11.965} & {\tiny{}11.965} & {\tiny{}4.241} & {\tiny{}90.00} & {\tiny{}90.00} & {\tiny{}90.00} & {\tiny{}12.14}\tabularnewline
\hline 
 & \multirow{2}{*}{{\tiny{}1 atm}} & \multirow{2}{*}{{\tiny{}$\mathrm{P\overline{3}m1}$(Z=1)}} & \multirow{2}{*}{{\tiny{}4.702}} & \multirow{2}{*}{{\tiny{}4.702}} & \multirow{2}{*}{{\tiny{}17.133}} & \multirow{2}{*}{{\tiny{}90.00}} & \multirow{2}{*}{{\tiny{}90.00}} & \multirow{2}{*}{{\tiny{}120.00}} & \multirow{2}{*}{{\tiny{}18.23}} &  & \multirow{2}{*}{{\tiny{}1 atm}} & {\tiny{}$\mathrm{P\bar{3}m1}$(Z=1)} & {\tiny{}4.681} & {\tiny{}4.681} & {\tiny{}8.501} & {\tiny{}90.00} & {\tiny{}90.00} & {\tiny{}120.00} & {\tiny{}17.93}\tabularnewline
\cline{13-20} 
 &  &  &  &  &  &  &  &  &  &  &  & {\tiny{}$\mathrm{Cmmm}$(Z=4)} & {\tiny{}9.812} & {\tiny{}13.887} & {\tiny{}4.728} & {\tiny{}90.00} & {\tiny{}90.00} & {\tiny{}90.00} & {\tiny{}17.90}\tabularnewline
\cline{2-10} \cline{12-20} 
{\tiny{}$\mathrm{\mathbf{Li_{13}Sn_{5}}}$} & {\tiny{}5 GPa} & {\tiny{}$\mathrm{P\overline{3}m1}$(Z=1)} & {\tiny{}4.495} & {\tiny{}4.495} & {\tiny{}16.458} & {\tiny{}90.00} & {\tiny{}90.00} & {\tiny{}120.00} & {\tiny{}16.00} & {\tiny{}$\mathrm{\mathbf{Li_{7}Sn_{2}}}$} & {\tiny{}5 GPa} & {\tiny{}$\mathrm{P\bar{3}m1}$(Z=1)} & {\tiny{}4.476} & {\tiny{}4.476} & {\tiny{}8.088} & {\tiny{}90.00} & {\tiny{}90.00} & {\tiny{}120.00} & {\tiny{}15.59}\tabularnewline
\cline{2-10} \cline{12-20} 
\textbf{\tiny{}(587)} & {\tiny{}10 GPa} & {\tiny{}$P\overline{3}m1$(Z=1)} & {\tiny{}4.358} & {\tiny{}4.358} & {\tiny{}16.020} & {\tiny{}90.00} & {\tiny{}90.00} & {\tiny{}120.00} & {\tiny{}14.64} & \textbf{\tiny{}(790)} & {\tiny{}10 GPa} & {\tiny{}$\mathrm{P\bar{3}m1}$(Z=1)} & {\tiny{}4.341} & {\tiny{}4.341} & {\tiny{}7.828} & {\tiny{}90.00} & {\tiny{}90.00} & {\tiny{}120.00} & {\tiny{}14.20}\tabularnewline
\cline{2-10} \cline{12-20} 
 & {\tiny{}20 GPa} & {\tiny{}$\mathrm{P\overline{3}m1}$(Z=1)} & {\tiny{}4.173} & {\tiny{}4.173} & {\tiny{}15.433} & {\tiny{}90.00} & {\tiny{}90.00} & {\tiny{}120.00} & {\tiny{}12.93} &  & {\tiny{}20 GPa} & {\tiny{}$\mathrm{P\bar{3}m1}$(Z=1)} & {\tiny{}4.160} & {\tiny{}4.160} & {\tiny{}7.485} & {\tiny{}90.00} & {\tiny{}90.00} & {\tiny{}120.00} & {\tiny{}12.46}\tabularnewline
\hline 
 & {\tiny{}1 atm} & {\tiny{}$\mathrm{R\overline{3}m}$(Z=3)} & {\tiny{}4.684} & {\tiny{}4.684} & {\tiny{}31.582} & {\tiny{}90.00} & {\tiny{}90.00} & {\tiny{}120.00} & {\tiny{}18.18} &  & {\tiny{}1 atm} & {\tiny{}$\mathrm{P2/m}$(Z=3)} & {\tiny{}6.885} & {\tiny{}4.623} & {\tiny{}7.060} & {\tiny{}90.00} & {\tiny{}104.07} & {\tiny{}90.00} & {\tiny{}18.16}\tabularnewline
\cline{2-10} \cline{12-20} 
\textbf{\tiny{}$\textbf{\ensuremath{\mathrm{\mathbf{Li_{8}Sn_{3}}}}}$} & {\tiny{}5 GPa} & {\tiny{}$\mathrm{R\overline{3}m}$(Z=3)} & {\tiny{}4.478} & {\tiny{}4.478} & {\tiny{}30.298} & {\tiny{}90.00} & {\tiny{}90.00} & {\tiny{}120.00} & {\tiny{}15.94} & {\tiny{}$\textit{\ensuremath{\mathrm{\mathbf{Li_{3}Sn_{1}}}}}$} & {\tiny{}5 GPa} & {\tiny{}$\mathrm{Fm\overline{3}m}$(Z=4)} & {\tiny{}6.305} & {\tiny{}6.305} & {\tiny{}6.305} & {\tiny{}90.00} & {\tiny{}90.00} & {\tiny{}90.00} & {\tiny{}15.67}\tabularnewline
\cline{2-10} \cline{12-20} 
\textbf{\tiny{}(602)} & {\tiny{}10 GPa} & {\tiny{}$\mathrm{R\overline{3}m}$(Z=3)} & {\tiny{}4.343} & {\tiny{}4.343} & {\tiny{}29.462} & {\tiny{}90.00} & {\tiny{}90.00} & {\tiny{}120.00} & {\tiny{}14.58} & \textbf{\tiny{}(677)} & {\tiny{}10 GPa} & {\tiny{}$\mathrm{Fm\overline{3}m}$(Z=4)} & {\tiny{}6.119} & {\tiny{}6.119} & {\tiny{}6.119} & {\tiny{}90.00} & {\tiny{}90.00} & {\tiny{}90.00} & {\tiny{}14.32}\tabularnewline
\cline{2-10} \cline{12-20} 
 & {\tiny{}20 GPa} & {\tiny{}$\mathrm{R\overline{3}m}$(Z=3)} & {\tiny{}4.161} & {\tiny{}4.161} & {\tiny{}28.337} & {\tiny{}9000} & {\tiny{}90.00} & {\tiny{}120.00} & {\tiny{}12.87} &  & {\tiny{}20 GPa} & {\tiny{}$\mathrm{Fm\overline{3}m}$(Z=4)} & {\tiny{}5.867} & {\tiny{}5.867} & {\tiny{}5.867} & {\tiny{}90.00} & {\tiny{}90.00} & {\tiny{}90.00} & {\tiny{}12.62}\tabularnewline
\hline 
\end{tabular}}
\end{table}

In Table \ref{tab:structural-parameters-1}, lattice parameters for
each of the stable and metastable Li-Sn compounds identified by our
calculations at 1 atm, 5, 10, and 20 GPa are tabulated. At 1 atm pressure,
except for $\textrm{L\ensuremath{i_{7}}S\ensuremath{n_{2}}}$, the
structural information predicted by our calculations for the experimentally
known stoichiometries like $\mathrm{Li_{2}Sn_{5}}$, $\mathrm{Li_{1}Sn_{1}}$,
$\mathrm{Li_{7}Sn_{3}}$, $\mathrm{Li_{5}Sn_{2}}$, and $\mathrm{Li_{13}Sn_{5}}$
are found to be in very good agreement with experimental results.
Our calculations also predicted the unknown structure of once predicted
stoichiometries like $\textrm{L\ensuremath{\textrm{i}_{8}}S\ensuremath{\textrm{n}_{3}}}$,\cite{Gasior_1996}
$\textrm{L\ensuremath{\textrm{i}_{3}}S\ensuremath{\textrm{n}_{1}}}$,\cite{Thackeray@2003}
and $\textrm{L\ensuremath{\textrm{i}_{4,}}S\ensuremath{\textrm{n}_{1}}},$\cite{Alblas_1984}
together with the structure of novel Li-rich compounds: $\textrm{L\ensuremath{\textrm{i}_{5}}S\ensuremath{\textrm{n}_{1}}}$
and $\textrm{L\ensuremath{\textrm{i}_{7}}S\ensuremath{\textrm{n}_{1}}}$.
The structure of already known Li-Sn stoichiometries are discussed
in detail in Section 5 of SI. While, here, we only discuss the structure
of all yet unexplored Li-Sn stoichiometries, predicted by our calculations
as metastable or stable chemical compositions of Li-Sn in the pressure
range of 1 atm\textendash 20 GPa. The crystallographic details of
each of these structures are given in Table T12 of SI.

\subsubsection{\textup{Unknown Structures of Already Predicted Li-Sn Compounds}}

\begin{figure}
\includegraphics[width=15cm]{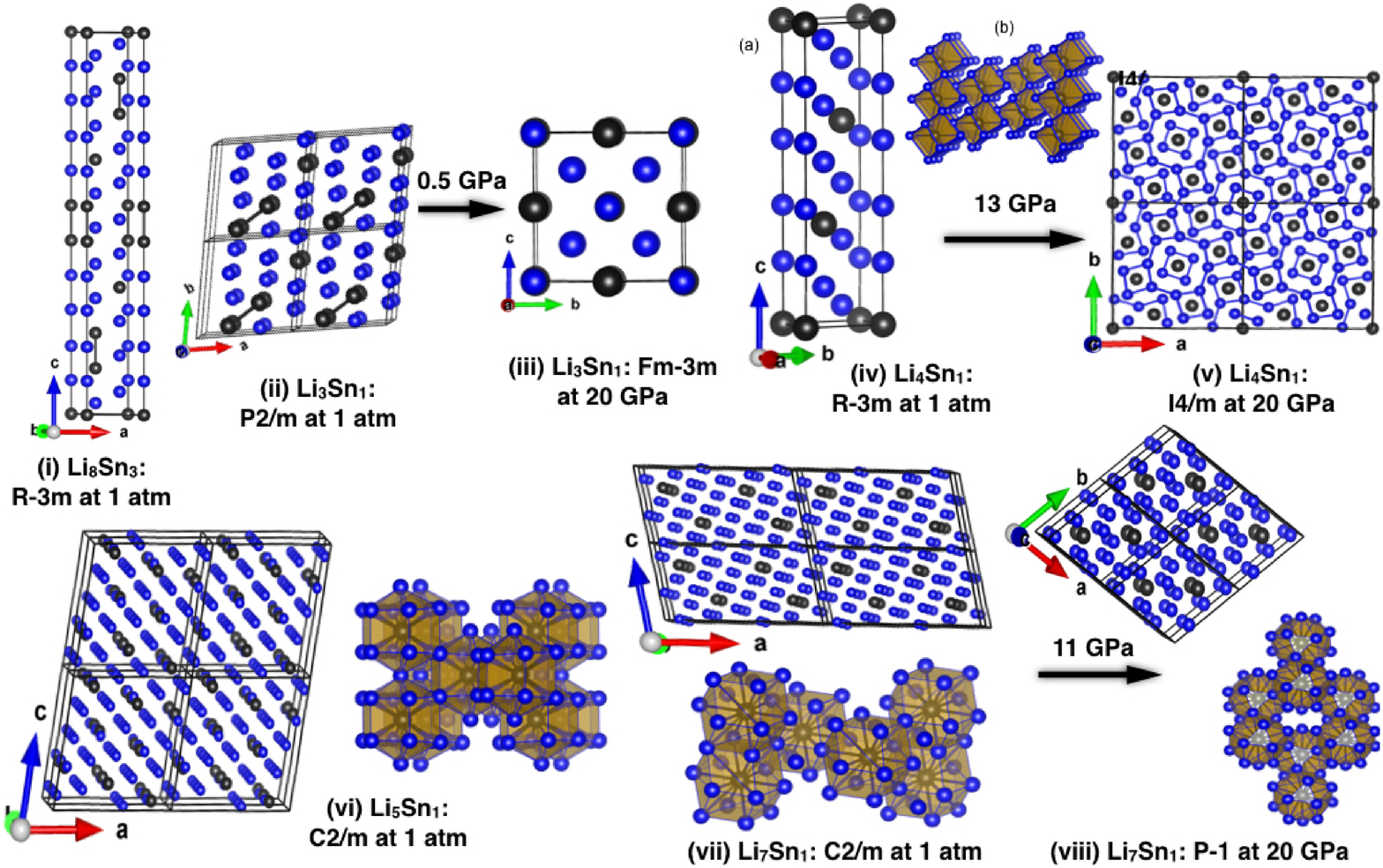}

\caption{\textbf{Crystal structures of the newly predicted Li-Sn compounds.}
(i) Crystal structure of $\mathrm{Li_{8}Sn_{3}}$ ($\mathrm{R\overline{3}m}$),
(ii \& iii) Phase transformation from $\mathrm{P2/m}$ to $\mathrm{Fm\overline{3}m}$
of $\mathrm{Li_{3}Sn_{1}}$, (iv-a \& v) Phase transformation from
$\mathrm{R\overline{3}m}$ to $\mathrm{I4/m}$ of $\mathrm{Li_{4}Sn_{1}}$,
(iv-b) Polyhedral representation of $\mathrm{R\overline{3}m}$-$\mathrm{Li_{4}Sn_{1}}$,
(vi ) Crystal structure of $\mathrm{Li_{5}Sn_{1}}$ ($\mathrm{C2/m}$)
and its corresponding polyhedral representation, and (vii \& viii)
Phase transformation from $\mathrm{C2/m}$ to $\mathrm{P\overline{1}}$
of $\mathrm{Li_{7}Sn_{1}}$ and their corresponding polyhedral representation.\label{fig:Unknown-structures}}
\end{figure}

\paragraph{$\textrm{\textbf{L\ensuremath{\textrm{i}_{8}}S\ensuremath{\textrm{n}_{3}}:}}$}

In 1996, Gasior et al.\cite{Gasior_1996} revealed a possible existence
of a new stoichiometry $\mathrm{Li_{8}Sn_{3}}$ along with the seven
previously reported compounds when they studied the Li\textendash Sn
system by means of electromotive force (EMF) method at the composition
range of 2.4\textendash 95.2 at\% of Li. However, to the best of our
knowledge the crystal structure of that newly predicted phase is yet
unknown and our calculations predict its structure for the first time.
We predicted $\mathrm{Li_{8}Sn_{3}}$ to be one of the most stable
Li-Sn compound (also at 1 atm pressure) and identify it to crystallize
in trigonal crystal system with $\mathrm{R\overline{3}m}$ (Z=3) space
group symmetry with $\mathrm{a=4.684}$ \AA and $\mathrm{c=31.582}$
\AA at ambient pressure (Fig.\ref{fig:Unknown-structures}(i) and
Table \ref{tab:structural-parameters-1}). Interestingly, the structure
of $\mathrm{Li_{8}Sn_{3}}$ is isostructure with $\mathrm{Li_{8}Ge_{3}}$
\cite{Morris_2014} and high pressure $\mathrm{Li_{8}C_{3}}$ structure.\cite{Lin@2015}
All these compounds have Li as cation and anion belonging to group
IV element of periodic table, thus, should have similar structure.
This gives us confidence about correctness of our predicted structure.
On investigating the structure in detail, we found that the arrangement
of atoms in $\mathrm{Li_{8}Sn_{3}}$ is very similar to $\mathrm{Li_{13}Sn_{5}}$
structure. The only difference is that dimers and monomers of Sn repeat
after every 4 Li atoms instead of 5 Li atoms in case of $\mathrm{Li_{13}Sn_{5}}$
structure. The dumbbell length of Sn atoms (\textasciitilde{}2.91
Å) however remains same as in $\mathrm{Li_{5}Sn_{2}}$ and $\mathrm{Li_{13}Sn_{5}}$
structures. With pressure the Li-Li, Li-Sn, and Sn-Sn distance changes
from 2.80 \AA, 2.80 \AA, and 2.91 \AA at 1 atm pressure to 2.47
\AA, 2.51 \AA, and 2.76 \AA, respectively, at 20 GPa (Fig. S12
and Table T11 in SI).

\paragraph{$\textrm{\textbf{L\ensuremath{\textrm{i}_{3}}S\ensuremath{\textrm{n}_{1}}:}}$}

More than a decade ago, Thackeray et. al.\cite{Thackeray@2003} predicted
existence of f.c.c.-$\mathrm{Li_{3}Sn_{1}}$ in the intermediate step
of lithiation in Sn anode on the basis of argument that it is the
basic building block of the experimentally known highest lithiated
phase of Sn, i.e., $\mathrm{Li_{4.4}Sn}$\cite{Li22Sn5} (corrected
stoichiometry is $\mathrm{Li_{4.25}Sn}$\cite{Li17Sn4_Gowar,Li17Sn4_Lipu}).
They proposed that the transition from $\mathrm{Li_{3}Sn_{1}}$ to
$\mathrm{Li_{4.4}Sn}$ can be stabilized by a small amount of residual
copper which helps them to crystallize in f.c.c. structure. Interestingly,
our calculations also predicted $\mathrm{Li_{3}Sn_{1}}$ to be thermodynamically
stable above 12 GPa pressure, where it crystallizes in $\mathrm{Fm\overline{3}m}$
(f.c.c.) structure with four formula unit (i.e., Z=4) and lattice
parameter as $\mathrm{a=5.867}$ \AA at 20 GPa. The structure is
isostructural to $\mathrm{Li_{3}Pb_{1}}$ structure, which exist as
a stable compound in the Li-Pb phase diagram.\cite{Zalkin@1956} $\mathrm{Fm\overline{3}m}$-$\mathrm{Li_{3}Sn_{1}}$
crystal compromises of ``...ABAB...'' planner stacking along any
of its crystallographic axis (\textcolor{blue}{Fig}.\ref{fig:Unknown-structures}(iii)).
Under this arrangement the formation of the structure can be interpreted
by considering each ``A'' plane consisting of pure Li atoms and
then the subsequent first neighbour planes (i.e., B plane) exhibit
both Li and Sn atoms which are arranged in the plane in alternative
manner. However, it should be noted here that at ambient pressure
(and 0 K temperature), our evolutionary structure search predicts
a distorted f.c.c. structure (\textcolor{blue}{Fig}.\ref{fig:Unknown-structures}(ii)),
space group $\mathrm{P2/m}$, Z = 3 with $\mathrm{a=6.885}$ \AA,
$\mathrm{b=4.623}$ \AA, $\mathrm{c=7.060}$ \AA and $\beta=104.07^{\circ}$)
to be energetically more favorable than the $\mathrm{Fm\overline{3}m}$
- $\mathrm{Li_{3}Sn_{1}}$ (which though transforms into latter structure
above 0.5 GPa pressure (\textcolor{blue}{Fig.}S7(a) in SI)), and this
leads to a slight discrepancy between our results and prediction made
by Thackeray et al.\cite{Thackeray@2003} We attribute this discrepancy
to the fact that they considered room temperature while our DFT calculations
are fundamentally restricted to deal with 0 K temperature only. Still,
in order to picturize the exact phase that may (metastably) exist
during lithiation of Sn at ambient pressure and temperature, we plotted
the Gibbs free energy for both the phases with temperature at 1 atm
(\textcolor{blue}{Fig.}S7(b) in SI). The contribution to free energy
at finite temperature comes from both electronic excitations as well
as lattice vibrations. But, the contribution of thermal excitation
of electrons to free energy is very small and can be neglected without
any loss of generality. Therefore, we only calculated the contribution
to free energy due to the lattice excitations under quasi-harmonic
approximation. On analyzing \textcolor{black}{Fig.S7(b) in SI}, phase
transition from $\mathrm{P2/m}$ to $\mathrm{Fm\overline{3}m}$ structure
near the room temperature ($\geq$311 K) can clearly be noticed. Thus,
in agreement with prediction made by Thackeray et al.,\cite{Thackeray@2003}
our study affirms the formation of $\mathrm{Fm\overline{3}m}$ - $\mathrm{Li_{3}Sn_{1}}$
phase during the room temperature lithiation of Sn anode in LIBs.
The shortest Li-Li, Li-Sn, and Sn-Sn distance in $\mathrm{Li_{3}Sn_{1}}$
are found to reduce from 2.77 \AA, 2.78 \AA, and 2.95 \AA in $\textrm{P2/m}$
phase at 1 atm pressure to 2.55 \AA, 2.55 \AA, and 4.15 \AA, respectively,
at 20 GPa in $\mathrm{Fm\overline{3}m}$ phase (Fig. S12 and Table
T11 in SI).

\paragraph{$\textrm{\textbf{L\ensuremath{\textrm{i}_{4}}S\ensuremath{\textrm{n}_{1}}:}}$}

In mid 80's, Alblas et al.\cite{Alblas_1984} investigated structure
of molten Li-Sn alloys using neutron diffraction measurement and discovered
a strong short range ordering behavior near $\mathrm{Li_{4}Sn_{1}}$
composition.\cite{Alblas_1984} Almost same phenomenon was later observed
from the extended \textit{\textcolor{black}{ab}} \textit{initio }molecular
dynamics (MD) simulations, as well.\cite{Genser_2001} However, to
the best of our knowledge no information about the phase and the structure
for the crystalline $\mathrm{Li_{4}Sn_{1}}$ is yet reported. For
the first time, our study reveal two crystalline phases for $\mathrm{Li_{4}Sn_{1}}$
in the investigated pressure range: trigonal crystal system with space
group $\mathrm{R\overline{3}m}$ (Z =3, at 1 atm pressure $\mathrm{a=4.742}$
\AA and $\mathrm{c=13.932}$ \AA) between 1 atm\textendash 13.4
GPa and tetragonal system with space group $\mathrm{I4/m}$ (Z = 10,
$\mathrm{a=11.965}$ and $\mathrm{c=4.241}$ \AA at 20 GPa) from
13.4 GPa to 20 GPa (Fig.\ref{Convex-hull-1}(b)). Though the compound
remains metastable till 16 GPa and gets thermodynamically stable only
above that but, a structural phase transition from $\mathrm{R\overline{3}m}$
to $\mathrm{I4/m}$ is observed at \textasciitilde{}13 GPa pressure
(\textcolor{black}{Fig.}S9 in SI). In $\mathrm{R\overline{3}m}$-$\mathrm{Li_{4}Sn_{1}}$,
one can notice atomic chains parallel to hexagonal axis (\textcolor{black}{Fig.}\textbf{\textcolor{blue}{\ref{fig:Unknown-structures}}}(iv)-(a)).
Each chain consists of periodic monomers of Sn after every 4 Li atoms.
However, in polyhedral representations of the same, (distorted) cubic
hexahedron (where length of each side of the distorted cubes is either
3.18 Å or 3.42 Å) with Li atoms at the vertex positions and Sn atoms
at the body center positions can be seen in the structure (\textcolor{black}{Fig.}\textbf{\textcolor{blue}{\ref{fig:Unknown-structures}}}(iv)-(b)).
In a cube, each of six Li atoms are connected with three nearby Sn
atoms from the adjacent cubes, while each of the other two Li-atoms
are connected with one Sn atom within the cube. Thus, the formation
of the structure can be interpreted on the basis of extended Zintl\textendash Klemm
principle where Sn atoms fulfill the ``octet'' by acquiring two
electrons from six Li-atoms and two more electrons from other two
Li-atoms, as expected. On the other hand, the high pressure $\mathrm{I4/m}$
phase compromises ``...ABAB...'' planner stacking along its crystallographic
$\mathrm{c}$-axis (\textcolor{black}{Fig.}\textbf{\textcolor{blue}{\ref{fig:Unknown-structures}}}(v)).
The atomic layers represented here by A and B are identical and have
the same 2D periodicity in the ab plane, but adjacent layers are displaced
relative to each other by $\mathrm{d_{+}=a(\frac{1}{2},}\frac{1}{2},0)$.
The shortest Li-Li, Li-Sn, and Sn-Sn distance in $\mathrm{R\overline{3}m}$-$\mathrm{Li_{3}Sn_{1}}$
phase reduce from 2.60 \AA, 2.86 \AA, and 4.75 \AA at 1 atm pressure
to 2.23 \AA, 2.48 \AA, and 4.19 \AA, respectively, at 20 GPa in
$\mathrm{I4/m}$ phase (Fig. S12 and Table T11 in SI).

\subsubsection{\textup{Newly Discovered Li-Sn Compounds}}

Experimentally, \emph{$\mathrm{Li_{17}Sn_{4}}$} with space group
$\mathrm{F\overline{4}3m}$ and Z = 20 is known as the highest lithiated
compound of Sn at ambient pressure. However, our \textit{\textcolor{black}{ab}}
\textit{initio} evolutionary algorithm search reveals the possible
existence of two extremely Li-rich compounds, i.e., $\mathrm{Li_{5}Sn_{1}}$
and $\mathrm{Li_{7}Sn_{1}}$, at high pressure. Here, the number of
Li atoms exceeds the highest formal valence of 4$^{-}$ for Sn. As
a consequence, both the systems are remarkably electron rich. Though
at ambient pressure both, $\mathrm{Li_{5}Sn_{1}}$ and $\mathrm{Li_{7}Sn_{1}}$
are found metastable but we believe that these stoichiometries might
be recovered during the quenching from high pressure to ambient conditions
as it does not exhibit modes of negative frequency at ambient pressure
(Fig. S1 in SI). If this can be achieved then the Sn anode with higher
specific capacity and probably better mechanical properties can be
realized.

\paragraph{$\textrm{\textbf{L\ensuremath{\textrm{i}_{5}}S\ensuremath{\textrm{n}_{1}}:}}$}

Our calculations predict $\mathrm{Li_{5}Sn_{1}}$ to crystallize in
monoclinic $\mathrm{C2/m}$ phase with Z = 8 through out the investigated
pressure range. The lattice parameters at 1 atm pressure are found
as: $\mathrm{a=15.918}$ \AA, $\mathrm{b=5.738}$ \AA, $\mathrm{c=12.043}$
\AA, and $\beta=128.86^{\circ}$. The atomic arrangement in $\mathrm{Li_{5}Sn_{1}}$
can be described as Li sharing 14 fold $\mathrm{Li_{14}Sn}$ elongated
hexagonal bipyramid with Sn atom located at the center position (\textcolor{black}{Fig.}\ref{fig:Unknown-structures}(vi)).
The Li-Sn distance in the cage varies from $2.88$ Å - $3.15$ Å,
while Sn atoms are well separated from each other by distance above
$4.66$ Å (Fig. S12 and Table T11 in SI). It should also be noted
that even at high pressure the Sn atoms remain isolated in $\mathrm{Li_{5}Sn_{1}}$. 

\paragraph{$\textrm{\textbf{L\ensuremath{\textrm{i}_{7}}S\ensuremath{\textrm{n}_{1}}:}}$}

As per our calculations $\mathrm{Li_{7}Sn_{1}}$ is found to exhibit
two possible phases in the pressure range of 1 atm \textendash{} 20
GPa. $\mathrm{Li_{7}Sn_{1}}$ first crystallizes in a monoclinic structure
with $\mathrm{C2/m}$ symmetry (Z= 4, $\mathrm{a=15.678}$ \AA, $\mathrm{b=4.826}$
\AA, $\mathrm{c=8.058}$ \AA, and $\beta=102.67^{\circ}$) at ambient
pressure, and later transforms to triclinic structure with $\mathrm{P\overline{1}}$
symmetry (Z = 1, $\mathrm{a=4.609}$\AA, $\mathrm{b=6.693}$\AA,
$\mathrm{c=6.697}$ \AA, and $\alpha=67.57^{\circ}$, $\beta=79.92^{\circ}$,
$\gamma=80.25^{\circ}$ at 20 GPa) above 10.7 GPa (\textcolor{black}{Fig.}S10
in SI). Similar to $\mathrm{Li_{5}Sn_{1}}$, at 1 atm pressure $\textrm{C2/m}-\mathrm{Li_{7}Sn_{1}}$
also possess Li sharing 14 fold $\mathrm{Li_{14}Sn}$ elongated hexagonal
bipyramid with Sn atom located at the center position (\textcolor{black}{Fig.}\ref{fig:Unknown-structures}(vii)).
The Li-Sn distance in the cage is found to vary from $2.83$ Å to
$3.38$ Å while Sn-Sn shortest distance is determined as $\sim4.83$
Å. The high pressure $\mathrm{P\overline{1}}$-$\mathrm{Li_{7}Sn_{1}}$
structure consists of face-sharing 16-fold $\mathrm{Li_{16}Sn}$ octadecahedrons
(\textcolor{black}{Fig.}\ref{fig:Unknown-structures}(viii)). Here
the coordination number of Sn atoms increases from 14 to 16 having
Li-Sn distance in the range of $2.65$ Å to $2.85$ Å at 20 GPa pressure
(Fig. S12 and Table T11 in SI).

\subsection{Volume Expansion}

It is known that the massive structural change and volume expansion
of the order of $\sim240\%$ during lithiation/delithiation of Sn
lead to severe cracking and pulverization of the electrode and capacity
fading, which restricts the use of Sn as an anode material in commonly
used Li-ion batteries.\cite{Zhang_2011,Chou_2011,Tian_2015} However,
application of pressure can help in reducing this drastic change in
volume during lithiation/delithiation.\cite{Zeng_2015} It is evident
from Figure \ref{fig:Volume-expansion}(a) that volume expands quadratically
with the increase in Li-concentration, which is in agreement with
the results of Li et al.,\cite{Li@2013} but simultaneously, it also
contracts with the increase in pressure. For example, as compared
to $\beta-$Sn the volume of the most stable Li$_{13}$Sn$_{5}$ compound
increases by $\sim131\%$, $101\%$, $84\%$, and $62\%$, when the
pressure is $1$ atm, $5$, $10$, and $20$ GPa, respectively. Similarly,
as compared to the volume of $\beta-$Sn at $1$ atm pressure, the
respective increase in the volume of Li-rich phases like Li$_{17}$Sn$_{4}$/Li$_{5}$Sn$_{1}$/Li$_{7}$Sn$_{1}$is
observed to be $\sim240/280/425\%,$ $191/220/334\%$, $165/190/288\%$,
$130/150/230\%$, when the pressure is $1$ atm, $5$, $10$, and
$20$ GPa. Thus, it is clear from our calculations that, in general,
pressure reduces the volume expansion as expected, and in particular,
a pressure of $20$ GPa helps in reducing this volume expansion by
roughly half as compared to $1$ atm pressure. This also leads to
increase in energy density. Pressure can be efficiently generated
in Li-Sn compounds by chemical doping of an impurity, by a proper
selection of substrate material such that it induces pressure due
to lattice mismatch on Sn anode deposited on it or, by imposing pressure
externally, to obtain pressure induced Li-rich phases. However, using
directly such high pressure (of the order of GPa) may not be suitable
for battery applications and therefore, we propose to use high pressure
phases quenched to ambient pressure. We believe that this may still
prove useful in producing relatively less change in volume during
lithiation/delithiation process and providing better mechanical integrity,
which is essentially required to improve performance of Li-Sn batteries.

\subsection{Electro-chemical Properties}

\paragraph{Gravimetric Capacity:}

The gravimetric capacity of all Li-Sn compounds explored in this work
are tabulated in Table \ref{tab:structural-parameters-1}. Since the
gravimetric capacity is directly proportional to the concentration
of Li (eq.\ref{eq:3}), it increases with the increase in Li concentration.
Interestingly our calculations predict several novel pressure-induced
high capacity Li-rich compositions ($\mathrm{Li_{3}Sn_{1}}$, $\mathrm{Li_{4}Sn_{1}}$,
$\mathrm{Li_{5}Sn_{1}}$, and $\mathrm{Li_{7}Sn_{1}}$) which are
although metastable at ambient pressure, but should be synthesizable
because of non-existence of any negative frequencies in phonon dispersion
curve at 1 atm (Fig.S1 in SI). Specifically, Li$_{5}$Sn$_{1}$ and
Li$_{7}$Sn$_{1}$are the two compounds that exhibit higher specific
capacity ($\sim1129$ and $\sim1580$ mAh g$^{-1}$) than yet known
most lithiated Li$_{17}$Sn$_{4}$ phase ($\sim960$ mAh g$^{-1}$).
In fact, Li$_{7}$Sn$_{1}$ exhibit much higher capacity, $\sim$5
times than LiC$_{6}$($\sim372$ mAh/g) and $\gtrsim$1.5 times than
Li$_{17}$Sn$_{4}$. Since pressure does not affect the capacity directly,
our results suggest that a Li-ion battery with higher capacity might
be formed if the quenched high pressure \textendash{} high capacity
phase is used as the pre-lithiated anode material. 

\begin{figure}
\includegraphics[width=7.5cm]{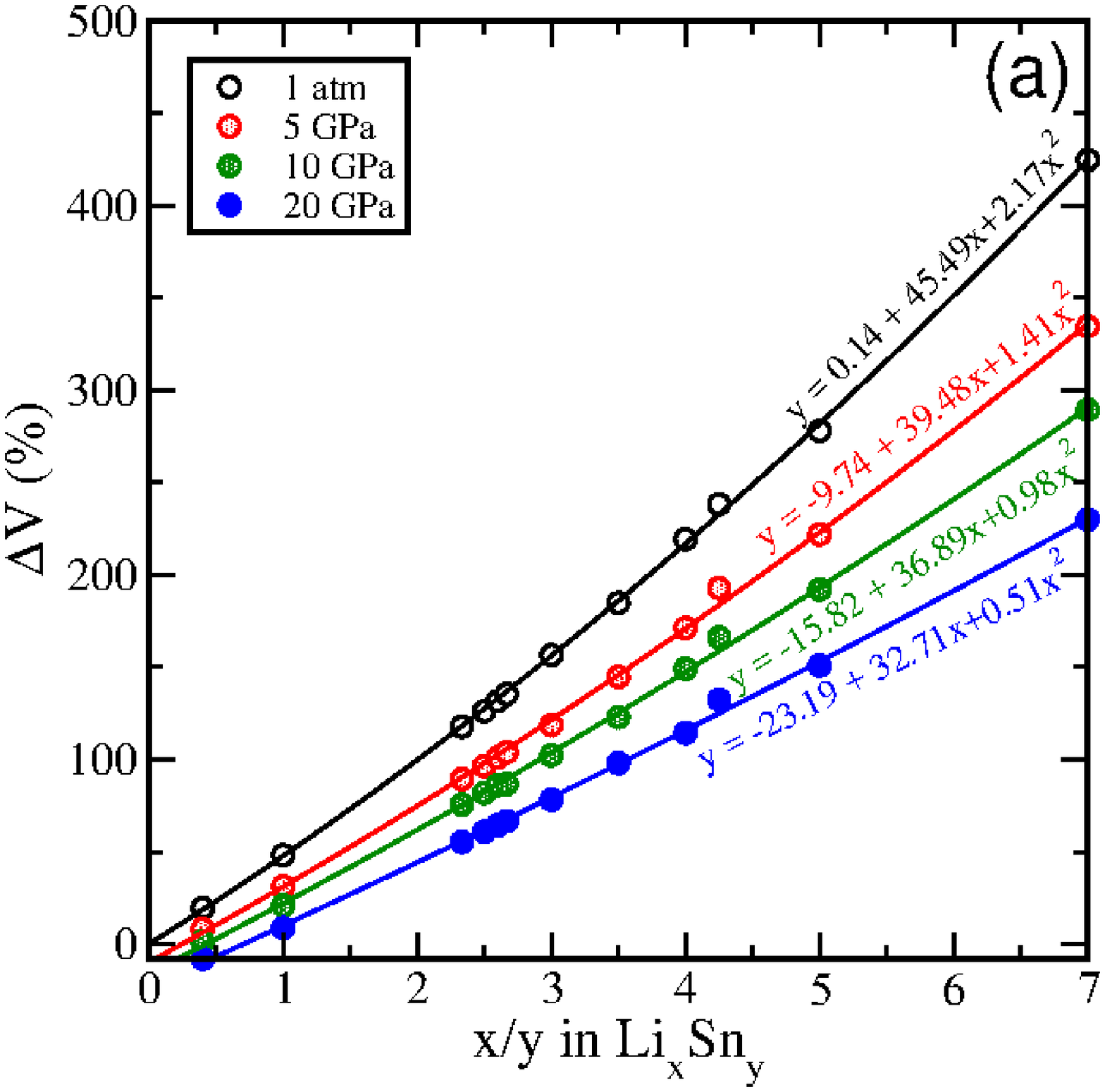}~\includegraphics[width=7.5cm]{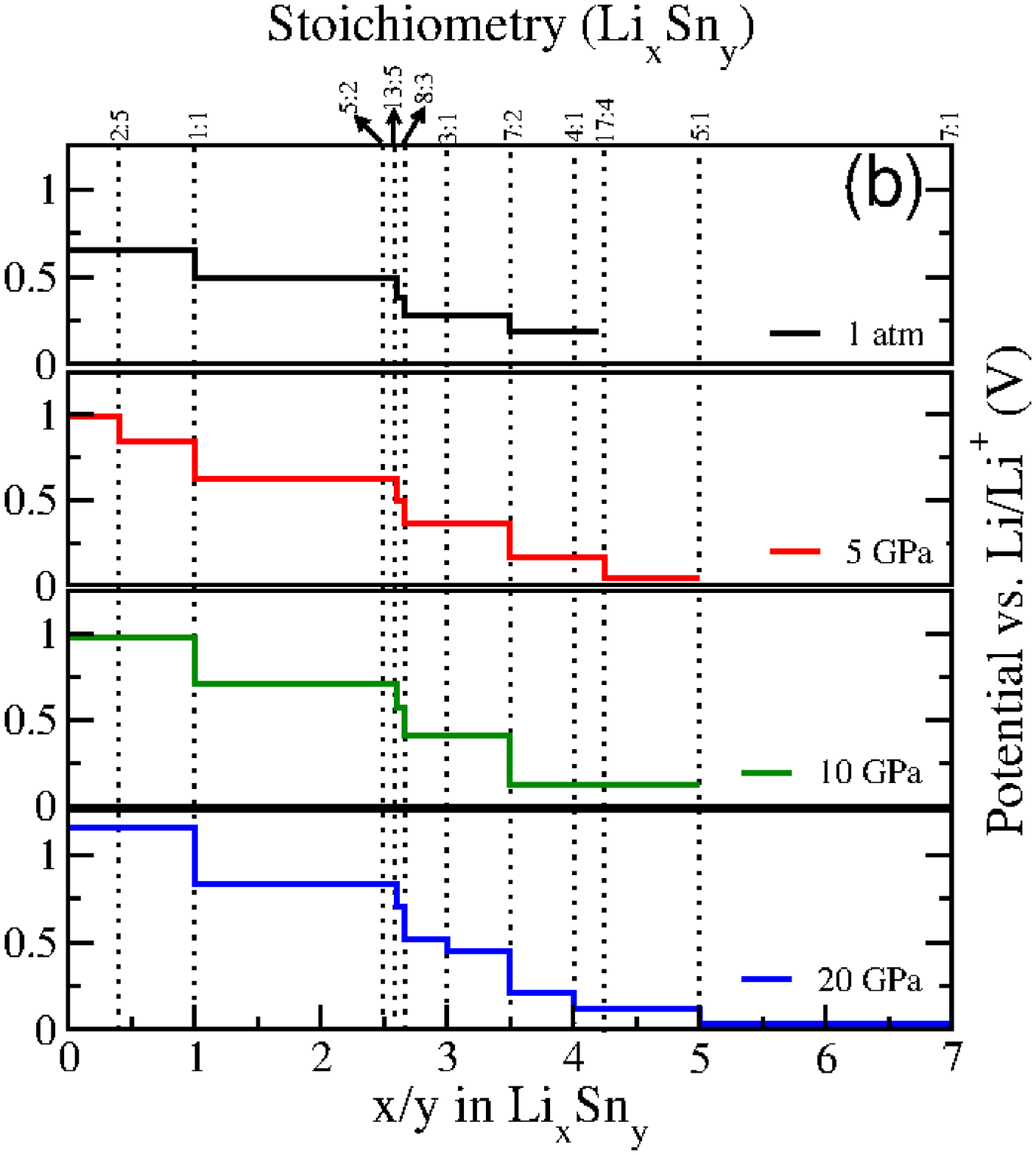}\caption{(a) Volume expansion of Li-Sn compounds at 1 atm, 5, 10, and 20 GPa
pressure with respect to the volume of $\beta$-Sn ({\scriptsize{}$\mathrm{I4_{1}/amd}$})
at 1 atm pressure. (b) Potential-composition curves at 1 atm, 5, 10,
and 20 GPa pressure for stable Li-Sn compounds that lie on the convex
hull in Fig.\ref{Convex-hull-1}(a).\label{fig:Volume-expansion}}
\end{figure}

\paragraph{Li-insertion Voltage:}

Average lithium insertion voltages in Sn anode of Li-ion battery are
presented in Figure \ref{fig:Volume-expansion}(b) for the $1$ atm,
$5$, $10$, and $20$ GPa pressure values. These are calculated using
DFT total energies, assuming that the displaced charge is due to Li
and that the reaction proceeds sequentially through the phases lie
on the tie-lines of the convex hull corresponding to respective pressure
value.\cite{Ceder@1997} On analysing Figure \ref{fig:Volume-expansion}(b),
we found that the average Li-insertion voltage increases with the
increase in pressure. At $1$ atm pressure the voltage drops from
$\sim0.65$ V to $0.18$ V, when Sn is fully lithiated to Li$_{17}$Sn$_{4}$phase.
Note that as per our calculations, at $1$ atm pressure (and $0$
K), Li$_{2}$Sn$_{5}$ does not lie on the tie-line of the convex
hull, thus, it is not included in the voltage-composition curve for
the $1$ atm pressure. The $5$ GPa curve includes two new phases,
viz. Li$_{2}$Sn$_{5}$ and Li$_{5}$Sn$_{1}$, which increases the
voltage range by $\sim50\%$, as the voltage drops from $1$ V to
$0.04$ V, when Sn is lithiated to Li$_{5}$Sn$_{1}$ phase. Compared
to $5$ GPa, the voltage decreases slightly in case of $10$ GPa,
as\textbf{ }Li$_{2}$Sn\textbf{$_{5}$ }and Li$_{17}$Sn$_{4}$ become
unstable at this pressure value. However, the voltage again increases
to $1.15$ V at $20$ GPa as several Li-rich phases like Li$_{5}$Sn$_{2}$,
Li$_{3}$Sn$_{1}$, Li$_{4}$Sn$_{1}$, Li$_{5}$Sn$_{1}$, and Li$_{7}$Sn$_{1}$
become stable, which increases the number of successive two-phase
reactions. Thus, the application of pressure not just increases the
energy density but also provides relatively higher insertion voltage
which is desired to make the battery safer, as it reduces the chance
of lithium plating that results into dendrites causing short-circuit
in the cell.\cite{Morris_2014} 

\subsection{Mechanical properties of Li-Sn compounds.}

A composition (having negative formation enthalpy) which does not
lie on the tie-line of the convex hull can be regarded as metastable,
only and only if, it satisfies the criteria of both, dynamical and
mechanical stability. \textcolor{black}{In above section we have shown
that all the new phases (stable or metastable) revealed in this work
are dynamically stable by calculating their phonon dispersion curves
(See Fig.S1 \& S2 in SI).} Now to check the mechanical stability of
the same, we used the Born criteria of mechanical stability described
by Wu et al. and Mouhat et al.,\cite{Wu@2007,Mouhat@2014} which highly
depends on the symmetry of the crystal. To check the criteria, the
elastic constants of all thermodynamically stable and metastable Li-Sn
compounds at pressure value of 1 atm, 5, 10, and 20 GPa are calculated
(Table T3, T4, T5, \& T6 of SI). Using the elastic constants corresponding
to the particular composition and pressure in the Born criteria, it
is interestingly found that all examined compounds are mechanically
stable. This further supports our prediction for the possible existence
of newly predicted compounds in the Li-Sn binary system. 

One of the major bottleneck of using Sn anode is the elastic softening
and mechanical degradation with lithiation, which leads to electrode
damage and fracture. In order to see the influence of pressure on
mechanical properties, we examined the elastic properties of all investigated
Li-Sn stoichiometries at various pressure values. Such studies may
prove helpful in restricting the failure mechanism of Li-ion batteries
during charging\textendash discharging process. However, it is noteworthy
here that DFT calculated elastic constants are for the single crystal,
while the Li-Sn compounds formed during lithiation in Li-ion batteries,
are usually in polycrystalline micro-structures state.\cite{Stournara@2012}
Thus, in order to characterize the mechanical performance of Li-Sn
compounds, we calculate the isotropic elastic constants for a polycrystalline
aggregate by averaging the computed anisotropic single crystal elastic
constants over all possible orientations of the grains in a polycrystal.
The continuum model based on Reuss and Voigt approaches are explicitly
used for such studies (\textcolor{black}{see SI} for more details).

\begin{figure}
\centering{}\includegraphics[width=4.5cm]{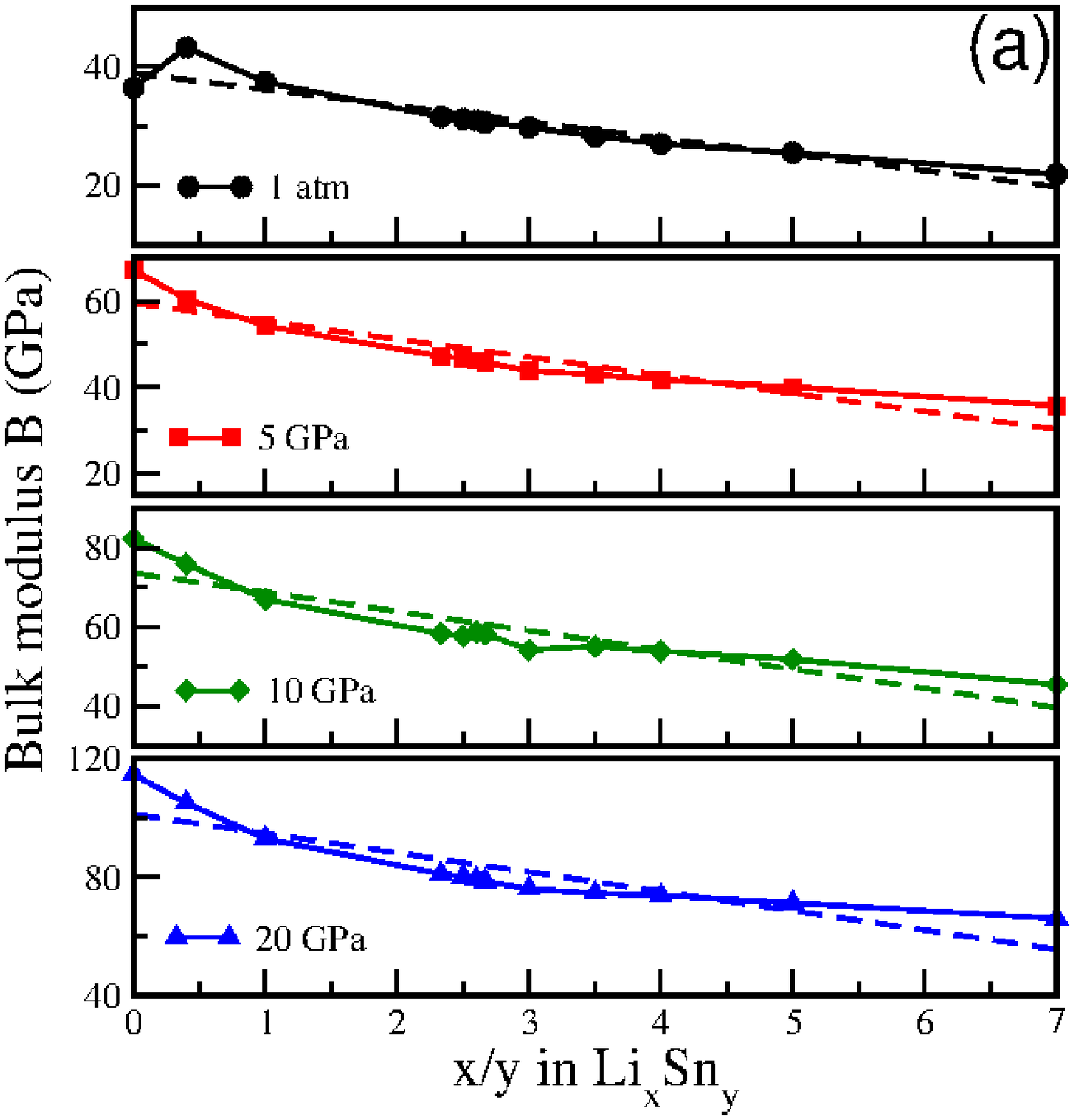}~\includegraphics[width=4.5cm]{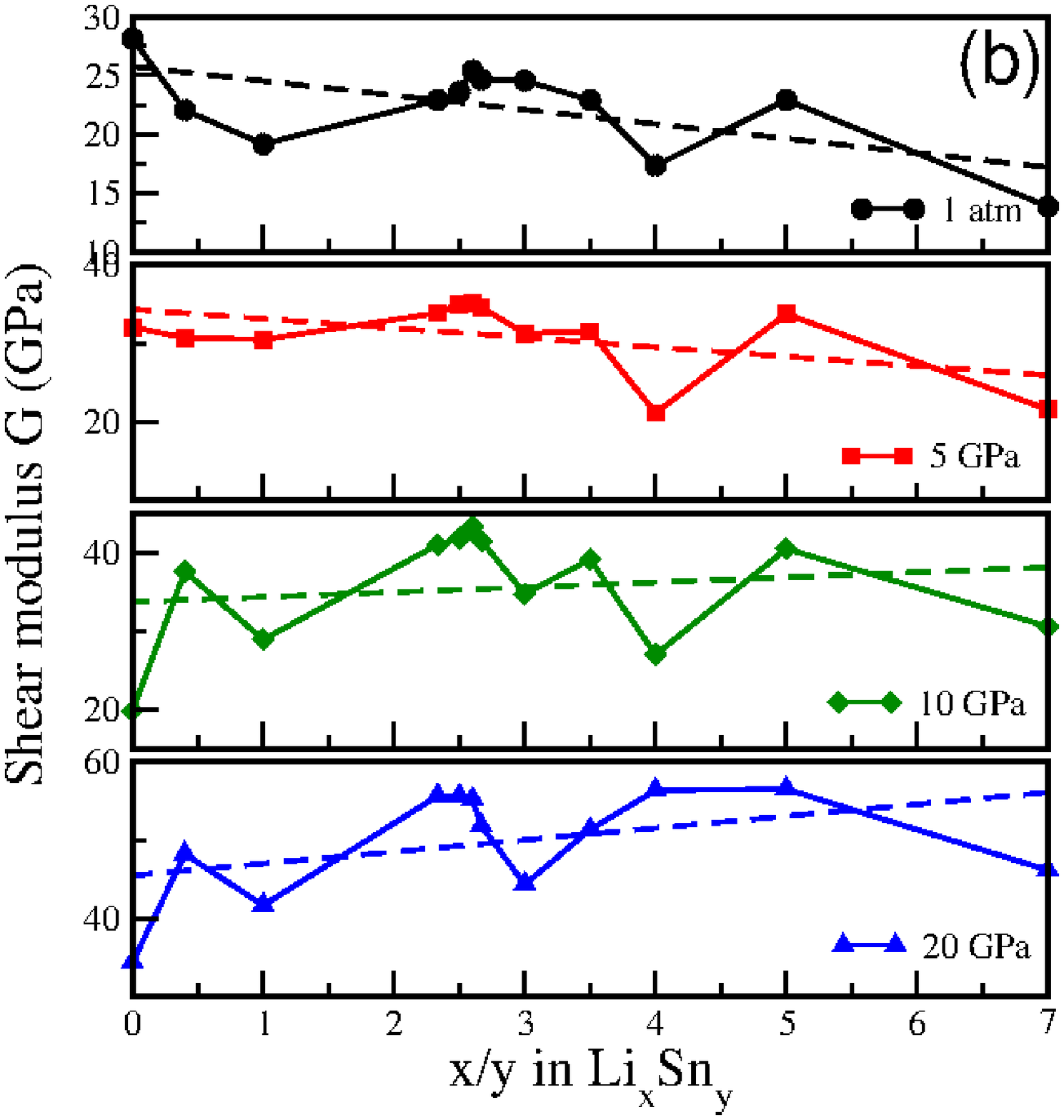}~\includegraphics[width=4.5cm]{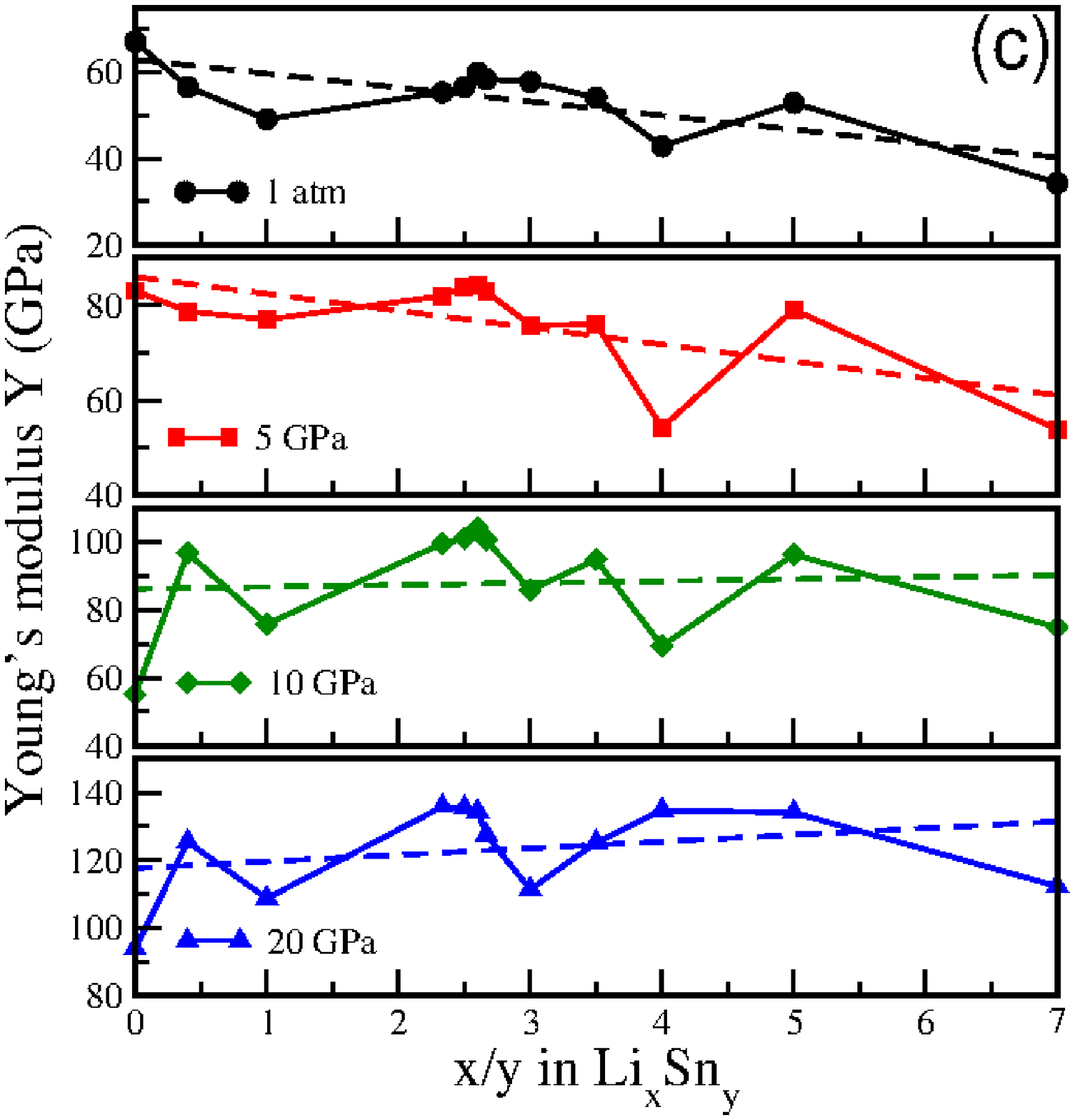}~\includegraphics[width=4.5cm]{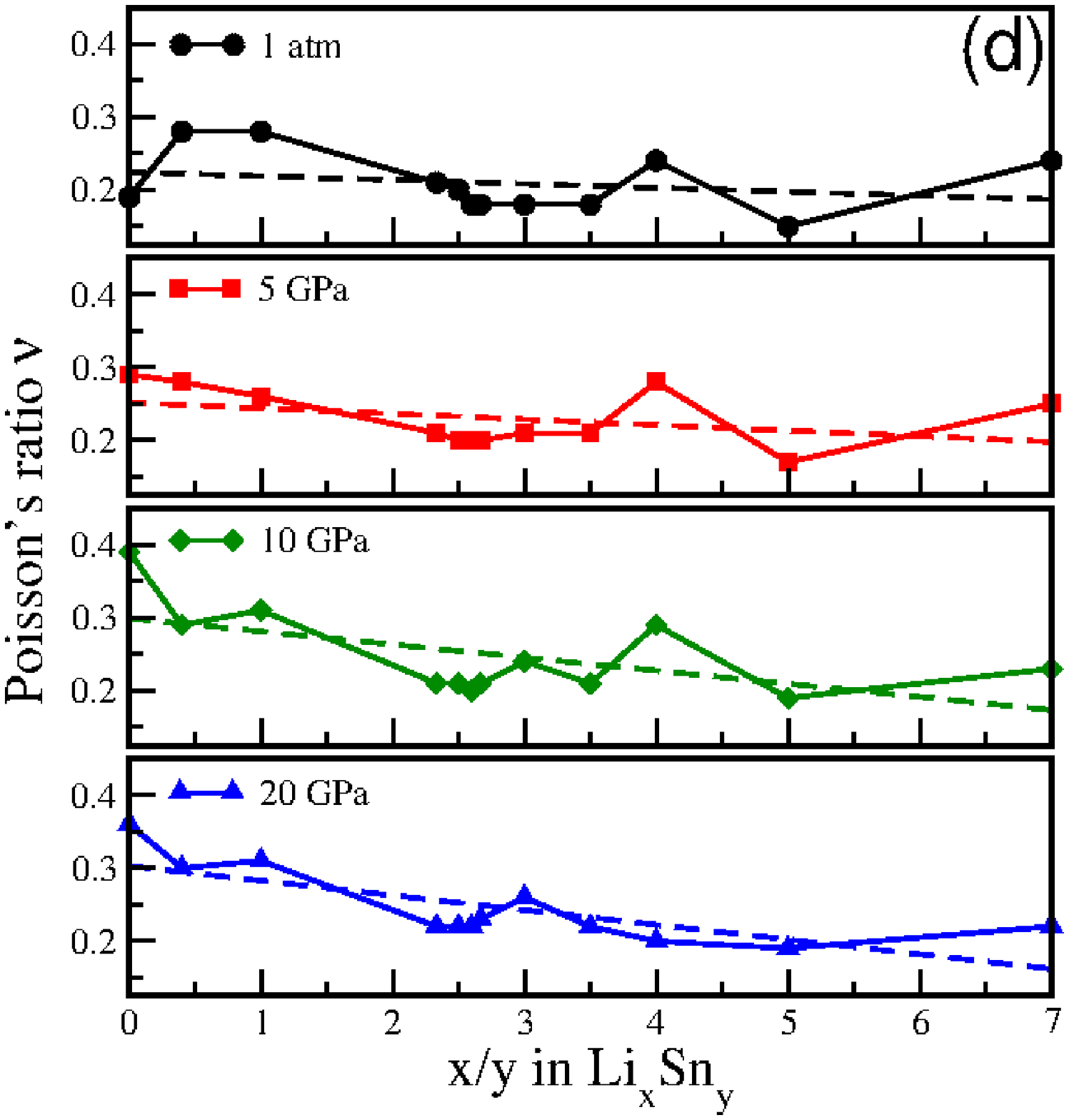}\caption{(a) Bulk modulus (B), (b) Shear modulus (G), (c) Young's modulus (Y),
and (d) Poisson's ratio ($\nu$) of Li-Sn compounds with respect to
lithium fraction x/y in Li$_{\mathrm{x}}$Sn$_{\mathrm{y}}$at pressure
value of 1 atm, 5, 10, and 20 GPa. \label{fig:Elastic-Properties}}
\end{figure}

The orientation averaged bulk modulus, Young's modulus, shear modulus,
and Poisson's ratio with increasing Li concentration in Sn are presented
in Figure \ref{fig:Elastic-Properties}. It can be seen that bulk,
shear, and Young's moduli depend significantly on Li concentration,
as well as pressure. Mostly these moduli decrease with the increase
in Li/Sn fraction (x/y), showing elastic softening in the Li-rich
phases, in agreement with previous theoretical and experimental findings.\cite{Zhang@2015,Stournara@2012,Chen@2008}
However, with the increase in pressure the average value of these
moduli increases substantially. While, as compared to Sn at ambient
pressure a decrease of $\sim$50\% is noticed in the bulk modulus
of most Li-rich phase, $\mathrm{Li_{7}Sn_{1}}$, this difference gets
reduced to $\sim40\%$ at high pressure. Moreover, instead of an approximate
linear decrease, shear and Young's moduli show a linear increase with
the increase in the Li-concentration beyond $\sim$10 GPa. This indicates
a pressure induced relative stiffness in each phase, which strengthens
the Li-Sn phases at high pressure as compared to ambient pressure.
Our calculated Poisson's ratio for all investigated compounds, lies
between 0.1 to 0.4 (i.e., well within the limit of a stable and linear
elastic material), further supports our conclusion, as it is evident
from Fig. \ref{fig:Elastic-Properties}(d) that the relative value
of Poisson ratio for most of the investigated materials increases
with the increase in pressure. 

\begin{table}[H]
\caption{Calculated universal anisotropy index ($A^{U}$) of Li, Sn, and Li-Sn
compounds at 1 atm, 5, 10, 20 GPa pressure. \label{tab:Anisotropy-Index}}

\centering{}%
\begin{tabular}{|c|c|c|c|c|c|c|c|c|c|c|c|c|c|}
\hline 
\multirow{2}{*}{{\tiny{}Pressure}} & \multicolumn{13}{c|}{{\tiny{}$A^{U}$}}\tabularnewline
\cline{2-14} 
 & {\tiny{}$\textrm{Sn}$} & {\tiny{}$\textrm{L\ensuremath{i_{2}}S\ensuremath{n_{5}}}$} & {\tiny{}$\textrm{L\ensuremath{i_{1}}S\ensuremath{n_{1}}}$} & {\tiny{}$\textrm{L\ensuremath{i_{7}}S\ensuremath{n_{3}}}$} & {\tiny{}$\textrm{L\ensuremath{i_{5}}S\ensuremath{n_{2}}}$} & {\tiny{}$\textrm{L\ensuremath{i_{13}}S\ensuremath{n_{5}}}$} & {\tiny{}$\textrm{L\ensuremath{i_{8}}S\ensuremath{n_{3}}}$} & {\tiny{}$\textrm{L\ensuremath{i_{3}}S\ensuremath{n_{1}}}$} & {\tiny{}$\textrm{L\ensuremath{i_{7}}S\ensuremath{n_{2}}}$} & {\tiny{}$\textrm{L\ensuremath{i_{4}}S\ensuremath{n_{1}}}$} & {\tiny{}$\textrm{L\ensuremath{i_{5}}S\ensuremath{n_{1}}}$} & {\tiny{}$\textrm{L\ensuremath{i_{7}}S\ensuremath{n_{1}}}$} & {\tiny{}$\textrm{Li}$}\tabularnewline
\hline 
\hline 
{\tiny{}1 atm} & {\tiny{}1.53} & {\tiny{}0.59} & {\tiny{}0.39} & {\tiny{}3.69} & {\tiny{}2.80} & {\tiny{}2.87} & {\tiny{}2.96} & {\tiny{}2.09} & {\tiny{}1.70} & {\tiny{}6.17} & {\tiny{}0.33} & {\tiny{}6.26} & {\tiny{}7.93}\tabularnewline
\hline 
{\tiny{}5 GPa} & {\tiny{}0.89} & {\tiny{}0.39} & {\tiny{}0.15} & {\tiny{}2.56} & {\tiny{}2.21} & {\tiny{}2.55} & {\tiny{}2.39} & {\tiny{}2.01} & {\tiny{}1.54} & {\tiny{}9.88} & {\tiny{}0.13} & {\tiny{}3.69} & {\tiny{}13.87}\tabularnewline
\hline 
{\tiny{}10 GPa} & {\tiny{}1.88} & {\tiny{}0.61} & {\tiny{}1.37} & {\tiny{}2.36} & {\tiny{}1.92} & {\tiny{}2.31} & {\tiny{}2.33} & {\tiny{}2.68} & {\tiny{}1.33} & {\tiny{}7.32} & {\tiny{}0.12} & {\tiny{}0.85} & {\tiny{}2.14}\tabularnewline
\hline 
{\tiny{}20 GPa} & {\tiny{}1.77} & {\tiny{}0.64} & {\tiny{}0.98} & {\tiny{}2.01} & {\tiny{}1.79} & {\tiny{}2.09} & {\tiny{}2.34} & {\tiny{}2.70} & {\tiny{}1.14} & {\tiny{}0.56} & {\tiny{}0.14} & {\tiny{}0.21} & {\tiny{}1.41}\tabularnewline
\hline 
\end{tabular}
\end{table}

To further check the mechanical robustness of compressed Li-Sn compounds,
we also examined the influence of elastic anisotropy with respect
to pressure on Li-Sn compounds, since it is well known that due to
anisotropy of materials micro-cracks can be induced in Sn anode during
charging and discharging.\cite{Zhang@2015} The universal anisotropy
index ($A^{U})$, proposed by Ranganathan et.al.,\cite{Ranganathan@2008}
is used in this work to quantify the elastic anisotropy:

\begin{equation}
A^{U}=5\frac{G_{V}}{G_{R}}+\frac{B_{V}}{B_{R}}-6.\label{eq:4}
\end{equation}
 Zero value of $A^{U}$ indicates the local isotropy, while, the more
deviation of the parameter from zero represents more elastic anisotropy
of the crystalline structure. The calculated elastic anisotropy index
for different pressure values listed in \textcolor{black}{Table}\textcolor{blue}{{}
\ref{tab:Anisotropy-Index}}, show an extremely opposite trend for
the two end members of the Li-Sn system at 1 atm and 5 GPa pressure.
While, $\mathrm{\alpha-Sn}$ at 1 atm and $\mathrm{\beta-Sn}$ at
5 GPa are nearly isotropic with $A^{U}$$=1.53$ and $0.89$, respectively,
a strong anisotropic property is found for b.c.c. Li ($A^{U}=7.93$
at 1 atm and $13.87$ at 5 GPa). The anisotropic behaviour of these
two elements are in good agreement with the previous theoretical and
experiment results.\cite{Ranganathan@2008,Zhang@2015,Felice@1977,House@1960}
However on the contrary, the f.c.c. Li at high pressure is quite isotropic
($A^{U}=2.14-1.14$) and it's isotropic nature improves with the increase
in pressure. It should also be noticed that at ambient pressure, the
medium intercalated-lithium phases (1 $<$ x $<$ 3.5) are moderately
anisotropic, while, except $\mathrm{Li_{5}Sn_{1}}$, the Li-rich phases
(3.5 $<$ x $\leq$ 7) are highly anisotropic. One can see from Table
\ref{tab:Anisotropy-Index} that the value of $A^{U}$ for two of
our newly predicted metastable compounds, i.e., $\mathrm{Li_{4}Sn_{1}}$
and $\mathrm{Li_{7}Sn_{1}}$, at ambient pressure reaches close to
the value of b.c.c Li, with $A^{U}=6.17$ and $6.26$, respectively,
while, $\mathrm{Li_{5}Sn_{1}}$ is found to be most isotropic among
all other Li-Sn compounds with anisotropy index of 0.33. This is not
very surprising as a similar trend was also noticed by Zhang et.al.,\cite{Zhang@2015}
as they observed $\mathrm{Li_{17}Sn_{4}}$ (a closer system to $\mathrm{Li_{5}Sn_{1}}$)
to be more isotropic than any other Li-Sn phases with $A^{U}\sim0$.
The most important thing to notice is that the isotropy improves with
the increase in pressure and even the Li-rich compounds like $\mathrm{Li_{4}Sn_{1}}$
and $\mathrm{Li_{7}Sn_{1}}$ become highly isotropic (equivalent to
$\mathrm{Li_{5}Sn_{1}}$) at high pressure. These results clearly
reveal that pressure improves the mechanical properties of Li-Sn compounds
and thus, we strongly believe that the quenched high pressure phases
may also exhibit superior mechanical properties, which can help in
resolving the current problem of fracture and pulverization in Sn
anode.

\subsection{Electronic Properties and Chemical Bonds.}

\begin{figure}
\includegraphics[width=5cm]{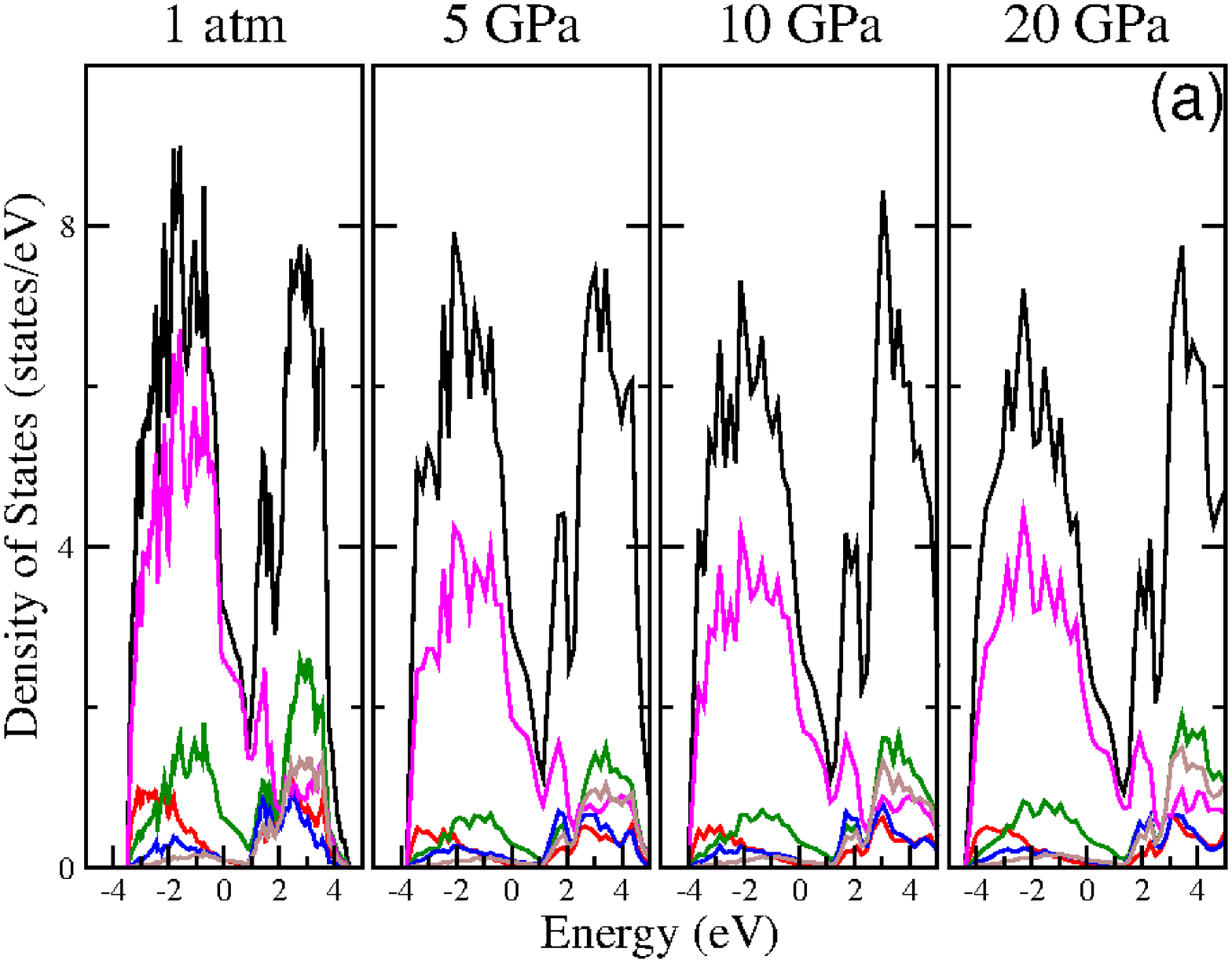}~\includegraphics[width=5cm]{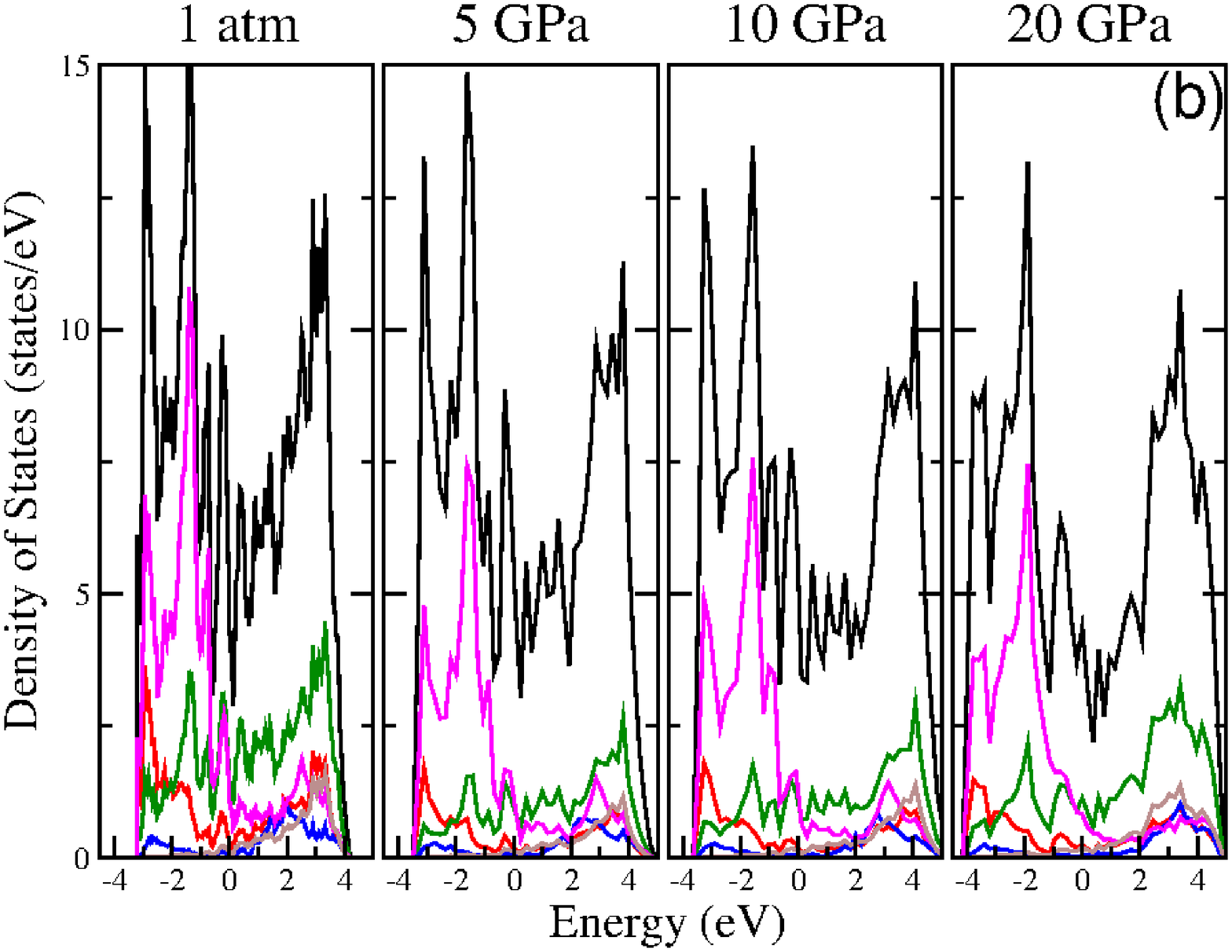}~\includegraphics[width=5cm]{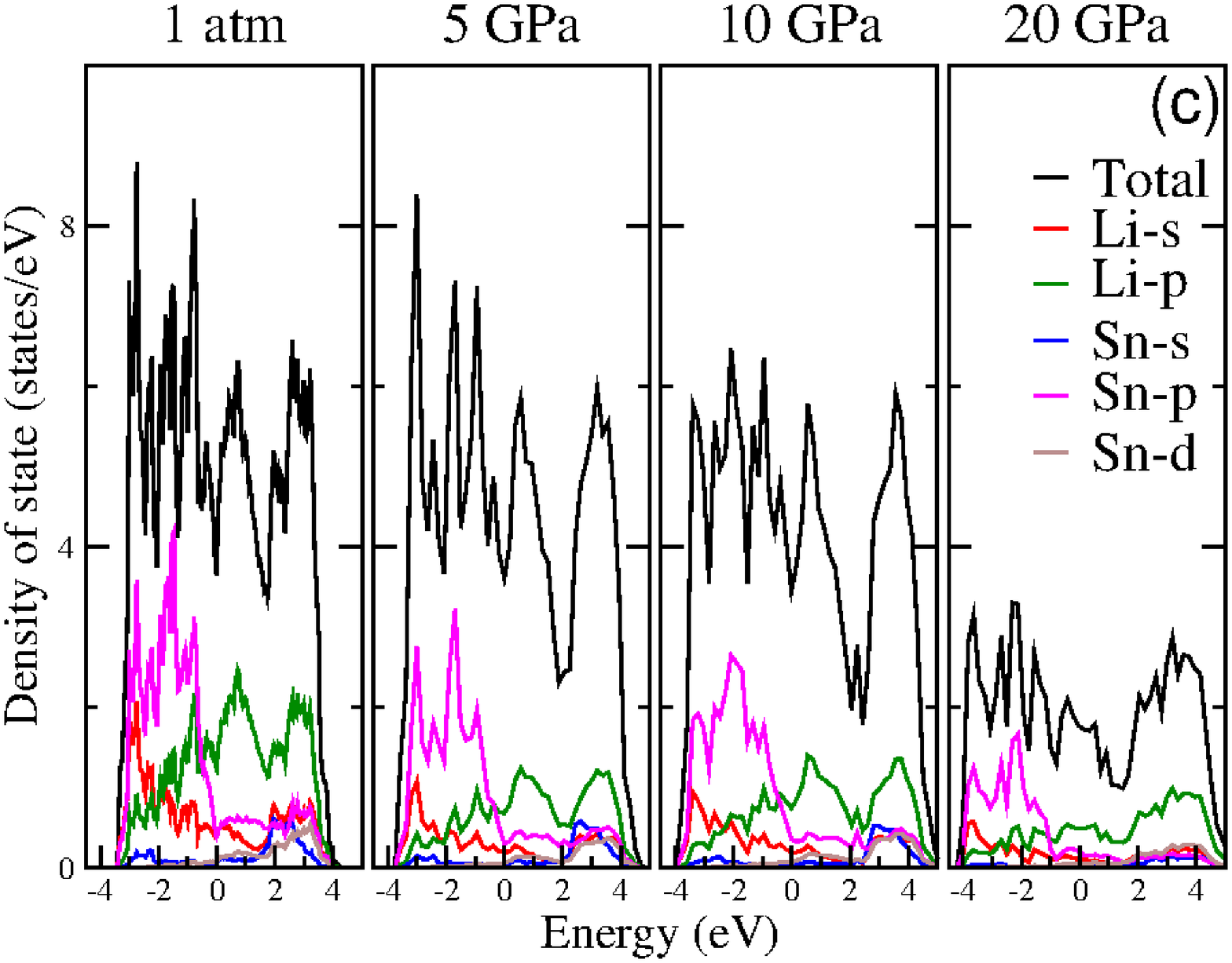}\caption{Total and partial density of states (DOS) of (a) $\mathrm{Li_{8}Sn_{3}}$,
(b) $\mathrm{Li_{5}Sn_{1}}$, and (c) $\mathrm{Li_{7}Sn_{1}}$ at
1 atm, 5, 10, and 20 GPa pressure. \label{fig:DOS}}
\end{figure}

Electronic properties, in general, and electrical conductivity, in
particular, of any electrode material also play crucial role in determining
performance of a Li-ion battery. \textcolor{black}{Fig.S11 in SI}\textcolor{blue}{{}
}depicts the calculated total electronic density of states (DOS) along
with the atom resolved partial density of states (PDOS) of all the
phases at 1 atm. The crossover of several bands at Fermi level ($\mathrm{E}_{\mathrm{F}}$,
set to zero) indicate all Li-Sn compounds to exhibit metallic nature,
thus, providing a better electrical conductivity. For most of the
Li-Sn phases, the PDOS show a large overlap between Sn-$5p$ and Li-$2s$
and/or Li-$2p$ states at the Fermi level, which illustrates the charge
transfer from Li-$2s$ and/or Li-$2p$ to Sn-$5p$. One can see that
the main contribution to total DOS near $\mathrm{E}_{\mathrm{F}}$
comes from the Sn-$5p$ state but, as expected this gets weaken with
the increase in Li-concentration, which is also in agreement with
the work of Zhang et al.\cite{Zhang@2015} on Li-Sn systems at ambient
conditions. Moreover, this is also very similar to what is observed
recently by Zhang et al.\cite{Zhang@2016} and Yang et al.\cite{Yang@2016}
for Li-Si and Li-Au systems at high pressure, respectively. Finally,
in Li-rich phases where $\mathrm{x\geq4}$, Li-$2s$ and Li-$2p$
states dominate and provide major contribution to the total DOS near
$\mathrm{E}_{\mathrm{F}}$, as compared to Sn-$5p$ state, owing to
the exceptionally high concentration of Li (\textcolor{black}{Fig.S11
in SI}). In order to determine the effect of pressure on density of
states, we present DOS for three of the yet unknown/unexplored compounds
($\textrm{L\ensuremath{\textrm{i}_{8}}S\ensuremath{\textrm{n}_{3}}},$
$\textrm{L\ensuremath{\textrm{i}_{5}}S\ensuremath{\textrm{n}_{1}}}$,
and $\textrm{L\ensuremath{\textrm{i}_{7}}S\ensuremath{\textrm{n}_{1}}}$)
at different pressure values in Fig.\ref{fig:DOS}. It is evident
from the figure that the nature of DOS is mainly insensitive to the
applied pressure. The application of pressure in GPa though decreases
the intensity of peaks (density/states) as compared to 1 atm pressure,
but other than that no distinct change is noticed for any particular
phase within the pressure value of 5\textendash 20 GPa. 

In order to relate the chemical, electronic, and mechanical properties,
the average electronegativity (EN) and the net charge on Li and Sn
atoms for each $\mathrm{Li_{x}Sn_{y}}$ compound are also calculated.
The average metallic EN is calculated using formula: $\textrm{\ensuremath{\frac{x(\textrm{EN})_{Li}+y(\textrm{EN})_{Sn}}{x+y}}}$,
where $\textrm{(EN\ensuremath{)_{\textrm{L}\textrm{i}}}}$ and $\textrm{(EN\ensuremath{)_{\textrm{Sn}}}}$
are metallic EN values of Li and Sn, respectively, and $\mathrm{x}$
and $\mathrm{y}$ are the numbers of Li and Sn atoms per chemical
formula. EN describes the ability of an atom or ion to attract bonding
valence electrons, and thus, the average EN of two bonded atoms can
well reflect the strength or nature of the chemical bonds.\cite{Li@2013}
A large value of average EN reflects a strong covalent bonding (Sn-Sn),
while intermediate to low values indicate weak ionic (Li-Sn) and metallic
bonds (Li-Li), respectively. It is evident from the above formula
that average EN does not depend on pressure but it is found that it
decreases monotonically with the increase in Li-concentration, that
also explains the elastic softening in Li-rich compounds. The average
EN for $\textrm{Sn}$, $\textrm{\textrm{L\ensuremath{\textrm{i}_{2}}S\ensuremath{\textrm{n}_{5}}}}$,
$\textrm{L\ensuremath{\textrm{i}_{1}}S\ensuremath{\textrm{n}_{1}}}$,
$\textrm{L\ensuremath{i_{7}}S\ensuremath{\textrm{n}_{3}}}$, $\textrm{L\ensuremath{\textrm{i}_{5}}S\ensuremath{n_{2}}}$,
$\textrm{L\ensuremath{\textrm{i}_{13}}S\ensuremath{\textrm{n}_{5}}}$,
$\textrm{L\ensuremath{i_{8}}S\ensuremath{n_{3}}}$, $\textrm{L\ensuremath{\textrm{i}_{3}}S\ensuremath{\textrm{n}_{1}}}$,
$\textrm{L\ensuremath{\textrm{i}_{7}}S\ensuremath{\textrm{n}_{2}}}$,
$\textrm{L\ensuremath{\textrm{i}_{4}}S\ensuremath{\textrm{n}_{1}}}$,
$\textrm{L\ensuremath{\textrm{i}_{17}}S\ensuremath{\textrm{n}_{4}}}$,
$\textrm{L\ensuremath{\textrm{i}_{5}}S\ensuremath{\textrm{n}_{1}}}$,
and $\textrm{L\ensuremath{\textrm{i}_{7}}S\ensuremath{\textrm{n}_{1}}}$
is calculated to be 2.860, 2.227, 1.753, 1.310, 1.279, 1.261, 1.250,
1.200, 1.138, 1.089, 1.068, 1.015, 0.923, and 0.646, respectively. 

Bader charge further justifies the increase of ionic character in
Li-Sn compounds with higher Li-concentration. In agreement with previous
calculations,\cite{Stournara@2012} our calculations show that when
$x<5$ in $\textrm{L\ensuremath{\textrm{i}_{x}}Sn}$, all Li atoms
donate electrons to Sn atoms and acquire $\sim+1$ charge, while Sn
atoms take $1-4$ electrons, leading to a net charge of roughly $-1$
to $-4$ compared to pristine Sn, depending on the amount of Li and
neighboring Sn atoms, as well as on the anisotropy of the phase.\cite{Stournara@2012}
\textcolor{black}{For example, in $\mathrm{Li_{8}Sn_{3}}$, Sn atoms
forming dumbbell exhibit a charge of -1.96, while monomer of Sn atoms
are found to have -2.68 charge at ambient pressure (Table 12 of SI).
Likewise, ${\color{blue}\mathrm{R\overline{3}m}}$-$\textrm{{\color{blue}L\ensuremath{\textrm{i}_{4}}S\ensuremath{\textrm{n}_{1}}}}$
that contains isolated Sn atoms, exhibit $\sim$-3.28 charge on each
Sn atom at ambient pressure. On the other hand, in $\mathrm{I\mathrm{4/m}}$
-$\textrm{L\ensuremath{\textrm{i}_{4}}S\ensuremath{\textrm{n}_{1}}}$,
there are two types of Sn atoms, one surrounded by 8 Li atoms, carrying
charge of \textminus 3.38 while, others that are enclosed by 7 Li
atoms possess -3.04 charge at 20 GPa pressure (Table 13 in SI).} Well,
in principle, Li can donate $1$ electron to acquire $+1$ charge
and Sn can accept upto $4$ electrons to attain maximum of $-4$ charge,
to form most lithiated Sn compounds like Li$_{4}$Sn$_{1}$or Li$_{17}$Sn$_{4}$.
However, our calculations reveal compounds with higher concentration
of Li, such as, Li$_{5}$Sn$_{1}$ and Li$_{7}$Sn$_{1}$, which suggest
possibility of Sn to earn $-5$ and $-7$ charge, respectively. But,
this seems difficult considering octet rule. Here, our Bader charge
analysis helped us to resolve this ambiguity by providing the charge
distribution in such Li-rich materials. Interestingly, our calculations
illustrate that instead of Sn acquiring more than $-4$ charge, few
of the Li atoms accept electrons and becomes negatively charged (Table
12 \& 13 of SI). In $\mathrm{Li_{5}Sn_{1}}$, Sn exhibits net charge
of $\sim-3.6$, while 36 Li atoms possess $\sim+0.82$ and 4 Li atoms
acquire $-0.03$ charge. Likewise, in $\textrm{L\ensuremath{\textrm{i}_{7}}S\ensuremath{\textrm{n}_{1}}}$,
out of 28 Li atoms in the conventional cell, 8 Li accept electrons
to become negatively charge with charge $\sim0.05-0.28$, at ambient
pressure (see Table 12 of SI). This result is not at all surprising
as the same has been noticed in the case of NaCl$_{7}$\cite{Zhang_2013}and
KCl$_{7}$\cite{Zhang_KCl@2016}, where one of the Cl atom acquires
+ve charge to maintain charge neutrality. The negative charge on few
of the Li also explains the reason of sudden increase in the contribution
of Li-states to the total DOS near Fermi-level. Moreover, except $\textrm{L\ensuremath{\textrm{i}_{7}}S\ensuremath{\textrm{n}_{1}}}$,
the net charge on each atom is not found to be much sensitive to pressure.
In $\mathrm{Li_{7}Sn_{1}}$, with application of moderate pressure
instead of 8 Li atoms at ambient pressure, only 4 atoms are found
to have negative charge, but at the same time the net charge increases
from $\sim-0.28$ to $\sim-0.79$. This possibly happens due to reduction
in anisotropicity and phase change in $\mathrm{Li_{7}Sn_{1}}$ at
20 GPa.

\section{Conclusions }

First-principles evolutionary algorithm based simulations are performed
to study the lithiation of Sn anode at ambient and moderately high
pressure of $\leq$20 GPa. At ambient pressure, our calculations not
only validate the well known existing Li-Sn compositions by correctly
predicting their phases, but also revealed  one of the most stable
and yet unexplored Li-Sn compound\textendash $\textrm{L}\textrm{i}_{8}\textrm{S}\textrm{n}_{3}$,
together with several Li-rich metastable stoichiometries. Other than
augmenting a new phase in ambient pressure phase diagram, our calculations
also highlight the role of pressure (compressive strain)  in stablizing
several unknown Li-rich compounds to enrich the moderate pressure
phase diagram with compositions like $\mathrm{Li_{3}Sn_{1}}$, $\mathrm{Li_{4}Sn_{1}}$,
$\mathrm{Li_{5}Sn_{1}}$, and $\mathrm{Li_{7}Sn_{1}}$. Interestingly,
discovery of $\mathrm{Li_{5}Sn_{1}}$ and $\mathrm{Li_{7}Sn_{1}}$
surpasses the theoretical gravimetric capacity of existing highest
lithiated phase, $\mathrm{Li_{17}Sn_{4}}$, and suggests an option
to enhance the capacity of a Li-Sn battery by almost 1.5 times ($\sim1580$
mAh g$^{-1}$) as compared to $\mathrm{Li_{17}Sn_{4}}$, by using
quenched high pressure \textendash{} high capacity phase as the pre-lithiated
anode material. Most importantly, besides providing a rich structural
diversity for the Li-Sn system, our calculations also suggest that
application of pressure helps in reducing the volume expansion by
$\sim$50\% at 20 GPa and improving the mechanical properties by substantially
increasing the average value of bulk, shear, and Young's modulii as
compared to ambient pressure. This will certainly help in overcoming
the deformation and fracture of Sn anode during charge-discharge process.
Moreover, the electronic properties are found to be less sensitive
to pressure, thus, essentially providing a good electrical conductivity
even at high pressure. Over all our study elucidates pressure to be
an extra variable for improving the mechanical and electrochemical
properties of Li-Sn compounds, which can help in providing a new dimension
to the research in Li-Sn batteries. We expect our predictions will
provide basis for future experimental investigations of the Li-Sn
system.
\begin{acknowledgments}
Authors thank Prof. Artem R. Oganov for fruitful discussion and critical
reading of this manuscript. P. J. would also like to acknowledge the
support provided by Grant No. SR/FTP/PS-052/2012 from Department of
Science and Technology (DST), Government of India. The high performance
computing facility and workstations available at the School of Natural
Sciences, Shiv Nadar University, was used to perform all calculations.
\end{acknowledgments}

\bibliographystyle{apsrev}
\addcontentsline{toc}{section}{\refname}\bibliography{refs}

\begin{thebibliography}{77}
\expandafter\ifx\csname natexlab\endcsname\relax\def\natexlab#1{#1}\fi
\expandafter\ifx\csname bibnamefont\endcsname\relax
  \def\bibnamefont#1{#1}\fi
\expandafter\ifx\csname bibfnamefont\endcsname\relax
  \def\bibfnamefont#1{#1}\fi
\expandafter\ifx\csname citenamefont\endcsname\relax
  \def\citenamefont#1{#1}\fi
\expandafter\ifx\csname url\endcsname\relax
  \def\url#1{\texttt{#1}}\fi
\expandafter\ifx\csname urlprefix\endcsname\relax\def\urlprefix{URL }\fi
\providecommand{\bibinfo}[2]{#2}
\providecommand{\eprint}[2][]{\url{#2}}

\bibitem[{\citenamefont{Nitta et~al.}(2015)\citenamefont{Nitta, Wu, Lee, and
  Yushin}}]{Nitta_2015}
\bibinfo{author}{\bibfnamefont{N.}~\bibnamefont{Nitta}},
  \bibinfo{author}{\bibfnamefont{F.}~\bibnamefont{Wu}},
  \bibinfo{author}{\bibfnamefont{J.}~\bibnamefont{Lee}}, \bibnamefont{and}
  \bibinfo{author}{\bibfnamefont{G.}~\bibnamefont{Yushin}},
  \bibinfo{journal}{Materials Today} \textbf{\bibinfo{volume}{18}},
  \bibinfo{pages}{252} (\bibinfo{year}{2015}).

\bibitem[{\citenamefont{Tarascon and Armand}(2001)}]{Tarascon_2001}
\bibinfo{author}{\bibfnamefont{J.-M.} \bibnamefont{Tarascon}} \bibnamefont{and}
  \bibinfo{author}{\bibfnamefont{M.}~\bibnamefont{Armand}},
  \bibinfo{journal}{Nature} \textbf{\bibinfo{volume}{414}},
  \bibinfo{pages}{359} (\bibinfo{year}{2001}).

\bibitem[{\citenamefont{Etacheri et~al.}(2011)\citenamefont{Etacheri, Marom,
  Elazari, Salitra, and Aurbach}}]{Etacheri_2011}
\bibinfo{author}{\bibfnamefont{V.}~\bibnamefont{Etacheri}},
  \bibinfo{author}{\bibfnamefont{R.}~\bibnamefont{Marom}},
  \bibinfo{author}{\bibfnamefont{R.}~\bibnamefont{Elazari}},
  \bibinfo{author}{\bibfnamefont{G.}~\bibnamefont{Salitra}}, \bibnamefont{and}
  \bibinfo{author}{\bibfnamefont{D.}~\bibnamefont{Aurbach}},
  \bibinfo{journal}{Energy Environ. Sci.} \textbf{\bibinfo{volume}{4}},
  \bibinfo{pages}{3243} (\bibinfo{year}{2011}).

\bibitem[{\citenamefont{Kang et~al.}(2006)\citenamefont{Kang, Meng, Br{\'e}ger,
  Grey, and Ceder}}]{Kang_2006}
\bibinfo{author}{\bibfnamefont{K.}~\bibnamefont{Kang}},
  \bibinfo{author}{\bibfnamefont{Y.~S.} \bibnamefont{Meng}},
  \bibinfo{author}{\bibfnamefont{J.}~\bibnamefont{Br{\'e}ger}},
  \bibinfo{author}{\bibfnamefont{C.~P.} \bibnamefont{Grey}}, \bibnamefont{and}
  \bibinfo{author}{\bibfnamefont{G.}~\bibnamefont{Ceder}},
  \bibinfo{journal}{Science} \textbf{\bibinfo{volume}{311}},
  \bibinfo{pages}{977} (\bibinfo{year}{2006}).

\bibitem[{\citenamefont{Armand and Tarascon}(2008)}]{Armand_2008}
\bibinfo{author}{\bibfnamefont{M.}~\bibnamefont{Armand}} \bibnamefont{and}
  \bibinfo{author}{\bibfnamefont{J.-M.} \bibnamefont{Tarascon}},
  \bibinfo{journal}{Nature} \textbf{\bibinfo{volume}{451}},
  \bibinfo{pages}{652} (\bibinfo{year}{2008}).

\bibitem[{\citenamefont{Whittingham}(2014)}]{Whittingham_2014}
\bibinfo{author}{\bibfnamefont{M.~S.} \bibnamefont{Whittingham}},
  \bibinfo{journal}{Chem. Rev.} \textbf{\bibinfo{volume}{114}},
  \bibinfo{pages}{11414} (\bibinfo{year}{2014}).

\bibitem[{\citenamefont{Park et~al.}(2010)\citenamefont{Park, Kim, Kim, and
  Sohn}}]{Park_csr_2010}
\bibinfo{author}{\bibfnamefont{C.-M.} \bibnamefont{Park}},
  \bibinfo{author}{\bibfnamefont{J.-H.} \bibnamefont{Kim}},
  \bibinfo{author}{\bibfnamefont{H.}~\bibnamefont{Kim}}, \bibnamefont{and}
  \bibinfo{author}{\bibfnamefont{H.-J.} \bibnamefont{Sohn}},
  \bibinfo{journal}{Chem. Soc. Rev.} \textbf{\bibinfo{volume}{39}},
  \bibinfo{pages}{3115} (\bibinfo{year}{2010}).

\bibitem[{\citenamefont{Zhang}(2011)}]{Zhang_2011}
\bibinfo{author}{\bibfnamefont{W.-J.} \bibnamefont{Zhang}},
  \bibinfo{journal}{Journal of Power Sources} \textbf{\bibinfo{volume}{196}},
  \bibinfo{pages}{13} (\bibinfo{year}{2011}).

\bibitem[{\citenamefont{Obrovac and Chevrier}(2014)}]{Obrovac_2014}
\bibinfo{author}{\bibfnamefont{M.~N.} \bibnamefont{Obrovac}} \bibnamefont{and}
  \bibinfo{author}{\bibfnamefont{V.~L.} \bibnamefont{Chevrier}},
  \bibinfo{journal}{Chem. Rev.} \textbf{\bibinfo{volume}{114}},
  \bibinfo{pages}{11444} (\bibinfo{year}{2014}).

\bibitem[{\citenamefont{Tian et~al.}(2015)\citenamefont{Tian, Xin, Wang, He,
  and Han}}]{Tian_2015}
\bibinfo{author}{\bibfnamefont{H.}~\bibnamefont{Tian}},
  \bibinfo{author}{\bibfnamefont{F.}~\bibnamefont{Xin}},
  \bibinfo{author}{\bibfnamefont{X.}~\bibnamefont{Wang}},
  \bibinfo{author}{\bibfnamefont{W.}~\bibnamefont{He}}, \bibnamefont{and}
  \bibinfo{author}{\bibfnamefont{W.}~\bibnamefont{Han}},
  \bibinfo{journal}{Journal of Materiomics} \textbf{\bibinfo{volume}{1}},
  \bibinfo{pages}{153} (\bibinfo{year}{2015}).

\bibitem[{\citenamefont{Chou et~al.}(2011)\citenamefont{Chou, Kim, and
  Hwang}}]{Chou_2011}
\bibinfo{author}{\bibfnamefont{C.-Y.} \bibnamefont{Chou}},
  \bibinfo{author}{\bibfnamefont{H.}~\bibnamefont{Kim}}, \bibnamefont{and}
  \bibinfo{author}{\bibfnamefont{G.~S.} \bibnamefont{Hwang}},
  \bibinfo{journal}{The Journal of Physical Chemistry C}
  \textbf{\bibinfo{volume}{115}}, \bibinfo{pages}{20018}
  (\bibinfo{year}{2011}).

\bibitem[{\citenamefont{Johari et~al.}(2011)\citenamefont{Johari, Qi, and
  Shenoy}}]{Priya_2011}
\bibinfo{author}{\bibfnamefont{P.}~\bibnamefont{Johari}},
  \bibinfo{author}{\bibfnamefont{Y.}~\bibnamefont{Qi}}, \bibnamefont{and}
  \bibinfo{author}{\bibfnamefont{V.~B.} \bibnamefont{Shenoy}},
  \bibinfo{journal}{Nano Lett.} \textbf{\bibinfo{volume}{11}},
  \bibinfo{pages}{5494} (\bibinfo{year}{2011}).

\bibitem[{\citenamefont{Larcher et~al.}(2007)\citenamefont{Larcher, Beattie,
  Morcrette, Edstrom, Jumas, and Tarascon}}]{Larcher@2007}
\bibinfo{author}{\bibfnamefont{D.}~\bibnamefont{Larcher}},
  \bibinfo{author}{\bibfnamefont{S.}~\bibnamefont{Beattie}},
  \bibinfo{author}{\bibfnamefont{M.}~\bibnamefont{Morcrette}},
  \bibinfo{author}{\bibfnamefont{K.}~\bibnamefont{Edstrom}},
  \bibinfo{author}{\bibfnamefont{J.-C.} \bibnamefont{Jumas}}, \bibnamefont{and}
  \bibinfo{author}{\bibfnamefont{J.-M.} \bibnamefont{Tarascon}},
  \bibinfo{journal}{J. Mater. Chem.} \textbf{\bibinfo{volume}{17}},
  \bibinfo{pages}{3759} (\bibinfo{year}{2007}).

\bibitem[{\citenamefont{Idota et~al.}(1997)\citenamefont{Idota, Kubota,
  Matsufuji, Maekawa, and Miyasaka}}]{Idota@1997}
\bibinfo{author}{\bibfnamefont{Y.}~\bibnamefont{Idota}},
  \bibinfo{author}{\bibfnamefont{T.}~\bibnamefont{Kubota}},
  \bibinfo{author}{\bibfnamefont{A.}~\bibnamefont{Matsufuji}},
  \bibinfo{author}{\bibfnamefont{Y.}~\bibnamefont{Maekawa}}, \bibnamefont{and}
  \bibinfo{author}{\bibfnamefont{T.}~\bibnamefont{Miyasaka}},
  \bibinfo{journal}{Science} \textbf{\bibinfo{volume}{276}},
  \bibinfo{pages}{1395} (\bibinfo{year}{1997}).

\bibitem[{\citenamefont{Na and Park}(2010)}]{Na@2010}
\bibinfo{author}{\bibfnamefont{S.-H.} \bibnamefont{Na}} \bibnamefont{and}
  \bibinfo{author}{\bibfnamefont{C.-H.} \bibnamefont{Park}},
  \bibinfo{journal}{Journal of the Korean Physical Society}
  \textbf{\bibinfo{volume}{56}}, \bibinfo{pages}{494} (\bibinfo{year}{2010}).

\bibitem[{\citenamefont{Zeng et~al.}(2015)\citenamefont{Zeng, Zeng, Liu,
  Oganov, Zeng, Cui, and Mao}}]{Zeng_2015}
\bibinfo{author}{\bibfnamefont{Z.}~\bibnamefont{Zeng}},
  \bibinfo{author}{\bibfnamefont{Q.}~\bibnamefont{Zeng}},
  \bibinfo{author}{\bibfnamefont{N.}~\bibnamefont{Liu}},
  \bibinfo{author}{\bibfnamefont{A.~R.} \bibnamefont{Oganov}},
  \bibinfo{author}{\bibfnamefont{Q.}~\bibnamefont{Zeng}},
  \bibinfo{author}{\bibfnamefont{Y.}~\bibnamefont{Cui}}, \bibnamefont{and}
  \bibinfo{author}{\bibfnamefont{W.~L.} \bibnamefont{Mao}},
  \bibinfo{journal}{Adv. Energy Mater.} \textbf{\bibinfo{volume}{5}},
  \bibinfo{pages}{1500214} (\bibinfo{year}{2015}).

\bibitem[{\citenamefont{Robert et~al.}(2007)\citenamefont{Robert, Lippens,
  Olivier-Fourcade, Jumas, Gillot, Morcrette, and Tarascon}}]{Robert_2007}
\bibinfo{author}{\bibfnamefont{F.}~\bibnamefont{Robert}},
  \bibinfo{author}{\bibfnamefont{P.}~\bibnamefont{Lippens}},
  \bibinfo{author}{\bibfnamefont{J.}~\bibnamefont{Olivier-Fourcade}},
  \bibinfo{author}{\bibfnamefont{J.-C.} \bibnamefont{Jumas}},
  \bibinfo{author}{\bibfnamefont{F.}~\bibnamefont{Gillot}},
  \bibinfo{author}{\bibfnamefont{M.}~\bibnamefont{Morcrette}},
  \bibnamefont{and} \bibinfo{author}{\bibfnamefont{J.-M.}
  \bibnamefont{Tarascon}}, \bibinfo{journal}{Journal of Solid State Chemistry}
  \textbf{\bibinfo{volume}{180}}, \bibinfo{pages}{339} (\bibinfo{year}{2007}).

\bibitem[{\citenamefont{Dunlap et~al.}(1999)\citenamefont{Dunlap, Small,
  MacNeil, Obrovac, and Dahn}}]{Dunlap_1999}
\bibinfo{author}{\bibfnamefont{R.}~\bibnamefont{Dunlap}},
  \bibinfo{author}{\bibfnamefont{D.}~\bibnamefont{Small}},
  \bibinfo{author}{\bibfnamefont{D.}~\bibnamefont{MacNeil}},
  \bibinfo{author}{\bibfnamefont{M.}~\bibnamefont{Obrovac}}, \bibnamefont{and}
  \bibinfo{author}{\bibfnamefont{J.}~\bibnamefont{Dahn}},
  \bibinfo{journal}{Journal of Alloys and Compounds}
  \textbf{\bibinfo{volume}{289}}, \bibinfo{pages}{135} (\bibinfo{year}{1999}).

\bibitem[{\citenamefont{Hansen and Chang}(1969)}]{Li2Sn5}
\bibinfo{author}{\bibfnamefont{D.}~\bibnamefont{Hansen}} \bibnamefont{and}
  \bibinfo{author}{\bibfnamefont{L.~J.} \bibnamefont{Chang}},
  \bibinfo{journal}{Acta Cryst.} \textbf{\bibinfo{volume}{B25}},
  \bibinfo{pages}{2392} (\bibinfo{year}{1969}).

\bibitem[{\citenamefont{M{\"u}ller and Sch{\"a}fer}(1973)}]{alpha_LiSn}
\bibinfo{author}{\bibfnamefont{W.}~\bibnamefont{M{\"u}ller}} \bibnamefont{and}
  \bibinfo{author}{\bibfnamefont{H.}~\bibnamefont{Sch{\"a}fer}},
  \bibinfo{journal}{Z. Naturforsch.} \textbf{\bibinfo{volume}{28b}},
  \bibinfo{pages}{246} (\bibinfo{year}{1973}).

\bibitem[{\citenamefont{Blase and Cordier}(1990)}]{beta_LiSn}
\bibinfo{author}{\bibfnamefont{W.}~\bibnamefont{Blase}} \bibnamefont{and}
  \bibinfo{author}{\bibfnamefont{G.}~\bibnamefont{Cordier}},
  \bibinfo{journal}{Zeitschrift f{\"u}r Kristallographie}
  \textbf{\bibinfo{volume}{193}}, \bibinfo{pages}{317} (\bibinfo{year}{1990}).

\bibitem[{\citenamefont{M{\"u}ller}(1974)}]{Li7Sn3}
\bibinfo{author}{\bibfnamefont{W.}~\bibnamefont{M{\"u}ller}},
  \bibinfo{journal}{Z. Naturforsch.} \textbf{\bibinfo{volume}{29b}},
  \bibinfo{pages}{304} (\bibinfo{year}{1974}).

\bibitem[{\citenamefont{Frank et~al.}(1975{\natexlab{a}})\citenamefont{Frank,
  M{\"u}ller, and Sch{\"a}fer}}]{Li5Sn2}
\bibinfo{author}{\bibfnamefont{U.}~\bibnamefont{Frank}},
  \bibinfo{author}{\bibfnamefont{W.}~\bibnamefont{M{\"u}ller}},
  \bibnamefont{and}
  \bibinfo{author}{\bibfnamefont{H.}~\bibnamefont{Sch{\"a}fer}},
  \bibinfo{journal}{Z. Naturforsch.} \textbf{\bibinfo{volume}{30b}},
  \bibinfo{pages}{1} (\bibinfo{year}{1975}{\natexlab{a}}).

\bibitem[{\citenamefont{Frank and M{\"u}ller}(1975)}]{Li13Sn5}
\bibinfo{author}{\bibfnamefont{U.}~\bibnamefont{Frank}} \bibnamefont{and}
  \bibinfo{author}{\bibfnamefont{W.}~\bibnamefont{M{\"u}ller}},
  \bibinfo{journal}{Z.Naturforsch} \textbf{\bibinfo{volume}{30b}},
  \bibinfo{pages}{316} (\bibinfo{year}{1975}).

\bibitem[{\citenamefont{Frank et~al.}(1975{\natexlab{b}})\citenamefont{Frank,
  M{\"u}ller, and Sch{\"a}fer}}]{Li7Sn2}
\bibinfo{author}{\bibfnamefont{U.}~\bibnamefont{Frank}},
  \bibinfo{author}{\bibfnamefont{W.}~\bibnamefont{M{\"u}ller}},
  \bibnamefont{and}
  \bibinfo{author}{\bibfnamefont{H.}~\bibnamefont{Sch{\"a}fer}},
  \bibinfo{journal}{Z. Naturforsch.} \textbf{\bibinfo{volume}{30b}},
  \bibinfo{pages}{6} (\bibinfo{year}{1975}{\natexlab{b}}).

\bibitem[{\citenamefont{Goward et~al.}(2001)\citenamefont{Goward, Taylor,
  Souza, and Nazar}}]{Li17Sn4_Gowar}
\bibinfo{author}{\bibfnamefont{G.}~\bibnamefont{Goward}},
  \bibinfo{author}{\bibfnamefont{N.}~\bibnamefont{Taylor}},
  \bibinfo{author}{\bibfnamefont{D.}~\bibnamefont{Souza}}, \bibnamefont{and}
  \bibinfo{author}{\bibfnamefont{L.}~\bibnamefont{Nazar}},
  \bibinfo{journal}{Journal of Alloys and Compounds}
  \textbf{\bibinfo{volume}{329}}, \bibinfo{pages}{82} (\bibinfo{year}{2001}).

\bibitem[{\citenamefont{Lupu et~al.}(2003)\citenamefont{Lupu, Mao, Rabalais,
  Guloy, and Richardson}}]{Li17Sn4_Lipu}
\bibinfo{author}{\bibfnamefont{C.}~\bibnamefont{Lupu}},
  \bibinfo{author}{\bibfnamefont{J.-G.} \bibnamefont{Mao}},
  \bibinfo{author}{\bibfnamefont{J.}~\bibnamefont{Rabalais}},
  \bibinfo{author}{\bibfnamefont{A.}~\bibnamefont{Guloy}}, \bibnamefont{and}
  \bibinfo{author}{\bibfnamefont{J.~J.} \bibnamefont{Richardson}},
  \bibinfo{journal}{Inorganic Chemistry} \textbf{\bibinfo{volume}{42}},
  \bibinfo{pages}{3765} (\bibinfo{year}{2003}).

\bibitem[{\citenamefont{Gladyshevskii et~al.}(1964)\citenamefont{Gladyshevskii,
  Oleksiv, and P.I.}}]{Li22Sn5}
\bibinfo{author}{\bibfnamefont{E.}~\bibnamefont{Gladyshevskii}},
  \bibinfo{author}{\bibfnamefont{G.}~\bibnamefont{Oleksiv}}, \bibnamefont{and}
  \bibinfo{author}{\bibfnamefont{K.}~\bibnamefont{P.I.}},
  \bibinfo{journal}{Kristallografiya} \textbf{\bibinfo{volume}{9}},
  \bibinfo{pages}{338} (\bibinfo{year}{1964}).

\bibitem[{\citenamefont{Matsuoka and Shimizu}(2009)}]{Matsuoka_2009}
\bibinfo{author}{\bibfnamefont{T.}~\bibnamefont{Matsuoka}} \bibnamefont{and}
  \bibinfo{author}{\bibfnamefont{K.}~\bibnamefont{Shimizu}},
  \bibinfo{journal}{Nature} \textbf{\bibinfo{volume}{458}},
  \bibinfo{pages}{186} (\bibinfo{year}{2009}).

\bibitem[{\citenamefont{Vnuk et~al.}(1984)\citenamefont{Vnuk, De~Monte, and
  Smith}}]{Vnuk_1984}
\bibinfo{author}{\bibfnamefont{F.}~\bibnamefont{Vnuk}},
  \bibinfo{author}{\bibfnamefont{A.}~\bibnamefont{De~Monte}}, \bibnamefont{and}
  \bibinfo{author}{\bibfnamefont{R.~W.} \bibnamefont{Smith}},
  \bibinfo{journal}{Journal of Applied Physics} \textbf{\bibinfo{volume}{55}},
  \bibinfo{pages}{4171} (\bibinfo{year}{1984}).

\bibitem[{\citenamefont{Zhang et~al.}(2013)\citenamefont{Zhang, Oganov,
  Goncharov, Zhu, Boulfelfel, Lyakhov, Stavrou, Somayazulu, Prakapenka, and
  Kon{\^o}pkov{\'a}}}]{Zhang_2013}
\bibinfo{author}{\bibfnamefont{W.}~\bibnamefont{Zhang}},
  \bibinfo{author}{\bibfnamefont{A.~R.} \bibnamefont{Oganov}},
  \bibinfo{author}{\bibfnamefont{A.~F.} \bibnamefont{Goncharov}},
  \bibinfo{author}{\bibfnamefont{Q.}~\bibnamefont{Zhu}},
  \bibinfo{author}{\bibfnamefont{S.~E.} \bibnamefont{Boulfelfel}},
  \bibinfo{author}{\bibfnamefont{A.~O.} \bibnamefont{Lyakhov}},
  \bibinfo{author}{\bibfnamefont{E.}~\bibnamefont{Stavrou}},
  \bibinfo{author}{\bibfnamefont{M.}~\bibnamefont{Somayazulu}},
  \bibinfo{author}{\bibfnamefont{V.~B.} \bibnamefont{Prakapenka}},
  \bibnamefont{and}
  \bibinfo{author}{\bibfnamefont{Z.}~\bibnamefont{Kon{\^o}pkov{\'a}}},
  \textbf{\bibinfo{volume}{342}}, \bibinfo{pages}{1502} (\bibinfo{year}{2013}).

\bibitem[{\citenamefont{Stearns et~al.}(2003)\citenamefont{Stearns, Gryko,
  Diefenbacher, Ramachandran, and McMillan}}]{Stearns@2003}
\bibinfo{author}{\bibfnamefont{L.~A.} \bibnamefont{Stearns}},
  \bibinfo{author}{\bibfnamefont{J.}~\bibnamefont{Gryko}},
  \bibinfo{author}{\bibfnamefont{J.}~\bibnamefont{Diefenbacher}},
  \bibinfo{author}{\bibfnamefont{G.~K.} \bibnamefont{Ramachandran}},
  \bibnamefont{and} \bibinfo{author}{\bibfnamefont{P.~F.}
  \bibnamefont{McMillan}}, \bibinfo{journal}{Journal of Solid State Chemistry}
  \textbf{\bibinfo{volume}{173}}, \bibinfo{pages}{251} (\bibinfo{year}{2003}).

\bibitem[{\citenamefont{Zhang et~al.}(2016{\natexlab{a}})\citenamefont{Zhang,
  Wang, Yang, and Ma}}]{Zhang@2016}
\bibinfo{author}{\bibfnamefont{S.}~\bibnamefont{Zhang}},
  \bibinfo{author}{\bibfnamefont{Y.}~\bibnamefont{Wang}},
  \bibinfo{author}{\bibfnamefont{G.}~\bibnamefont{Yang}}, \bibnamefont{and}
  \bibinfo{author}{\bibfnamefont{Y.}~\bibnamefont{Ma}}, \bibinfo{journal}{ACS
  Applied Materials \& Interfaces} \textbf{\bibinfo{volume}{8}},
  \bibinfo{pages}{16761} (\bibinfo{year}{2016}{\natexlab{a}}).

\bibitem[{\citenamefont{Zhao et~al.}(2014)\citenamefont{Zhao, Lu, Liu, Lee, T.,
  and Cui}}]{Zhao_2014}
\bibinfo{author}{\bibfnamefont{J.}~\bibnamefont{Zhao}},
  \bibinfo{author}{\bibfnamefont{Z.}~\bibnamefont{Lu}},
  \bibinfo{author}{\bibfnamefont{N.}~\bibnamefont{Liu}},
  \bibinfo{author}{\bibfnamefont{H.-W.} \bibnamefont{Lee}},
  \bibinfo{author}{\bibfnamefont{M.~M.} \bibnamefont{T.}}, \bibnamefont{and}
  \bibinfo{author}{\bibfnamefont{Y.}~\bibnamefont{Cui}},
  \bibinfo{journal}{Nature Comm.} \textbf{\bibinfo{volume}{5}},
  \bibinfo{pages}{5088} (\bibinfo{year}{2014}).

\bibitem[{\citenamefont{Cloud et~al.}(2014)\citenamefont{Cloud, Wang, Yoder,
  Taylor, and Yang}}]{Cloud_2014}
\bibinfo{author}{\bibfnamefont{J.~E.} \bibnamefont{Cloud}},
  \bibinfo{author}{\bibfnamefont{Y.}~\bibnamefont{Wang}},
  \bibinfo{author}{\bibfnamefont{T.~S.} \bibnamefont{Yoder}},
  \bibinfo{author}{\bibfnamefont{L.~W.} \bibnamefont{Taylor}},
  \bibnamefont{and} \bibinfo{author}{\bibfnamefont{Y.}~\bibnamefont{Yang}},
  \bibinfo{journal}{Angewandte Chemie International Edition}
  \textbf{\bibinfo{volume}{53}}, \bibinfo{pages}{14527} (\bibinfo{year}{2014}).

\bibitem[{\citenamefont{Oganov and Glass}(2006)}]{Oganov_2006}
\bibinfo{author}{\bibfnamefont{A.}~\bibnamefont{Oganov}} \bibnamefont{and}
  \bibinfo{author}{\bibfnamefont{C.}~\bibnamefont{Glass}},
  \bibinfo{journal}{J.Chem. Phys.} \textbf{\bibinfo{volume}{124}},
  \bibinfo{pages}{244704} (\bibinfo{year}{2006}).

\bibitem[{\citenamefont{Oganov et~al.}(2011)\citenamefont{Oganov, Lyakhov, and
  Valley}}]{Oganov_2011}
\bibinfo{author}{\bibfnamefont{A.}~\bibnamefont{Oganov}},
  \bibinfo{author}{\bibfnamefont{A.~O.} \bibnamefont{Lyakhov}},
  \bibnamefont{and} \bibinfo{author}{\bibfnamefont{M.}~\bibnamefont{Valley}},
  \bibinfo{journal}{Acc. Chem. Res.} \textbf{\bibinfo{volume}{44}},
  \bibinfo{pages}{227} (\bibinfo{year}{2011}).

\bibitem[{\citenamefont{Lyakhov et~al.}(2013)\citenamefont{Lyakhov, Oganov,
  Stokes, and Zhu}}]{Lyakhov_2013}
\bibinfo{author}{\bibfnamefont{A.~O.} \bibnamefont{Lyakhov}},
  \bibinfo{author}{\bibfnamefont{A.~R.} \bibnamefont{Oganov}},
  \bibinfo{author}{\bibfnamefont{H.}~\bibnamefont{Stokes}}, \bibnamefont{and}
  \bibinfo{author}{\bibfnamefont{Q.}~\bibnamefont{Zhu}},
  \bibinfo{journal}{Comput. Phys. Commun.} \textbf{\bibinfo{volume}{184}},
  \bibinfo{pages}{1172} (\bibinfo{year}{2013}).

\bibitem[{\citenamefont{Pickard and Needs}(2011)}]{Pickard_2011}
\bibinfo{author}{\bibfnamefont{C.~J.} \bibnamefont{Pickard}} \bibnamefont{and}
  \bibinfo{author}{\bibfnamefont{R.~J.} \bibnamefont{Needs}},
  \bibinfo{journal}{Journal of Physics: Condensed Matter}
  \textbf{\bibinfo{volume}{23}}, \bibinfo{pages}{053201}
  (\bibinfo{year}{2011}).

\bibitem[{\citenamefont{Goedecker}(2004)}]{Goedecker_2004}
\bibinfo{author}{\bibfnamefont{S.}~\bibnamefont{Goedecker}},
  \bibinfo{journal}{The Journal of Chemical Physics}
  \textbf{\bibinfo{volume}{120}}, \bibinfo{pages}{9911} (\bibinfo{year}{2004}).

\bibitem[{\citenamefont{Tipton et~al.}(2013)\citenamefont{Tipton, Bealing,
  Mathew, and Hennig}}]{Tipton_2013}
\bibinfo{author}{\bibfnamefont{W.~W.} \bibnamefont{Tipton}},
  \bibinfo{author}{\bibfnamefont{C.~R.} \bibnamefont{Bealing}},
  \bibinfo{author}{\bibfnamefont{K.}~\bibnamefont{Mathew}}, \bibnamefont{and}
  \bibinfo{author}{\bibfnamefont{R.~G.} \bibnamefont{Hennig}},
  \bibinfo{journal}{Phys. Rev. B} \textbf{\bibinfo{volume}{87}},
  \bibinfo{pages}{184114} (\bibinfo{year}{2013}).

\bibitem[{\citenamefont{Morris et~al.}(2014)\citenamefont{Morris, Grey, and
  Pickard}}]{Morris_2014}
\bibinfo{author}{\bibfnamefont{A.~J.} \bibnamefont{Morris}},
  \bibinfo{author}{\bibfnamefont{C.~P.} \bibnamefont{Grey}}, \bibnamefont{and}
  \bibinfo{author}{\bibfnamefont{C.~J.} \bibnamefont{Pickard}},
  \bibinfo{journal}{Phys. Rev. B} \textbf{\bibinfo{volume}{90}},
  \bibinfo{pages}{054111} (\bibinfo{year}{2014}).

\bibitem[{\citenamefont{Valencia-Jaime
  et~al.}(2016)\citenamefont{Valencia-Jaime, Sarmiento-P{\'e}rez, Botti,
  Marques, Amsler, Goedecker, and Romero}}]{Valencia-Jaime_2016}
\bibinfo{author}{\bibfnamefont{I.}~\bibnamefont{Valencia-Jaime}},
  \bibinfo{author}{\bibfnamefont{R.}~\bibnamefont{Sarmiento-P{\'e}rez}},
  \bibinfo{author}{\bibfnamefont{S.}~\bibnamefont{Botti}},
  \bibinfo{author}{\bibfnamefont{M.}~\bibnamefont{Marques}},
  \bibinfo{author}{\bibfnamefont{M.}~\bibnamefont{Amsler}},
  \bibinfo{author}{\bibfnamefont{S.}~\bibnamefont{Goedecker}},
  \bibnamefont{and} \bibinfo{author}{\bibfnamefont{A.}~\bibnamefont{Romero}},
  \bibinfo{journal}{Journal of Alloys and Compounds}
  \textbf{\bibinfo{volume}{655}}, \bibinfo{pages}{147} (\bibinfo{year}{2016}).

\bibitem[{\citenamefont{Badding et~al.}(1995)\citenamefont{Badding, Parker, and
  Nesting}}]{Badding@1995}
\bibinfo{author}{\bibfnamefont{J.}~\bibnamefont{Badding}},
  \bibinfo{author}{\bibfnamefont{L.}~\bibnamefont{Parker}}, \bibnamefont{and}
  \bibinfo{author}{\bibfnamefont{D.}~\bibnamefont{Nesting}},
  \bibinfo{journal}{Journal of Solid State Chemistry}
  \textbf{\bibinfo{volume}{117}}, \bibinfo{pages}{229} (\bibinfo{year}{1995}).

\bibitem[{\citenamefont{Sun et~al.}(2016)\citenamefont{Sun, Dacek, Ong,
  Hautier, Jain, Richards, Gamst, Persson, and Ceder}}]{Sun@2016}
\bibinfo{author}{\bibfnamefont{W.}~\bibnamefont{Sun}},
  \bibinfo{author}{\bibfnamefont{S.~T.} \bibnamefont{Dacek}},
  \bibinfo{author}{\bibfnamefont{S.~P.} \bibnamefont{Ong}},
  \bibinfo{author}{\bibfnamefont{G.}~\bibnamefont{Hautier}},
  \bibinfo{author}{\bibfnamefont{A.}~\bibnamefont{Jain}},
  \bibinfo{author}{\bibfnamefont{W.~D.} \bibnamefont{Richards}},
  \bibinfo{author}{\bibfnamefont{A.~C.} \bibnamefont{Gamst}},
  \bibinfo{author}{\bibfnamefont{K.~A.} \bibnamefont{Persson}},
  \bibnamefont{and} \bibinfo{author}{\bibfnamefont{G.}~\bibnamefont{Ceder}},
  \bibinfo{journal}{Science Advances} \textbf{\bibinfo{volume}{2}},
  \bibinfo{pages}{1} (\bibinfo{year}{2016}).

\bibitem[{\citenamefont{Kresse and Furthmuller}(1996{\natexlab{a}})}]{vasp-1}
\bibinfo{author}{\bibfnamefont{G.}~\bibnamefont{Kresse}} \bibnamefont{and}
  \bibinfo{author}{\bibfnamefont{J.}~\bibnamefont{Furthmuller}},
  \bibinfo{journal}{Phys. Rev. B} \textbf{\bibinfo{volume}{54}},
  \bibinfo{pages}{11169} (\bibinfo{year}{1996}{\natexlab{a}}).

\bibitem[{\citenamefont{Kresse and Furthmuller}(1996{\natexlab{b}})}]{vasp-2}
\bibinfo{author}{\bibfnamefont{G.}~\bibnamefont{Kresse}} \bibnamefont{and}
  \bibinfo{author}{\bibfnamefont{J.}~\bibnamefont{Furthmuller}},
  \bibinfo{journal}{Computational Materials Science}
  \textbf{\bibinfo{volume}{6}}, \bibinfo{pages}{15}
  (\bibinfo{year}{1996}{\natexlab{b}}).

\bibitem[{\citenamefont{Bl{\"o}chl}(1994)}]{PAW2}
\bibinfo{author}{\bibfnamefont{P.~E.} \bibnamefont{Bl{\"o}chl}},
  \bibinfo{journal}{Phys. Rev. B} \textbf{\bibinfo{volume}{50}},
  \bibinfo{pages}{17953} (\bibinfo{year}{1994}).

\bibitem[{\citenamefont{Perdew et~al.}(1996)\citenamefont{Perdew, Burke, and
  Ernzerhof}}]{PBE}
\bibinfo{author}{\bibfnamefont{J.~P.} \bibnamefont{Perdew}},
  \bibinfo{author}{\bibfnamefont{K.}~\bibnamefont{Burke}}, \bibnamefont{and}
  \bibinfo{author}{\bibfnamefont{M.}~\bibnamefont{Ernzerhof}},
  \bibinfo{journal}{Phys. Rev. Lett.} \textbf{\bibinfo{volume}{77}},
  \bibinfo{pages}{3865} (\bibinfo{year}{1996}).

\bibitem[{\citenamefont{Togo et~al.}(2008)\citenamefont{Togo, Oba, and
  Tanaka}}]{Phonopy}
\bibinfo{author}{\bibfnamefont{A.}~\bibnamefont{Togo}},
  \bibinfo{author}{\bibfnamefont{F.}~\bibnamefont{Oba}}, \bibnamefont{and}
  \bibinfo{author}{\bibfnamefont{I.}~\bibnamefont{Tanaka}},
  \bibinfo{journal}{Phys. Rev. B} \textbf{\bibinfo{volume}{78}},
  \bibinfo{pages}{134106} (\bibinfo{year}{2008}).

\bibitem[{\citenamefont{Ceder et~al.}(1997)\citenamefont{Ceder, Aydinol, and
  Kohan}}]{Ceder@1997}
\bibinfo{author}{\bibfnamefont{G.}~\bibnamefont{Ceder}},
  \bibinfo{author}{\bibfnamefont{M.}~\bibnamefont{Aydinol}}, \bibnamefont{and}
  \bibinfo{author}{\bibfnamefont{A.}~\bibnamefont{Kohan}},
  \bibinfo{journal}{Computational Materials Science}
  \textbf{\bibinfo{volume}{8}}, \bibinfo{pages}{161} (\bibinfo{year}{1997}).

\bibitem[{\citenamefont{Hill}(1952)}]{Hill_1951}
\bibinfo{author}{\bibfnamefont{R.}~\bibnamefont{Hill}},
  \bibinfo{journal}{Proceedings of the Physical Society. Section A}
  \textbf{\bibinfo{volume}{65}}, \bibinfo{pages}{349} (\bibinfo{year}{1952}).

\bibitem[{\citenamefont{Hanfland et~al.}(1999)\citenamefont{Hanfland, Loa,
  Syassen, Schwarz, and Takemura}}]{Hanfland@1999}
\bibinfo{author}{\bibfnamefont{M.}~\bibnamefont{Hanfland}},
  \bibinfo{author}{\bibfnamefont{I.}~\bibnamefont{Loa}},
  \bibinfo{author}{\bibfnamefont{K.}~\bibnamefont{Syassen}},
  \bibinfo{author}{\bibfnamefont{U.}~\bibnamefont{Schwarz}}, \bibnamefont{and}
  \bibinfo{author}{\bibfnamefont{K.}~\bibnamefont{Takemura}},
  \bibinfo{journal}{Solid State Communications} \textbf{\bibinfo{volume}{112}},
  \bibinfo{pages}{123} (\bibinfo{year}{1999}).

\bibitem[{\citenamefont{Neaton and Ashcroft}(1999)}]{Neaton@1999}
\bibinfo{author}{\bibfnamefont{J.~B.} \bibnamefont{Neaton}} \bibnamefont{and}
  \bibinfo{author}{\bibfnamefont{N.~W.} \bibnamefont{Ashcroft}},
  \bibinfo{journal}{Nature} \textbf{\bibinfo{volume}{400}},
  \bibinfo{pages}{141} (\bibinfo{year}{1999}).

\bibitem[{\citenamefont{Yu et~al.}(2006)\citenamefont{Yu, Liu, Lu, and
  Chen}}]{Yu@2006}
\bibinfo{author}{\bibfnamefont{C.}~\bibnamefont{Yu}},
  \bibinfo{author}{\bibfnamefont{J.}~\bibnamefont{Liu}},
  \bibinfo{author}{\bibfnamefont{H.}~\bibnamefont{Lu}}, \bibnamefont{and}
  \bibinfo{author}{\bibfnamefont{J.}~\bibnamefont{Chen}},
  \bibinfo{journal}{Solid State Communications} \textbf{\bibinfo{volume}{140}},
  \bibinfo{pages}{538} (\bibinfo{year}{2006}).

\bibitem[{\citenamefont{Barnett et~al.}(1963)\citenamefont{Barnett, Bennion,
  and Hall}}]{Barnett@1963}
\bibinfo{author}{\bibfnamefont{J.}~\bibnamefont{Barnett}},
  \bibinfo{author}{\bibfnamefont{R.}~\bibnamefont{Bennion}}, \bibnamefont{and}
  \bibinfo{author}{\bibfnamefont{H.~T.} \bibnamefont{Hall}},
  \bibinfo{journal}{Science} \textbf{\bibinfo{volume}{141}},
  \bibinfo{pages}{1041} (\bibinfo{year}{1963}).

\bibitem[{\citenamefont{Genser and Hafner}(2001)}]{Genser_2001}
\bibinfo{author}{\bibfnamefont{O.}~\bibnamefont{Genser}} \bibnamefont{and}
  \bibinfo{author}{\bibfnamefont{J.}~\bibnamefont{Hafner}},
  \bibinfo{journal}{Phys. Rev. B} \textbf{\bibinfo{volume}{63}},
  \bibinfo{pages}{144204} (\bibinfo{year}{2001}).

\bibitem[{\citenamefont{Saal et~al.}(2013)\citenamefont{Saal, Kirklin, Aykol,
  Meredig, and Wolverton}}]{Saal@2013}
\bibinfo{author}{\bibfnamefont{J.~E.} \bibnamefont{Saal}},
  \bibinfo{author}{\bibfnamefont{S.}~\bibnamefont{Kirklin}},
  \bibinfo{author}{\bibfnamefont{M.}~\bibnamefont{Aykol}},
  \bibinfo{author}{\bibfnamefont{B.}~\bibnamefont{Meredig}}, \bibnamefont{and}
  \bibinfo{author}{\bibfnamefont{C.}~\bibnamefont{Wolverton}},
  \bibinfo{journal}{JOM} \textbf{\bibinfo{volume}{65}}, \bibinfo{pages}{1501}
  (\bibinfo{year}{2013}).

\bibitem[{\citenamefont{Kirklin et~al.}(2015)\citenamefont{Kirklin, Saal,
  Meredig, Thompson, Doak, Aykol, R{\"u}hl, and Wolverton}}]{Kirklin@2015}
\bibinfo{author}{\bibfnamefont{S.}~\bibnamefont{Kirklin}},
  \bibinfo{author}{\bibfnamefont{J.}~\bibnamefont{Saal}},
  \bibinfo{author}{\bibfnamefont{B.}~\bibnamefont{Meredig}},
  \bibinfo{author}{\bibfnamefont{A.}~\bibnamefont{Thompson}},
  \bibinfo{author}{\bibfnamefont{J.}~\bibnamefont{Doak}},
  \bibinfo{author}{\bibfnamefont{M.}~\bibnamefont{Aykol}},
  \bibinfo{author}{\bibfnamefont{S.}~\bibnamefont{R{\"u}hl}}, \bibnamefont{and}
  \bibinfo{author}{\bibfnamefont{C.}~\bibnamefont{Wolverton}},
  \bibinfo{journal}{npj Computational Materials} \textbf{\bibinfo{volume}{1}},
  \bibinfo{pages}{15010} (\bibinfo{year}{2015}).

\bibitem[{\citenamefont{Gasior et~al.}(1996)\citenamefont{Gasior, Moser, and
  Zakulski}}]{Gasior_1996}
\bibinfo{author}{\bibfnamefont{W.}~\bibnamefont{Gasior}},
  \bibinfo{author}{\bibfnamefont{Z.}~\bibnamefont{Moser}}, \bibnamefont{and}
  \bibinfo{author}{\bibfnamefont{W.}~\bibnamefont{Zakulski}},
  \bibinfo{journal}{Journal of Non-Crystalline Solids}
  \textbf{\bibinfo{volume}{205}}, \bibinfo{pages}{379} (\bibinfo{year}{1996}).

\bibitem[{\citenamefont{Lu et~al.}(2016)\citenamefont{Lu, Kim, Gao, Wu, Shao,
  Li, Zhou, Sun, Akinwande, Xing et~al.}}]{Lu@2016}
\bibinfo{author}{\bibfnamefont{P.}~\bibnamefont{Lu}},
  \bibinfo{author}{\bibfnamefont{J.-S.} \bibnamefont{Kim}},
  \bibinfo{author}{\bibfnamefont{H.}~\bibnamefont{Gao}},
  \bibinfo{author}{\bibfnamefont{J.}~\bibnamefont{Wu}},
  \bibinfo{author}{\bibfnamefont{D.}~\bibnamefont{Shao}},
  \bibinfo{author}{\bibfnamefont{B.}~\bibnamefont{Li}},
  \bibinfo{author}{\bibfnamefont{D.}~\bibnamefont{Zhou}},
  \bibinfo{author}{\bibfnamefont{J.}~\bibnamefont{Sun}},
  \bibinfo{author}{\bibfnamefont{D.}~\bibnamefont{Akinwande}},
  \bibinfo{author}{\bibfnamefont{D.}~\bibnamefont{Xing}}, \bibnamefont{et~al.},
  \bibinfo{journal}{Phys. Rev. B} \textbf{\bibinfo{volume}{94}},
  \bibinfo{pages}{224512} (\bibinfo{year}{2016}).

\bibitem[{\citenamefont{Darwiche et~al.}(2012)\citenamefont{Darwiche, Marino,
  Sougrati, Fraisse, Stievano, and Monconduit}}]{Darwiche@2012}
\bibinfo{author}{\bibfnamefont{A.}~\bibnamefont{Darwiche}},
  \bibinfo{author}{\bibfnamefont{C.}~\bibnamefont{Marino}},
  \bibinfo{author}{\bibfnamefont{M.~T.} \bibnamefont{Sougrati}},
  \bibinfo{author}{\bibfnamefont{B.}~\bibnamefont{Fraisse}},
  \bibinfo{author}{\bibfnamefont{L.}~\bibnamefont{Stievano}}, \bibnamefont{and}
  \bibinfo{author}{\bibfnamefont{L.}~\bibnamefont{Monconduit}},
  \bibinfo{journal}{Journal of the American Chemical Society}
  \textbf{\bibinfo{volume}{134}}, \bibinfo{pages}{20805}
  (\bibinfo{year}{2012}).

\bibitem[{\citenamefont{Thackeray et~al.}(2003)\citenamefont{Thackeray,
  Vaughey, Johnson, Kropf, Benedek, Fransson, and Edstrom}}]{Thackeray@2003}
\bibinfo{author}{\bibfnamefont{M.}~\bibnamefont{Thackeray}},
  \bibinfo{author}{\bibfnamefont{J.}~\bibnamefont{Vaughey}},
  \bibinfo{author}{\bibfnamefont{C.}~\bibnamefont{Johnson}},
  \bibinfo{author}{\bibfnamefont{A.}~\bibnamefont{Kropf}},
  \bibinfo{author}{\bibfnamefont{R.}~\bibnamefont{Benedek}},
  \bibinfo{author}{\bibfnamefont{L.}~\bibnamefont{Fransson}}, \bibnamefont{and}
  \bibinfo{author}{\bibfnamefont{K.}~\bibnamefont{Edstrom}},
  \bibinfo{journal}{Journal of Power Sources} \textbf{\bibinfo{volume}{113}},
  \bibinfo{pages}{124} (\bibinfo{year}{2003}).

\bibitem[{\citenamefont{Alblas et~al.}(1984)\citenamefont{Alblas, Lugt,
  Dijkstra, and Dijk}}]{Alblas_1984}
\bibinfo{author}{\bibfnamefont{B.~P.} \bibnamefont{Alblas}},
  \bibinfo{author}{\bibfnamefont{W.~v.~d.} \bibnamefont{Lugt}},
  \bibinfo{author}{\bibfnamefont{J.}~\bibnamefont{Dijkstra}}, \bibnamefont{and}
  \bibinfo{author}{\bibfnamefont{C.~v.} \bibnamefont{Dijk}},
  \bibinfo{journal}{Journal of Physics F: Metal Physics}
  \textbf{\bibinfo{volume}{14}}, \bibinfo{pages}{1995} (\bibinfo{year}{1984}).

\bibitem[{\citenamefont{Lin et~al.}(2015)\citenamefont{Lin, Strobel, and
  Cohen}}]{Lin@2015}
\bibinfo{author}{\bibfnamefont{Y.}~\bibnamefont{Lin}},
  \bibinfo{author}{\bibfnamefont{T.~A.} \bibnamefont{Strobel}},
  \bibnamefont{and} \bibinfo{author}{\bibfnamefont{R.~E.} \bibnamefont{Cohen}},
  \bibinfo{journal}{Phys. Rev. B} \textbf{\bibinfo{volume}{92}},
  \bibinfo{pages}{214106} (\bibinfo{year}{2015}).

\bibitem[{\citenamefont{Zalkin and Ramsey}(1956)}]{Zalkin@1956}
\bibinfo{author}{\bibfnamefont{A.}~\bibnamefont{Zalkin}} \bibnamefont{and}
  \bibinfo{author}{\bibfnamefont{W.~J.} \bibnamefont{Ramsey}},
  \bibinfo{journal}{The Journal of Physical Chemistry}
  \textbf{\bibinfo{volume}{60}}, \bibinfo{pages}{234} (\bibinfo{year}{1956}).

\bibitem[{\citenamefont{Li et~al.}(2013)\citenamefont{Li, Xie, Liu, Ma, Zhou,
  and Xue}}]{Li@2013}
\bibinfo{author}{\bibfnamefont{K.}~\bibnamefont{Li}},
  \bibinfo{author}{\bibfnamefont{H.}~\bibnamefont{Xie}},
  \bibinfo{author}{\bibfnamefont{J.}~\bibnamefont{Liu}},
  \bibinfo{author}{\bibfnamefont{Z.}~\bibnamefont{Ma}},
  \bibinfo{author}{\bibfnamefont{Y.}~\bibnamefont{Zhou}}, \bibnamefont{and}
  \bibinfo{author}{\bibfnamefont{D.}~\bibnamefont{Xue}},
  \bibinfo{journal}{Phys. Chem. Chem. Phys.} \textbf{\bibinfo{volume}{15}},
  \bibinfo{pages}{17658} (\bibinfo{year}{2013}).

\bibitem[{\citenamefont{Wu et~al.}(2007)\citenamefont{Wu, Zhao, Xiang, Hao,
  Liu, and Meng}}]{Wu@2007}
\bibinfo{author}{\bibfnamefont{Z.-J.} \bibnamefont{Wu}},
  \bibinfo{author}{\bibfnamefont{E.-J.} \bibnamefont{Zhao}},
  \bibinfo{author}{\bibfnamefont{H.-P.} \bibnamefont{Xiang}},
  \bibinfo{author}{\bibfnamefont{X.-F.} \bibnamefont{Hao}},
  \bibinfo{author}{\bibfnamefont{X.-J.} \bibnamefont{Liu}}, \bibnamefont{and}
  \bibinfo{author}{\bibfnamefont{J.}~\bibnamefont{Meng}},
  \bibinfo{journal}{Phys. Rev. B} \textbf{\bibinfo{volume}{76}},
  \bibinfo{pages}{054115} (\bibinfo{year}{2007}).

\bibitem[{\citenamefont{Mouhat and Coudert}(2014)}]{Mouhat@2014}
\bibinfo{author}{\bibfnamefont{F.}~\bibnamefont{Mouhat}} \bibnamefont{and}
  \bibinfo{author}{\bibfnamefont{F.-X.} \bibnamefont{Coudert}},
  \bibinfo{journal}{Phys. Rev. B} \textbf{\bibinfo{volume}{90}},
  \bibinfo{pages}{224104} (\bibinfo{year}{2014}).

\bibitem[{\citenamefont{Stournara et~al.}(2012)\citenamefont{Stournara, Guduru,
  and Shenoy}}]{Stournara@2012}
\bibinfo{author}{\bibfnamefont{M.~E.} \bibnamefont{Stournara}},
  \bibinfo{author}{\bibfnamefont{P.~R.} \bibnamefont{Guduru}},
  \bibnamefont{and} \bibinfo{author}{\bibfnamefont{V.~B.}
  \bibnamefont{Shenoy}}, \bibinfo{journal}{Journal of Power Sources}
  \textbf{\bibinfo{volume}{208}}, \bibinfo{pages}{165} (\bibinfo{year}{2012}).

\bibitem[{\citenamefont{Zhang et~al.}(2015)\citenamefont{Zhang, Ma, Wang, Zou,
  Lei, Pan, and Lu}}]{Zhang@2015}
\bibinfo{author}{\bibfnamefont{P.}~\bibnamefont{Zhang}},
  \bibinfo{author}{\bibfnamefont{Z.}~\bibnamefont{Ma}},
  \bibinfo{author}{\bibfnamefont{Y.}~\bibnamefont{Wang}},
  \bibinfo{author}{\bibfnamefont{Y.}~\bibnamefont{Zou}},
  \bibinfo{author}{\bibfnamefont{W.}~\bibnamefont{Lei}},
  \bibinfo{author}{\bibfnamefont{Y.}~\bibnamefont{Pan}}, \bibnamefont{and}
  \bibinfo{author}{\bibfnamefont{C.}~\bibnamefont{Lu}}, \bibinfo{journal}{RSC
  Adv.} \textbf{\bibinfo{volume}{5}}, \bibinfo{pages}{36022}
  (\bibinfo{year}{2015}).

\bibitem[{\citenamefont{Chen et~al.}(2008)\citenamefont{Chen, Bull, Roy,
  Mukaibo, Nara, Momma, Osaka, and Shacham-Diamand}}]{Chen@2008}
\bibinfo{author}{\bibfnamefont{J.}~\bibnamefont{Chen}},
  \bibinfo{author}{\bibfnamefont{S.~J.} \bibnamefont{Bull}},
  \bibinfo{author}{\bibfnamefont{S.}~\bibnamefont{Roy}},
  \bibinfo{author}{\bibfnamefont{H.}~\bibnamefont{Mukaibo}},
  \bibinfo{author}{\bibfnamefont{H.}~\bibnamefont{Nara}},
  \bibinfo{author}{\bibfnamefont{T.}~\bibnamefont{Momma}},
  \bibinfo{author}{\bibfnamefont{T.}~\bibnamefont{Osaka}}, \bibnamefont{and}
  \bibinfo{author}{\bibfnamefont{Y.}~\bibnamefont{Shacham-Diamand}},
  \bibinfo{journal}{Journal of Physics D: Applied Physics}
  \textbf{\bibinfo{volume}{41}}, \bibinfo{pages}{025302}
  (\bibinfo{year}{2008}).

\bibitem[{\citenamefont{Ranganathan and
  Ostoja-Starzewski}(2008)}]{Ranganathan@2008}
\bibinfo{author}{\bibfnamefont{S.~I.} \bibnamefont{Ranganathan}}
  \bibnamefont{and}
  \bibinfo{author}{\bibfnamefont{M.}~\bibnamefont{Ostoja-Starzewski}},
  \bibinfo{journal}{Phys. Rev. Lett.} \textbf{\bibinfo{volume}{101}},
  \bibinfo{pages}{055504} (\bibinfo{year}{2008}).

\bibitem[{\citenamefont{Felice et~al.}(1977)\citenamefont{Felice, Trivisonno,
  and Schuele}}]{Felice@1977}
\bibinfo{author}{\bibfnamefont{R.~A.} \bibnamefont{Felice}},
  \bibinfo{author}{\bibfnamefont{J.}~\bibnamefont{Trivisonno}},
  \bibnamefont{and} \bibinfo{author}{\bibfnamefont{D.~E.}
  \bibnamefont{Schuele}}, \bibinfo{journal}{Phys. Rev. B}
  \textbf{\bibinfo{volume}{16}}, \bibinfo{pages}{5173} (\bibinfo{year}{1977}).

\bibitem[{\citenamefont{House and Vernon}(1960)}]{House@1960}
\bibinfo{author}{\bibfnamefont{D.~G.} \bibnamefont{House}} \bibnamefont{and}
  \bibinfo{author}{\bibfnamefont{E.}~\bibnamefont{Vernon}},
  \bibinfo{journal}{British Journal of Applied Physics}
  \textbf{\bibinfo{volume}{11}}, \bibinfo{pages}{254} (\bibinfo{year}{1960}).

\bibitem[{\citenamefont{Yang et~al.}(2016)\citenamefont{Yang, Wang, Peng,
  Bergara, and Ma}}]{Yang@2016}
\bibinfo{author}{\bibfnamefont{G.}~\bibnamefont{Yang}},
  \bibinfo{author}{\bibfnamefont{Y.}~\bibnamefont{Wang}},
  \bibinfo{author}{\bibfnamefont{F.}~\bibnamefont{Peng}},
  \bibinfo{author}{\bibfnamefont{A.}~\bibnamefont{Bergara}}, \bibnamefont{and}
  \bibinfo{author}{\bibfnamefont{Y.}~\bibnamefont{Ma}},
  \bibinfo{journal}{Journal of the American Chemical Society}
  \textbf{\bibinfo{volume}{138}}, \bibinfo{pages}{4046} (\bibinfo{year}{2016}).

\bibitem[{\citenamefont{Zhang et~al.}(2016{\natexlab{b}})\citenamefont{Zhang,
  Oganov, Zhu, Zhu, Stavrou, and Goncharov}}]{Zhang_KCl@2016}
\bibinfo{author}{\bibfnamefont{W.}~\bibnamefont{Zhang}},
  \bibinfo{author}{\bibfnamefont{A.~R.} \bibnamefont{Oganov}},
  \bibinfo{author}{\bibfnamefont{Q.}~\bibnamefont{Zhu}},
  \bibinfo{author}{\bibfnamefont{Q.}~\bibnamefont{Zhu}},
  \bibinfo{author}{\bibfnamefont{E.}~\bibnamefont{Stavrou}}, \bibnamefont{and}
  \bibinfo{author}{\bibfnamefont{A.~F.} \bibnamefont{Goncharov}},
  \bibinfo{journal}{Scientific Reports} \textbf{\bibinfo{volume}{6}},
  \bibinfo{pages}{26265} (\bibinfo{year}{2016}{\natexlab{b}}).

\end{thebibliography}

\end{document}